\documentclass[twocolumn]{aastex631}
\usepackage{graphicx}
\usepackage{enumerate, enumitem}
\usepackage{amssymb, amsmath, amsfonts}
\usepackage{gensymb}
\usepackage{mathtools}
\usepackage{hyperref}
\usepackage{xspace}
\usepackage{comment}
\usepackage{xcolor}
\usepackage{ulem}
\usepackage{lipsum} 

\newcommand{\masyr}{\hbox{mas\,yr$^{-1}$}}

\newcommand{\Teff}{\mbox{$T_{\rm eff}$}}
\newcommand{\Lsun}{\mbox{$L_{\sun}$}}
\newcommand{\Lbol}{\mbox{$L_{\rm bol}$}}
\newcommand{\Msun}{\mbox{$M_{\sun}$}}

\newcommand{\Mjup}{\mbox{$M_{\rm Jup}$}}

\newcommand{\RHK}{\mbox{$R^{\prime}_{\rm HK}$}}
\newcommand{\CHs}{\mbox{${\it CH}_4s$}}
\newcommand{\pmoffs}[2]{^{+ #1}_{- #2}}

\newcommand{\HST}{\textsl{HST}\xspace}
\newcommand{\Hipparcos}{\textsl{Hipparcos}\xspace}
\newcommand{\hipparcos}{\textsl{Hipparcos}\xspace}
\newcommand{\Gaia}{\textsl{Gaia}\xspace}
\newcommand{\gaia}{\textsl{Gaia}\xspace}

\newcommand{\addition}[1]{\textcolor{black}{#1}} 

\newcommand{\orbitcodename}{{\tt orvara}\xspace}
\newcommand{\htofcodename}{\texttt{htof}\xspace}
\newcommand{\htofversion}{\href{https://github.com/gmbrandt/HTOF/tree/0.3.4}{0.3.4}\xspace}
\newcommand{\orbitcodeversion}{githash \href{https://github.com/t-brandt/orbit3d/commit/d66fce10cf8a08efd9fa4cdbf48e8e9a279d7386}{d66fce10cf8a08efd9fa4cdbf48e8e9a279d7386}}

\accepted{Sep 14, 2021}

\shorttitle{Improved Dynamical Masses with Hipparcos--Gaia EDR3 Accelerations}
\shortauthors{G. M. Brandt et al.}
\submitjournal{AJ}

\begin{document}

\title{
Improved Dynamical Masses for Six Brown Dwarf Companions Using \hipparcos and \Gaia EDR3}

\author[0000-0003-0168-3010]{G.~Mirek Brandt}
\altaffiliation{NSF Graduate Research Fellow}
\affiliation{Department of Physics, University of California, Santa Barbara, Santa Barbara, CA 93106, USA}

\author[0000-0001-9823-1445]{Trent J.~Dupuy}
\affiliation{Institute for Astronomy, University of Edinburgh, Royal Observatory, Blackford Hill, Edinburgh, EH9 3HJ, UK}

\author[0000-0002-6845-9702]{Yiting Li}
\affiliation{Department of Physics, University of California, Santa Barbara, Santa Barbara, CA 93106, USA}

\author{Minghan Chen}
\affiliation{Department of Physics, University of California, Santa Barbara, Santa Barbara, CA 93106, USA}

\author[0000-0003-2630-8073]{Timothy D.~Brandt}
\affiliation{Department of Physics, University of California, Santa Barbara, Santa Barbara, CA 93106, USA}

\author[0000-0001-9195-7390]{Tin Long Sunny Wong}
\affiliation{Department of Physics, University of California, Santa Barbara, Santa Barbara, CA 93106, USA}

\author[0000-0002-7405-3119]{Thayne Currie}
\affiliation{Subaru Telescope, National Astronomical Observatory of Japan, 
650 North A`oh$\bar{o}$k$\bar{u}$ Place, Hilo, HI  96720, USA}
\affiliation{NASA-Ames Research Center, Moffett Blvd., Moffett Field, CA, USA}
\affiliation{Eureka Scientific, 2452 Delmer Street Suite 100, Oakland, CA, USA}

\author[0000-0003-2649-2288]{Brendan P. Bowler}
\affiliation{Department of Astronomy, The University of Texas at Austin, 2515 Speedway, C1400, Austin, TX 78712, USA}

\author[0000-0003-2232-7664]{Michael C. Liu}
\affiliation{Institute for Astronomy, University of Hawaii at Manoa, Honolulu, HI 96822, USA}

\author[0000-0003-0562-1511]{William M. J. Best}
\affiliation{Department of Astronomy, The University of Texas at Austin, 2515 Speedway, C1400, Austin, TX 78712, USA}

\author[0000-0001-6041-7092]{Mark W. Phillips}
\affiliation{Astrophysics Group, University of Exeter, Physics Building, Stocker Road, Devon EX4 4QL, UK}

\begin{abstract}
We present comprehensive orbital analyses and dynamical masses for the substellar companions Gl~229~B, Gl~758~B, HD~13724~B, HD~19467~B, HD~33632~Ab, and HD~72946~B. Our dynamical fits incorporate radial velocities, relative astrometry, and most importantly calibrated \hipparcos-\Gaia EDR3 accelerations. For HD~33632~A and HD~72946 we perform three-body fits that account for their outer stellar companions. We present new relative astrometry of Gl~229~B with Keck/NIRC2, extending its observed baseline to 25 years. We obtain a $<$1\% mass measurement of $71.4 \pm 0.6\,\Mjup$ for the first T dwarf Gl~229~B and a 1.2\% mass measurement of its host star ($0.579 \pm 0.007\,\Msun$) that \addition{agrees} with the high-mass-end of the M dwarf mass-luminosity relation. We perform a homogeneous analysis of the host stars' ages and use them, along with the companions' measured masses and luminosities, to test substellar evolutionary models. Gl~229~B is the most discrepant, as models predict that an object this massive cannot cool to such a low luminosity within a Hubble time, implying that it may be an unresolved binary. The other companions are generally consistent with models, except for HD~13724~B that has a host-star activity age 3.8$\sigma$ older than its substellar cooling age. Examining our results in context with other mass--age--luminosity benchmarks, we find no trend with spectral type but instead note that younger or lower-mass brown dwarfs are over-luminous compared to models, while older or higher-mass brown dwarfs are under-luminous. \addition{The presented mass measurements for some companions are so precise that the stellar host ages, not the masses, limit the analysis.}

\end{abstract}

\keywords{---}




\section{Introduction}\label{sec:intro}
Brown dwarfs (BDs) are substellar objects with masses below the hydrogen-fusion mass limit of 75--$80\,\Mjup$ \citep{Burrows_2001RvMP...73..719B, Dupuy+Liu_2017}. Sufficiently massive to fuse deuterium but not hydrogen, they cool as they age. A BD has a convective interior coupled to an atmosphere that contains chemically diverse clouds with detailed interactions and opacities \citep{Marley_2015_BD_review}. The atmosphere modulates the BD's cooling, affecting its present-day spectrum, effective temperature, and luminosity \citep[e.g.,][]{Saumon+Marley_2008}.

A rich variety of atmospheric and evolutionary models have been constructed that predict the radii, spectra, and luminosities of BDs as functions of their age and composition (e.g.,~\citealt[]{2001AMESCOND, Baraffe+Chabrier+Barman+etal_2003, Saumon+Marley_2008, Spiegel+Burrows_2012, phillips_2020_atmo_models}). BDs all have similar Jupiter-sized radii after initial contraction finishes. However, a fundamental degeneracy exists whereby older and more massive BDs can have similar temperatures and luminosities to younger and less massive BDs \citep{1997ApJ...482..442B, Marleau+Cumming_2014}.  
This degeneracy between age, luminosity, and mass has to be broken to test evolutionary models. Possessing all three allows one to constrain BD properties and/or the physics of their cooling. Independent measures of planet age, luminosity, and mass for young ($\lesssim$500 Myr) giant planets, like those in $\beta$~Pictoris or HR~8799, allows one to potentially constrain their initial entropy at formation \citep{Marley+Fortney+Hubickyj_2007, Marleau+Cumming_2014}.

Direct-imaging instruments such as Subaru/CHARIS \citep{CHARIS_INST_PAPER1, CHARIS_INST_PAPER2}, VLT/SPHERE \citep{SPHERE_INST_PAPER}, Gemini/GPI \citep{GPI_INST_PAPER}, Keck/NIRC2 \citep{NIRC2_INST_PAPER, KECK_AO_System}, and recently interferometers like VLT/GRAVITY (e.g., \citealp{AMLagrange2020betapicc_direct_detection, Nowak_2020_beta_pic_c_direct_detection}) allow the measurement of spectra and luminosities with newfound precision for BD companions to nearby stars.
Because their host stars are bright and nearby, they often have well-measured metallicities and age indicators. 
However, these BDs are typically on wide orbits with long orbital periods. Radial velocity (RV) time series typically cover a small fraction of the orbit. Thus, the most difficult quantity to measure is usually the mass of the BD.

In recent years, observations of the acceleration of the host star in the plane of the sky, a.k.a. its astrometric acceleration, have allowed precise mass measurements for systems with long-period ($\gtrsim$5 years) giant planets and BDs \citep{2001ApJ_Zucker_Shay, 2010A&A...509A.103S, 2011A&A...527A.140R, Sahlmann+Segransan+Queloz_2011, Brandt_Dupuy_Bowler_2018, Calissendorff+Janson_2018, brandt_gliese_229b_mass_htof, Dupuy+Brandt+Kratter+etal_2019, 2019AA_Kervella, Nielson+DeRosa+etal+betapicc2019, 2019A&A...622A..96C, 2020ApJ_Currie_Thayne_HD33632, Brandt2020betapicbc, 2020AA_Maire_HD19467, Bowler2021_HD47127}.
Differences in position and proper motion between the \hipparcos \citep{HIP_TYCHO_ESA_1997,vanLeeuwen_2007} and \gaia \citep{Gaia_General_2016} missions can detect accelerations as small as several $\mu$as\,yr$^{-2}$
\citep{2016IAUFM..29A.217S, brandt_cross_cal_gaia_2018}. 
\hipparcos-\gaia accelerations can measure precise masses and orbits from first principles, even for long-period systems, and break the mass-inclination degeneracy inherent in RV analyses.

The latest \gaia data release, \Gaia EDR3, yields proper motions that are on average a factor of 3--4 times more precise than those from \gaia DR2 \citep{2020GaiaEDR3_catalog_summary,Lindegren+Klioner+Hernandez+etal_2020, BrandtGaiaEDR3HGCA}. This precision improvement allows for even stronger mass constraints for most directly imaged sources. 
In this work, we use the improved astrometry from \gaia EDR3, as published and calibrated in the EDR3 version of the \Hipparcos-\Gaia Catalog of Accelerations (HGCA; \citealt{BrandtGaiaEDR3HGCA}), to produce the most precise orbits and companion masses to date for six systems: Gl~229, Gl~758, HD~13924, HD~19467, HD~33632, and HD~72946. We use the Markov-Chain Monte-Carlo (MCMC) orbit-fitting code \orbitcodename \citep{TimOrbitFitTemporary}.

These six systems all have directly imaged BD companions on $\approx$15 to 500 year orbital periods, long-term precision RVs, and significant astrometric accelerations \citep{brandt_cross_cal_gaia_2018, BrandtGaiaEDR3HGCA}.

We structure the paper as follows. In Section \ref{sec:stellar_ages}, we compute Bayesian activity-based age estimates for all the stars in the sample except Gl~229~A, which only has upper and lower age limits. Section \ref{sec:rv_imaging} provides an overview of the RV and imaging data sets we use. Section \ref{sec:gaia_astrometry} reviews the absolute astrometry from \gaia and \hipparcos, as calibrated and published in the \gaia EDR3 version of the HGCA \citep{BrandtGaiaEDR3HGCA}. Section \ref{sec:orbitfitting} introduces our orbit fitting procedure, models, and priors. We discuss the orbit-fitting results in Section \ref{sec:results} and compare dynamical and model masses for HD~13724~A and Gl~229~A in Section \ref{sec:stellar_masses_from_mesa}.
Section \ref{sec:bdmodels} compares our BD companion dynamical masses to predictions from evolutionary models. We conclude in Section \ref{sec:conclusions}.

We denote posteriors by $m\pmoffs{u}{l}$, where $u$ and $l$ give the 68.3\% confidence interval about the median value $m$. We report $m \pm 1\sigma$ if $u$ and $l$ are approximately equal within the quoted precision.
HGCA v.EDR3 refers to the \gaia EDR3 version of the catalog, and HGCA v.DR2 to the original \gaia DR2 version \citep{brandt_cross_cal_gaia_2018}.

\section{Stellar Ages}\label{sec:stellar_ages}

Five of our six targets are main sequence, approximately solar-mass stars: Gl~758~A, HD~13724~A, HD~19467~A, HD~33632~A, and HD~72946~A.
In this Section we present uniform analyses of their ages based on activity and rotation.  Stellar ages, and therefore companion ages, will enable us to compare BD observables with predictions from evolutionary models at our measured dynamical masses.

\addition{A star's age can be constrained with gyrochronology}; G and K dwarfs lose angular momentum through their magnetized winds as they age \citep{gyro_review1, gyro_review2}. Activity indices tied to stellar rotation constrain the Rossby number and thereby the age. We convert the Rossby number to a rotation period using the convective overturn time given in \cite{Noyes_1984}.  Finally, we convert the rotation period to an age according to the calibration of \cite{Mamajek+Hillenbrand_2008}.

\addition{We adopt the Bayesian activity-age method that is described in} detail in \cite{Brandt+Kuzuhara+McElwain+etal_2014}, and further explained in \citet{2021_Li_Yiting_nine_masses_RV_planets}. Our method is identical to the latter work, but we summarize it here and the data involved. We use both the chromospheric activity index \RHK\ and the X-ray activity index $R_X$ to infer a Rossby number. The $R_X$ measurements come from the \textsl{ROSAT} all-sky survey catalogs \citep{Voges_1999,Voges+Aschenbach+Boller+etal_2000}. Some stars have only upper limits on X-ray fluxes; we compute these as $5\sigma$ values assuming the uncertainty from the nearest detection in the ROSAT faint source catalog \citep{Voges+Aschenbach+Boller+etal_2000}. The Ca\,{\sc ii} S-indices are from \cite{Pace_2013} and references therein (most sources have multiple measured S-indices). The method of \cite{Brandt+Kuzuhara+McElwain+etal_2014} uses the average of the maximum and minimum S-indices found in the literature (in the Mt. Wilson system). We convert these indices to Mt.~Wilson \RHK\ with the relations from \cite{Noyes_1984}. The \RHK\ and $R_X$ values are tabulated in Table \ref{tab:stellar_params}.

\begin{deluxetable*}{lccccccc}
\tablewidth{0pt}
    \tablecaption{Input stellar parameters for the Bayesian age analyses.\label{tab:stellar_params}}
    \tablehead{\colhead{Identifier} & \colhead{$B_{\rm T}$}  &  \colhead{$\sigma_{B_{\rm T}}$} & \colhead{$V_{\rm T}$} & \colhead{$\sigma_{V_{\rm T}}$} & \colhead{$\log{R_X}$} & \colhead{$\log{\RHK}$} & \colhead{$P_{\rm rot}$} \\ 
    \colhead{} & \colhead{(mag)}   & \colhead{(mag)}  & \colhead{(mag)}  & \colhead{(mag)}  & \colhead{(dex)} & \colhead{(dex)} & \colhead{(days)}
    }
    \startdata
    Gl 758 & 7.374 & 0.015 & 6.447 & 0.010 &  $<-5.04$ &  $-5.05$ & \nodata \\
    HD 19467 & 7.788 & 0.015 &  7.043 & 0.010 &  $<-4.75$ &  $-4.97$ & \nodata \\
    HD 13724 & 8.712 & 0.017 &  7.948 & 0.012 &  $<-5.13$ &  $-4.78$ & 21, 25.76 \\
    HD 33632A & 7.102 & 0.015 &  6.530 & 0.010 &   $-5.55$ &  $-4.83$ & \nodata \\
    HD 72946 & 7.933 & 0.017 &  7.159 & 0.011 &  $-4.80$ &  $-4.68$ & \nodata \\
    \enddata
        \tablecomments{\, $<$ denotes a $5\sigma$ upper bound on $R_X$.}
\end{deluxetable*}

\addition{In Table \ref{tab:stellar_params}, each source's $B_{\rm T}$-band and $V_{\rm T}$-band magnitude comes from the Tycho-2 catalogue \citep{Hog_2000}}. We denote the errors on the magnitudes with $\sigma$, e.g., $\sigma_{V_{\rm T}}$. \addition{We convert the B and V Tycho filters to Johnson B and V, using the transformations from Volume 1 of \citealp{HIP_TYCHO_ESA_1997}. We then use B-V and the activity indices to deduce a stellar age as in \citet{Brandt+Kuzuhara+McElwain+etal_2014}, providing a stellar rotation period when available.} Only HD~13724~A has measured periodic, photometric variability, with rotation periods ranging from 21~days \citep{Arriagada_2011} to 25.76~days \citep{2018Cat_Oelkers_HD13724_rotationPeriod}. Directly measured rotation periods do not require estimates of the Rossby number or convective overturn time and enable tighter constraints on the stellar age \citep[e.g.][]{Mamajek+Hillenbrand_2008, Brandt_Dupuy_Bowler_2018}. We incorporate these rotation periods as described by \cite{Brandt+Kuzuhara+McElwain+etal_2014}.

\begin{figure*}
    \centering
    \includegraphics[width=0.3\textwidth]{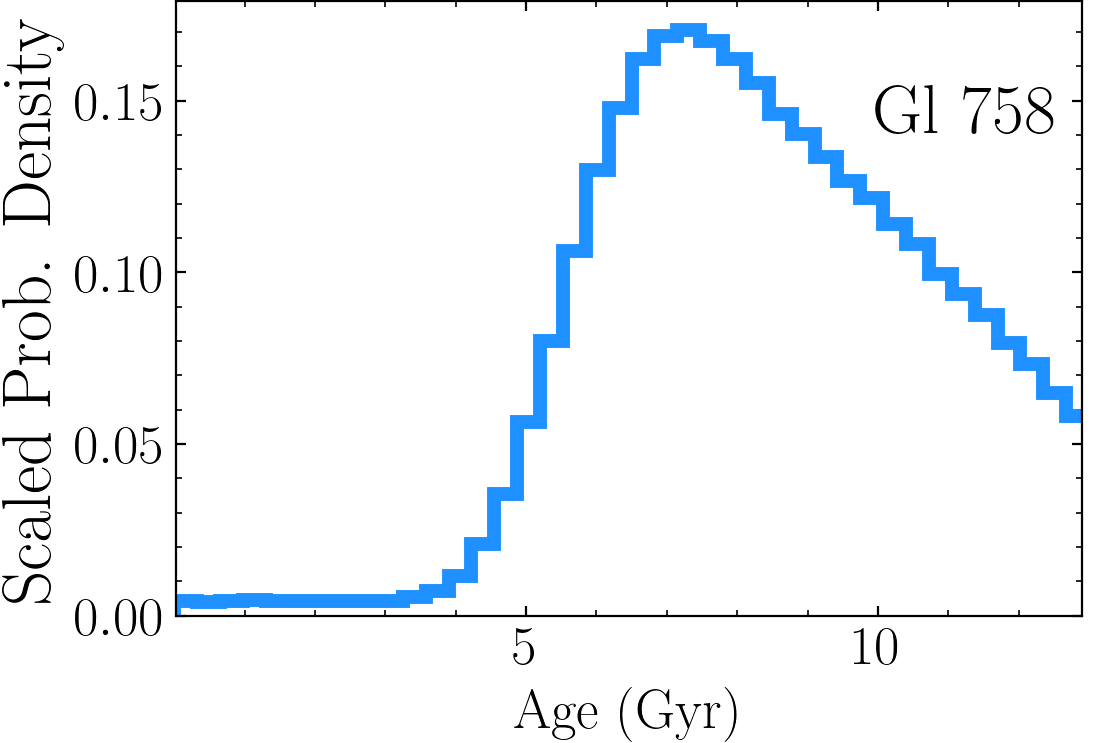}
    \includegraphics[width=0.275\textwidth]{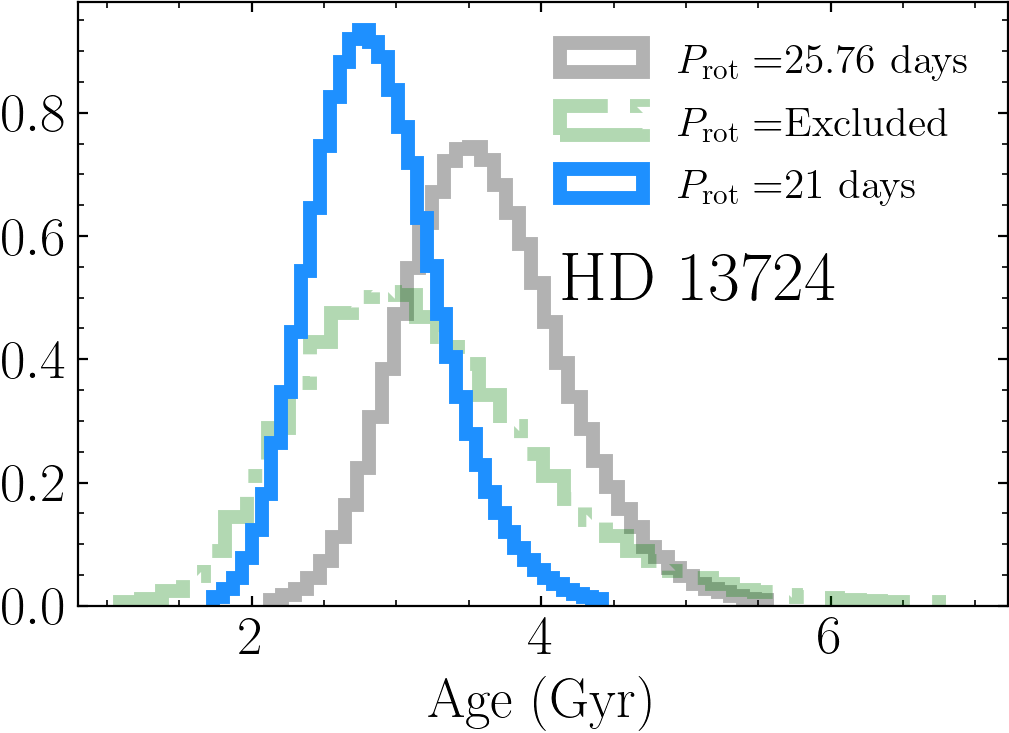}
    \includegraphics[width=0.3\textwidth]{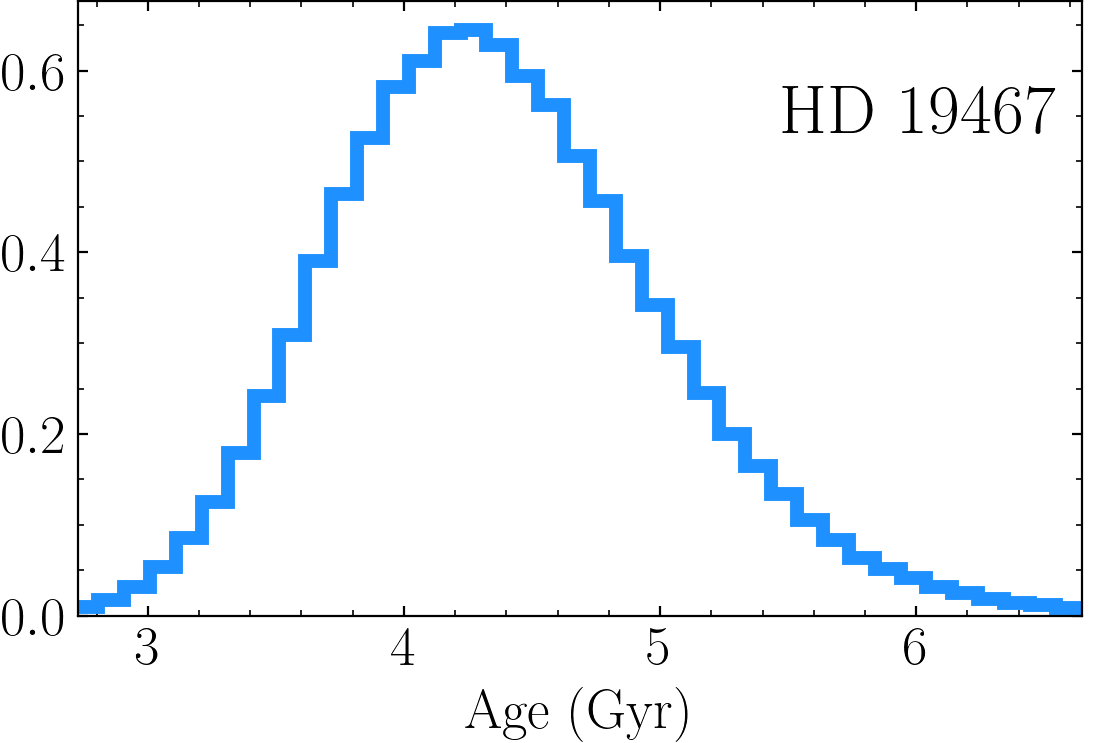}
    \includegraphics[width=0.3\textwidth]{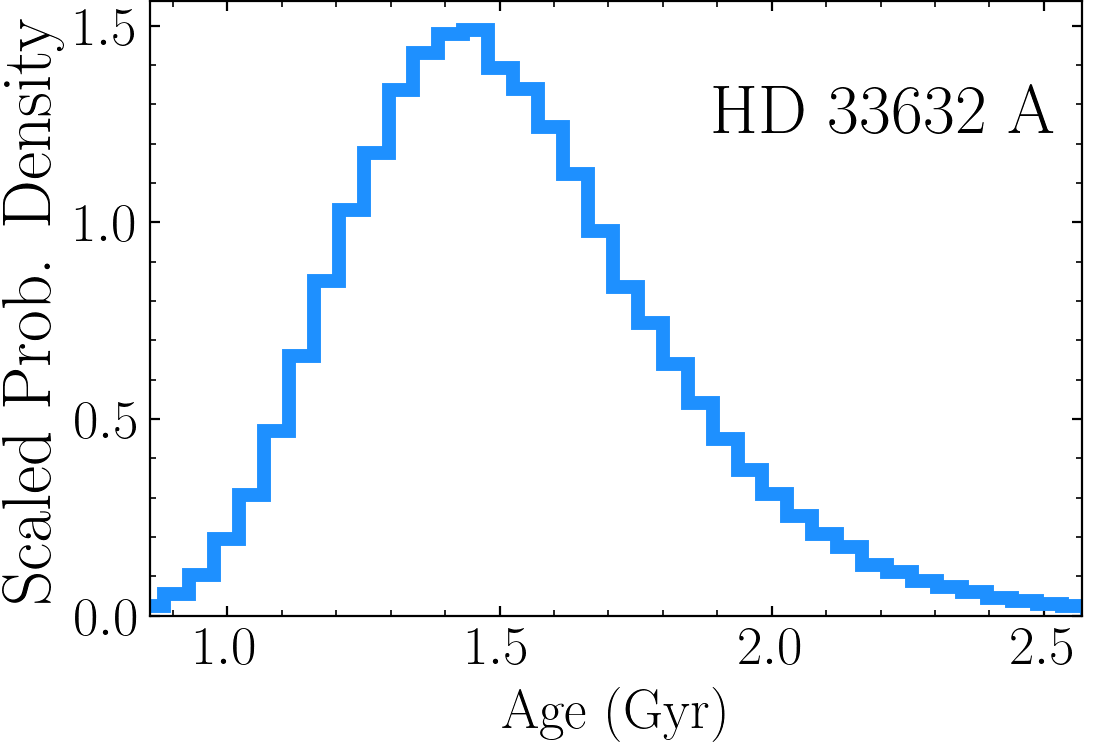}
    \includegraphics[width=0.3\textwidth]{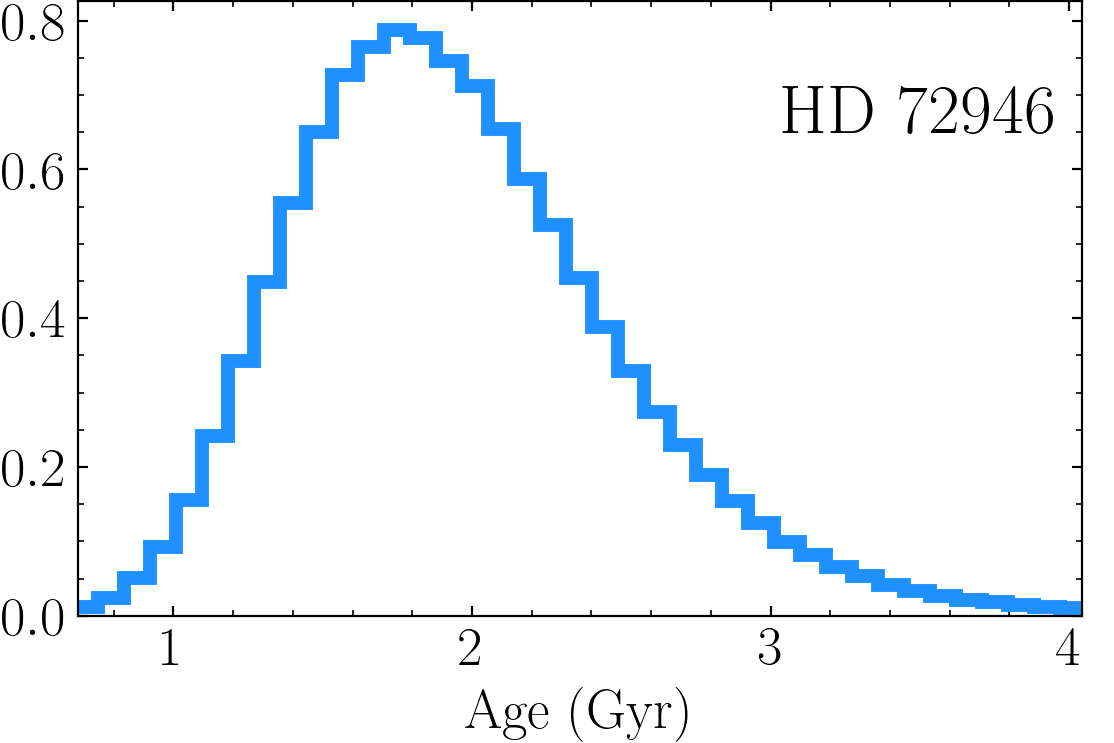}
    \caption{Posteriors from the Bayesian stellar age analyses using the method of \cite{Brandt+Kuzuhara+McElwain+etal_2014}. The median values and confidence intervals are listed in Table \ref{tab:stellar_ages}. The parameters used in the analyses are listed in Table \ref{tab:stellar_params}. HD~13724 (middle panel of the top row) had three stellar rotation periods considered: 21 days from \cite{Arriagada_2011}, 25.76 days from \cite{2018Cat_Oelkers_HD13724_rotationPeriod}, and a case where period information was neglected (labelled ``Excluded'').}
    \label{fig:stellar_ages}
\end{figure*}

The resulting stellar age posteriors are shown in Figure \ref{fig:stellar_ages}. We tabulate the median and 68.3\% confidence intervals in Table \ref{tab:stellar_ages}. In the following subsection, we compare our age estimates with other results in the literature.

\begingroup
\renewcommand{\arraystretch}{1.5}
\begin{deluxetable}{cCc}
\tablewidth{0pt}
    \tablecaption{Posterior values from the Bayesian stellar age estimates.\label{tab:stellar_ages}}
    \tablehead{\colhead{Identifier} & \colhead{Age Posterior (Gyr)} & \colhead{Notes}}
    \startdata
    Gl 758~A & 8.3\pmoffs{2.7}{2.1} & \nodata \\
    HD 13724~A & 2.8\pmoffs{0.5}{0.4} & \textit{a} \\
    HD 13724~A & 3.6\pmoffs{0.6}{0.5} & \textit{b} \\
    HD 13724~A & 3.1\pmoffs{0.9}{0.7} & \textit{c} \\
    HD 72946~A & 1.9\pmoffs{0.6}{0.5} & \nodata\\
    HD 33632 A & 1.7\pm 0.4 & \nodata \\
    HD 19467~A & 5.4\pmoffs{1.9}{1.3} & \nodata \\
    \enddata
        \tablenotetext{a}{This is our fiducial case using the \cite{Arriagada_2011} 21-day  rotation period.}
        \tablenotetext{b}{This uses the 25.76 day rotation period from \citet{2018Cat_Oelkers_HD13724_rotationPeriod}.}
        \tablenotetext{c}{This \addition{estimate does not involve a} rotation period as an input parameter.}
\end{deluxetable}
\endgroup

\subsection{Discussion on the ages of individual stars}

The ages and masses of our six BD host stars have been extensively studied (see for instance \citealt[]{2011AA...530A.138C}, \citealt[]{2021AA...646A..77G} and references therein). Here, we place our results within the context of previous age estimates.  We begin with a discussion of the age of Gl~229.  As an early-M dwarf, we excluded it from our re-analysis.

\paragraph{\it Gl~229 }
The ages of M dwarfs, like Gl~229, are hard to determine because of their extremely long main-sequence lifetimes (see, e.g., \citealt[]{2008AJ.West.135..785W}). 
\cite{brandt_gliese_229b_mass_htof} suggested an age of $2.6 \pm 0.5$ Gyr based on stellar activity but noted that the activity-age relation is poorly calibrated for M dwarfs.  They ultimately adopted a pair of wide uniform priors on the age (considering ages between 1 and 10 Gyr). In Section \ref{sec:bdmodels}, we reconsider the age in light of our new dynamical mass and adopt a prior uniform between 1 and 10 Gyr. This is to highlight the significant disagreement between modern models and Gl~229~B's high mass, at all reasonable ages.

\paragraph{\it Gl 758 }
This G8V star \citep{Maldonado+Eiroa+Villaver+etal_2012} is favored to be old. \cite{Brandt_Dupuy_Bowler_2018} inferred an activity-based age (using the same \citealt{Brandt+Kuzuhara+McElwain+etal_2014} method as we do now) that favored old ages $\gtrsim$6 Gyr with a long tail to $\approx$13 Gyr. We infer here a nearly identical posterior of $8.3\pmoffs{2.7}{2.1}$ Gyr and between 0.7~Gyr and 11.5~Gyr with 99.7\% confidence. This age is broadly consistent with all values in the literature.
\cite{2011AA...530A.138C} found the age between 4.53 and 12.06 Gyr (16\% and 84\% confidence intervals) with Padova isochrones \citep{2008AA...484..815B_padova1, 2009AA...508..355B_padova2} and a significantly older age between 8.5 and 13.4\,Gyr using BASTI isochrones \citep{basti_1_2004ApJ...612..168P, basti_2_2006ApJ...642..797P, basti_3_2009ApJ...697..275P}. \cite{Pace_2013} adopted an age of $11.16 \pm 2.46$ Gyr, albeit based on \cite{2011AA...530A.138C}. More recently, \cite{Luck_2017} re-examined the age of Gl~758 and explored the best-fit age using a wide variety of isochrones. They inferred 6.42 Gyr using the earlier, \cite{1994AAS..106..275B} isochrones; 5.31 Gyr using the \cite{2004ApJS..155..667D} isochrones that implemented (at the time) an improved prescription for convective core-overshoot; and 7.72 Gyr with the Dartmouth stellar evolution database \citep{2008ApJS..178...89D}. Most recently, \cite{Bowler_2020_Blunt_Nielsen} argued for minimum and maximum ages of 6 and 10 Gyr, respectively, from various age determinations in the literature.
We adopt the age prior shown in Figure \ref{fig:stellar_ages} for the BD model analysis in Section~\ref{sec:bdmodels}.

\paragraph{\it HD 13724 }
HD~13724~A is a G3/G5V dwarf \citep{2001KFNT...17..409K}. HD~13724~A's measured \RHK\ and $R_X$, combined with the 21-day period from \cite{Arriagada_2011} (consistent with the most-recent $20.2 \pm 1.2$\,day period derived by \citealp{2019AA_RickmanHD13724_discovery}), yield a precise age of $2.8\pmoffs{0.5}{0.4}$\,Gyr. We use this age in our comparisons to BD models in Section~\ref{sec:bdmodels}.

We infer a slightly older age of $3.6\pmoffs{0.6}{0.5}$ Gyr if we instead adopt the 25.76 day rotation period from \cite{2018Cat_Oelkers_HD13724_rotationPeriod}. The measured rotation periods favors old ages, however, the star's activity (neglecting any rotation period information) gives a similarly old age of $3.1\pmoffs{0.9}{0.7}$ Gyr. 
All three of our age estimates are much older than the $1.04 \pm 0.88$ Gyr found by \cite{Rickman2020_HD13724}, which was inferred from grids of Geneva stellar models \citep{Ekstrom_etal_2012, Georgy_etal_2013}. 

There is a lack of consensus within the literature on the age of HD~13724, albeit younger ages seem to be favored. Most recently, \cite{2021AA...646A..77G} report a posterior of $0.47 \pm 0.36$ Gyr, and \cite{2019AA...624A..78D_HD13724} report an age posterior of $1.11 \pm 0.98$ Gyr. Results from \cite{2011AA...530A.138C} are more consistent with our analyses that include the rotation period. \cite{2011AA...530A.138C} use the Padova isochrones to constrain the age to between 0.94 and 5.51 Gyr (16\% and 84\% confidence intervals)--- a wide range that encompasses all of the aforementioned age posteriors. Their 5\% and 95\% confidence intervals on the age are 0.27 and 7.18 Gyr. \cite{2011AA...530A.138C} found similarly wide posteriors using the BASTI isochrones. \cite{Stanford-Moore+Nielsen+DeRosa+etal_2020} infer an old age, similar to our own, based on stellar activity: centered on 5 Gyr, and between 1.4 and 12 Gyr with 95\% confidence. If HD~13724 is young, it is an unusually inactive star and a slow rotator for its age.

HD~13724 could have an anomalously high surface metallicity that skews stellar-evolution inferred ages to young values (gravitational settling depletes surface metallicity as Solar-type stars age; \citealp{Thoul1994}). However, we show in Section \ref{sec:bdmodels} that the BD age constraints favor a $\approx$1 Gyr age for HD~13724~A (close to that assumed by \citealt{Rickman2020_HD13724}), and that our rotation period-informed age of $2.8\pmoffs{0.5}{0.4}$ Gyr is 3$\sigma$ inconsistent with the inferred BD age. This would make its rotation period of 20-30 days much slower than that expected from gyrochronology, and render HD~13724~A an interesting test case for gyrochronology in G dwarfs. 


\paragraph{\it HD 19467}
This G3V star \citep{2021AA...646A..77G} has only an upper bound on its X-ray activity index and a chromospheric activity slightly less than Solar \citep{2010ApJ...725..875I, 2021AA...646A..77G}. The activity of this solar-type star (e.g., \citealt[]{2017AA...604A.108M}) points to HD~19467~A being nearly a solar twin. We infer an activity-age of $5.4\pmoffs{1.9}{1.3}$ Gyr, with a 95\% confidence intervals of 3.4 to 9.2 Gyr. This agrees well with the gyrochronology estimate of $5.6 \pm 0.8$ Gyr derived by \cite{2020AA_Maire_HD19467} from ASAS photometry. 

Our activity-based age is slightly younger than most isochronal estimates in the literature, but generally consistent within 1--2$\sigma$. For example, \cite{2011AA...530A.138C} report $8.7 \pm 3.4$ Gyr, and \cite{2018AA...619A..73L} give an activity age of $8.8 \pm 0.3$ Gyr. However, some estimates prefer even older ages that would be modestly inconsistent with our analysis, e.g., $10.5 \pm 1.9$\,Gyr by \cite{2018AA...614A..55A}. 
Our age agrees with the best-fit activity age of 6.18 Gyr by \cite{2010ApJ...725..875I}. For contrast, the recent dynamical analysis by \cite{2020AA_Maire_HD19467} adopted an age of $8 \pmoffs{2}{1}$ Gyr, favoring isochronal estimates. The median of that estimate is older than what we adopt but within our 95\% confidence interval.



\paragraph{\it HD 33632~A }
HD~33632~A is an F8V star \citep{2012AstL...38..331A_extended_hipcompilation} that is similarly as active as the Sun \citep{Pace_2013, 2017ApJ...835...25E_solar_RHK}. HD~33632~A may be slightly more massive than the Sun (e.g., $1.01 \pm 0.05\,\Msun$ from \citealt[]{2017AA...604A.108M}; $1.03 \pm 0.04\,\Msun$ from \citealt[]{2012ApJ...756...46R}; $1.10\,\Msun$ from \citealt[]{2011AA...530A.138C}). The activity of HD~33632~A implies a young age; magnetic braking has not yet slowed the star significantly. \cite{2010ApJ...725..875I} estimated, from activity, a fast rotation period of 9 days. Combining the $R_X$ activity index and the chromospheric activity, we infer an age of $1.7 \pm 0.4$ Gyr for HD~33632~A.

Other activity age estimates range from 2--5 Gyr, e.g., 3.5 Gyr from \cite{2010ApJ...725..875I} and $3.9^{+3.3}_{-1.8}$ Gyr from \cite{Stanford-Moore+Nielsen+DeRosa+etal_2020}. Isochronal ages favor $\gtrsim2$\,Gyr; with posteriors that are consistent with 1.5\,Gyr. For instance, \cite{2012ApJ...756...46R} determined a best-fit age of $4.8 \pm 1.4$\,Gyr. The analyses by \cite{2011AA...530A.138C} found maximum likelihood ages for HD~33632~A of 2.2\,Gyr and 3.2\,Gyr using Padova and BASTI ischrones, respectively. The 16\% and 84\% confidence interval ages were 1--4.15\,Gyr with Padova and 1.5--4.5\,Gyr using BASTI --- both fully consistent with our $1.7 \pm 0.4$\,Gyr age estimate. The abundance of neutron capture elements provides age estimates of HD~33632~A near $\approx$1.5--2.5\,Gyr \citep{Spina_et_al2017_neutroncapture, 2020ApJ_Currie_Thayne_HD33632}. The recent analysis of HD~33632~Ab by \cite{2020ApJ_Currie_Thayne_HD33632} adopted an age prior of $1.5 \pmoffs{3.0}{0.7}$ Gyr, fully consistent with our $1.7\pm 0.4$\,Gyr age.

\paragraph{\it HD 72946}
This G5V star \citep{2007AN....328..889K} has a variety of age constraints, including from isochrones and lithium abundances \citep{2012ApJ...756...46R, Luck_2017, 2018AA...614A..55A}, that place it anywhere from $0.5$\,Gyr to as old as $\approx$9\,Gyr. There is no measured photometric rotation period for the star, but there are measurements of the X-ray emission \citep{Voges_1999} and of the chromospheric activity \citep{Pace_2013, 2016AA_Bouchy2016_HD72946}. The latter indicates a star marginally more active, and therefore younger, than the Sun. Combining the X-ray and chromospheric activity indices, we infer an age of $1.9\pmoffs{0.6}{0.5}$\,Gyr from our Bayesian analysis. This younger age is consistent with estimates in the literature, albeit literature estimates span a wide range.

\cite{2020AA_Maire_HD72946} is the most similar (to our method) and the most recent age analysis. \cite{2020AA_Maire_HD72946} used the \cite{Mamajek+Hillenbrand_2008} relations with an average chromospheric activity ($\log{\RHK} = -4.60$\,dex, which is slightly more active and thereby younger than our adopted \cite{Gray+Corbally+Garrison+etal_2003, Pace_2013} index of $-4.68$\,dex) to infer a 15-day rotation period, implying an age near 1\,Gyr. However, using an average projected rotational activity, they placed a more stringent upper bound on the rotation period of 12\,days, excluding ages older than 1\,Gyr, or $\approx$1.5\,Gyr given liberal uncertainties. They ultimately chose to adopt  0.8--3\,Gyr as the range of probable ages, which is in excellent agreement with our inferred $1.9\pmoffs{0.6}{0.5}$\,Gyr age.

For further comparison, \cite{2012ApJ...756...46R} derived an isochronal age of $4.9 \pmoffs{4.6}{2.1}$\,Gyr. \cite{2011AA...530A.138C} inferred an age between 1.09 and 9.27\,Gyr (16\% and 84\% confidence intervals) with the Padova isochrones. They found similar results, 1.20 and 9.64\,Gyr, with BASTI isochrones. \cite{2018AA...614A..55A} report an isochronal age of $8.7 \pmoffs{2.6}{4.2}$\,Gyr --- favoring an age much older than our estimate and that by \cite{2020AA_Maire_HD72946} but still marginally consistent with both estimates.

\section{Radial Velocities and Relative Astrometry}\label{sec:rv_imaging}
All six systems have both direct imaging of the BD companions and radial velocity (RV) measurements of the host star. In this Section, we summarize the direct imaging and RV data for each of the sources and present new Keck/NIRC2 imaging of Gl~229~B. Table \ref{tab:relative_astrometry} lists the sources of the relative astrometry we use to fit each system. 

We retrieve the RV data for every source from Vizier.\footnote{\url{https://vizier.u-strasbg.fr/viz-bin/VizieR}} For many of the sources, a large fraction of the RVs come from the HIRES instrument on Keck \citep{Vogt+Allen+Bigelow+etal_1994}, originally published by \cite{Butler+Vogt+Laughlin+etal_2017}. We use the recently recalibrated HIRES data from \cite{TalOrHIRES}. Many other RV measurements come from the HARPS instrument at the European Southern Observatory (ESO) La Silla 3.6-m telescope \citep{2003MsngrHARPS_instrumentpaper}. We use the recently recalibrated HARPS data produced by \cite{TrifonovHarps}. For every source presented in this work except for HD~72946, the RVs do not cover a full orbital period. \hipparcos-\gaia absolute astrometry thus plays a crucial role in constraining the companion's mass and orbit.

\begin{deluxetable}{ccc}
\tablewidth{0pt}
    \tablecaption{Summary of the relative astrometry that we use in our orbital analyses. \label{tab:relative_astrometry}}
    \tablehead{\colhead{Identifiers} & \colhead{Reference} & \colhead{Measurements\tablenotemark{a}}}
    \startdata
    Gl 229 A/B & TB20 & 7 \\
    Gl 229 A/B & Table \ref{tab:relative_astrometry_Gl229New} & 2 \\
    Gl 758 A/B & BB18 & 4 \\
    HD 13724 A/B & R20 & 9 \\
    HD 19467 A/B & TRENDSV & 5 \\
    HD 19467 A/B & C15 & 1 \\
    HD 19467 A/B & BB20 & 1 \\
    HD 72946 A/B & M20 & 2 \\
    HD 72946/\,HD 72945 & \gaia EDR3 & 1 \\
    HD 33632 A/B & \gaia EDR3 & 1 \\
    HD 33632 A/Ab & Table \ref{tab:relative_astrometry_HD33632} & 1 \\
    \enddata
        \tablecomments{(A/B) refers to relative astrometry between A and B. For example, HD~72946~A/B refers to relative astrometry of HD 72946 B about HD~72946~A. The data reference points to the publication where the data are retrievable either in print or through a data source (e.g., Vizier) clearly linked to that publication. We do not reproduce the data here so that data remain consolidated within their original published source.}
        \tablenotetext{a}{The number of pairs of position angle/separation measurements.}
        \tablerefs{\gaia EDR3 -- \cite{Lindegren+Klioner+Hernandez+etal_2020}, M20 -- \cite{2020AA_Maire_HD72946}, TB20 -- \cite{brandt_gliese_229b_mass_htof}, BB18 -- \cite{Bowler+Dupuy+Endl+etal_2018}, R20 -- \cite{Rickman2020_HD13724}, TRENDSV -- \cite{2014Crepp_Johnson_TrendsV}, C15 -- \cite{2015Crepp_Rice_HD19467_relast}, BB20 -- \cite{Bowler_2020_Blunt_Nielsen}}
\end{deluxetable}

\subsection{Gl~229}
We adopt HIRES and HARPS radial velocities using the recently calibrated data sets by \cite{TalOrHIRES} and \cite{TrifonovHarps}, respectively. The combined RV data set consists of 248 observations spanning twenty years. We add nine new HIRES observations \citep{Rosenthal+Fulton+Hirsch+etal_2021} of Gl~229~A, spanning 2018 through early 2021. These additional HIRES RV data are summarized in Table \ref{tab:Gl229_RVs_new}. The additional three years of RVs show slight curvature in the RV time series of Gl~229~A; this curvature is consistent with that expected from the previous best fit orbits of Gl~229~B.

\begin{deluxetable}{ccc}
\tablewidth{0pt}
    \tablecaption{Summary of the additional Gl~229 RVs from HIRES. \label{tab:Gl229_RVs_new}}
    \tablehead{
    \colhead{Epoch} & \colhead{RV} & \colhead{RV error} \\
    \colhead{BJD} & \colhead{m/s} & \colhead{m/s}
    }
    \startdata
    2458116.862 & 8.72  &  1.23  \\
    2458117.852 & 8.71  &  1.24  \\
    2458396.142 & 6.98  &  1.18  \\
    2458777.037 & 16.16 &   1.08 \\
    2458794.994 & 4.30  &  1.14  \\
    2458880.798 & 15.76 &   1.05 \\
    2458907.830 & 12.75 &   0.95 \\
    2459101.122 & 6.79  &  1.15  \\
    2459267.794 & 19.94 &   1.03 \\
    \enddata
    \tablerefs{\cite{Rosenthal+Fulton+Hirsch+etal_2021}, A.~Howard, priv.~commun.}
\end{deluxetable}


We use the relative astrometry from \cite{brandt_gliese_229b_mass_htof} that consists of six observations between 1995 and 2000 using the Wide Field
and Planetary Camera 2 (WFPC2) aboard the {\sl Hubble
Space Telescope} (\HST) and one 2012 observation from the Subaru telescope with HiCIAO \citep{Suzuki+Kudo+Hashimoto+etal_2010}. 
As suggested in \cite{brandt_gliese_229b_mass_htof}, we double the formal PA errors on the 1995 November and 1996 November \HST observations (two epochs).  These two epochs used different guide stars than the other four \HST epochs.

We also present new relative astrometry of Gl~229~B, extending the direct imaging baseline to twenty-five years. We observed Gl~229 on 2020~October~24~UT and 2021~January~5~UT with NIRC2 in narrow-camera mode and the natural guide star adaptive optics system at the Keck~II telescope \citep{Wizinowich2000_PASP_112_315,KECK_AO_System}. In order to obtain high-S/N, unsaturated images of both the host star and companion, we alternated taking shallow and deep exposures. All data were taken using an $864\times120$-pixel subarray to reduce the minimum allowable exposure time. The images of Gl~229~B were obtained with an exposure time per coadd of 0.5\,s, 100 coadds, and the \CHs\ filter ($\lambda_C = 1.592$\,\micron\ and $\Delta\lambda = 0.126$\,\micron). For unsaturated images of Gl~229~A, we used different filters and exposure times at the two epochs. At the first epoch, the image quality was poorer, so we used the \CHs\ filter, exposure time per coadd of 0.01\,s, and 100 coadds. At the second epoch, we used the narrower $H_{\rm cont}$ filter ($\lambda_C = 1.580$\,\micron\ and $\Delta\lambda = 0.023$\,\micron), exposure time per coadd of 0.5\,s, and 100 coadds.

In shallow images, Gl~229~A is unsaturated while the companion is undetected. In deep images, the companion is clearly resolved while the primary is saturated (though the wings of the star's point-spread-function, PSF, are usable). This poses a challenge in measuring the separations of the system. Furthermore, the adaptive optics (AO) corrections for the observations are imperfect and time-varying, especially for the images observed in October 2020. 

\begin{figure*}
    \centering
    \includegraphics[width=\textwidth]{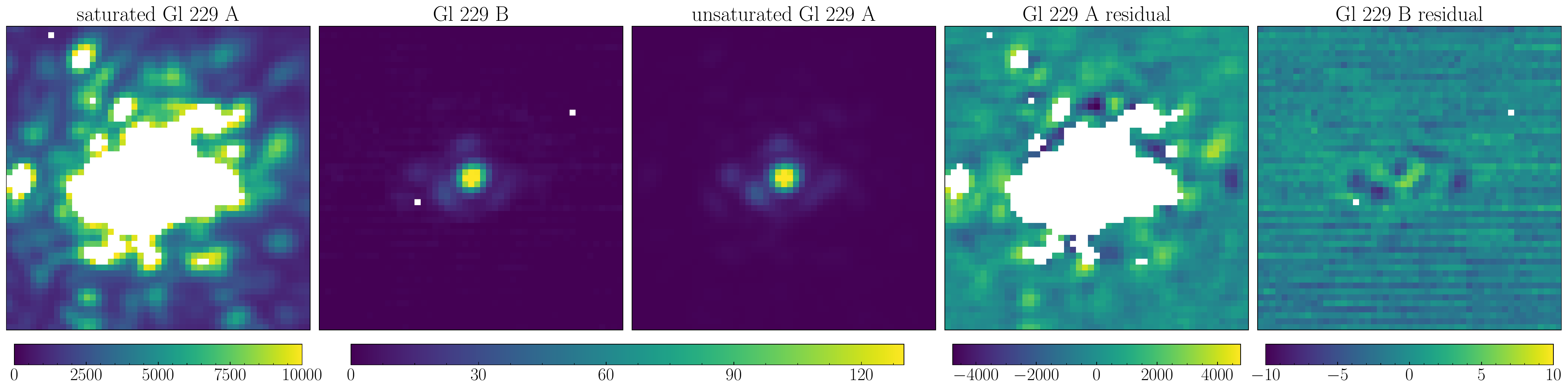}
\caption{PSF fitting for Gl~229~A/B. We use the unsaturated PSF of Gl~229~A (third panel from left) as a template to fit Gl~229~B (second from left) and the saturated PSF of Gl~229~A (left).  We mask saturated and hot pixels; these are shown in white.  
The right two panels show the corresponding residuals.}
    \label{fig:Gl229 PSF fit}
\end{figure*}

To obtain relative astrometry, we implement a least-squares PSF-fitting algorithm that uses the unsaturated PSFs of Gl~229~A as templates to fit for the positions of both Gl~229~A and Gl~229~B in deep images.  Gl~229~A is saturated in the deep images.  We mask hot and saturated pixels and fit the outer wings and speckles of Gl~229~A's diffraction pattern.  Figure \ref{fig:Gl229 PSF fit} shows three representative PSFs, the masked pixels, and the residuals from this procedure.  The outer speckles are sufficiently well-measured in the shallow images that they can centroid well the saturated PSF.

For every deep image, we use all shallow images from the same night to fit for a relative separation and position angle (PA) of the system. Thus for every image where the companion is detected, we obtain $\approx$15--20 templates that allow us to estimate the mean and uncertainty of the result. 
This gives very precise relative offsets in detector coordinates $x$ and $y$ ($\sim$0.01\,pix). 

We convert from detector coordinates into sky coordinates using the same method as \citet{Dupuy+Liu_2017} and \citet{Bowler+Dupuy+Endl+etal_2018}. This accounts for differential atmospheric refraction and aberration. We correct for differential chromatic refraction, although this effect is negligible compared to the uncertainties. We use the calibration of \citet{Service+Lu+Campbell+etal_2016} to correct for distortion; we subtract $0\fdg262\pm0\fdg002$ from the PA of the $y$-axis of NIRC2 given in the header; and account for the pixel scale and its uncertainty ($9.971\pm0.004$\,mas\,pix$^{-1}$).\footnote{\url{https://github.com/jluastro/nirc2_distortion/wiki}}
The 1.1\,mas uncertainty in the \citet{Service+Lu+Campbell+etal_2016} distortion solution implies a relative astrometry noise floor of 1.1\,mas in separation and $0\fdg013$ in PA for Gl~229~AB, which we add in quadrature to the other errors. Table~\ref{tab:relative_astrometry_Gl229New} lists our relative astrometry for Gl~229~B. These measurements are only 2--3$\sigma$ different than what the \cite{brandt_gliese_229b_mass_htof} orbital fit predicts.

\begin{deluxetable*}{lccccc}
\tablewidth{0pt}
    \tablecaption{New Keck/NIRC2 relative astrometry of Gl~229~A/B.\label{tab:relative_astrometry_Gl229New}}
    \tablehead{ 
    \colhead{Date} & \colhead{Sep} & \colhead{$\sigma_{\rm Sep}$ } &
    \colhead{PA} & \colhead{$\sigma_{\rm PA}$ } & \colhead{Filter}\\ 
    \colhead{(UT)} & \colhead{(mas)} & \colhead{(mas)} &
    \colhead{($\deg$)} & \colhead{($\deg$)} & \colhead{}
    }
    \startdata
    2020 Oct 24 & 4922.1 & 2.3 & 179.564 & 0.024 & \CHs \\
    2021 Jan  5 & 4890.5 & 2.4 & 179.735 & 0.024 & $H_{\rm cont}$ + \CHs
    \enddata
\end{deluxetable*}

\subsection{Gl~758}
We use the four epochs of Keck/NIRC2 direct imaging from \cite{Bowler+Dupuy+Endl+etal_2018}. Like \cite{Bowler+Dupuy+Endl+etal_2018}, we use RVs from the Automated Planet Finder (APF) at Lick Observatory, HIRES, and RVs from the Tull Coud\'e spectrograph \citep{Tull+MacQueen+Sneden+etal_1995}. The only difference in this RV data set between our analysis and that of \cite{Bowler+Dupuy+Endl+etal_2018}, is that we are able to use the new calibrated HIRES RVs from \cite{TalOrHIRES}. The entire RV data set from the three instruments consists of 526 measurements spanning nearly twenty years.

\subsection{HD~13724}
The companion to HD~13724 was first discovered with RVs by \cite{2019AA_RickmanHD13724_discovery}, using CORALIE \citep{CORALIE_INST_PAPER}. \cite{Rickman2020_HD13724} followed up with high-contrast imaging and measured the first dynamical mass for the companion. We adopt the same relative astrometry as \cite{Rickman2020_HD13724}. 
Like \cite{Rickman2020_HD13724}, we include HARPS and CORALIE radial velocities. The only difference in the HARPS dataset is that we use the newly calibrated \cite{TrifonovHarps} data. The overall RV baseline of the combined HARPS and CORALIE dataset is roughly twenty years and comprises 170 measurements. We are unable to include $\approx$5 unpublished CORALIE RVs from $\approx$2020 that were shown in \cite{Rickman2020_HD13724}.

Any two RV instruments will almost never agree on a measure of the RV offset (also known as the RV zero point) of a star due to unique systematics in the data processing pipeline or instrument. Instrument upgrades and small changes in a data reduction pipeline can also perturb the RV zero point. The CORALIE instrument was upgraded in June 2007  \citep{2010AA...511A..45SSegransan} and again in November 2014. 
We follow \cite{Rickman2020_HD13724} and \cite{Cheetham+Segransan+Peretti+etal_2018} and treat the CORALIE pre- and post-upgraded instruments as independent RV instruments, thereby splitting the CORALIE dataset in three: CORALIE-98 (before the 2007 upgrade), CORALIE-07 (between the 2007 and 2014 upgrades), and CORALIE-14 (after the 2014 upgrade). Accordingly, our fits to HD~13724 include four RV offsets: one for HARPS and one for each CORALIE state.


\subsection{HD~19467}
We adopt the relative astrometry used in the recent work by \cite{2020AA_Maire_HD19467}. This consists of 7 measurements total, spanning $\approx$8 years between 2011 and 2018, from \cite{2014Crepp_Johnson_TrendsV}, \cite{Crepp+Rice+Veicht+etal_2015}, and \cite{Bowler_2020_Blunt_Nielsen}.

The radial velocities of HD~19467~A consist of HARPS measurements calibrated by \cite{TrifonovHarps} and HIRES measurements calibrated by \cite{TalOrHIRES}. The RVs span a baseline of more than twenty years, although this is less than one-tenth of the orbital period of HD~19467~B.

\subsection{HD~33632~A \& B}

HD~33632~Ab was discovered by \cite{2020ApJ_Currie_Thayne_HD33632} with direct imaging from Subaru/CHARIS \citep{CHARIS_INST_PAPER1, CHARIS_INST_PAPER2} and Keck/NIRC2. We adopt the same relative astrometry here. These data were \addition{presented in cartesian coordinates} in \cite{2020ApJ_Currie_Thayne_HD33632}; we present them in \addition{polar coordinates in} Table~\ref{tab:relative_astrometry_HD33632}.

We use RVs from the Lick planet search with the Hamilton spectrograph \citep{2014ApJ_Fischer_LickRvs}. The RVs for HD~33632~A span roughly eleven years, a small fraction of the nearly 100-year period of HD~33632~Ab.

HD~33632 has a co-moving M dwarf companion HD~33632~B (2MASS J05131845+3720463) that is resolved in \gaia EDR3 \citep{2016A&A...587A..51S, 2020GaiaEDR3_catalog_summary}. The companion has a projected separation of $33\farcs99086\pm0\farcs00003$. 
We convert the correlated \gaia EDR3 positions into correlated relative astrometry (separation and PA). The resulting separations and PAs for HD~33632~B about HD~33632~A are in Table \ref{tab:relative_astrometry_stellar_companions}.


\begin{deluxetable*}{ccccccc}
\tablewidth{0pt}
    \tablecaption{Relative astrometry of HD~33632~A/Ab.\label{tab:relative_astrometry_HD33632}}
    \tablehead{ 
    \colhead{Date} & \colhead{Sep} & \colhead{$\sigma_{\rm Sep}$ } & \colhead{PA} & \colhead{$\sigma_{\rm PA}$ } & \colhead{Instrument} & \colhead{Filter} \\ 
    \colhead{(UT)} & \colhead{(mas)} & \colhead{(mas)} &
    \colhead{($\deg$)} & \colhead{($\deg$)} & \colhead{} & \colhead{}
    }
    \startdata
    2018-10-18 & 781  &  5  &  257.0  &   0.4  &  CHARIS & $JHK$ \\
    2018-11-01 & 774  &  5  &  256.7  &   0.4  &  NIRC2  & ${L}^{\prime}$ \\
    2020-08-31 & 746  &  5  &  262.8  &   0.4  &  CHARIS & $JHK$ \\
    2020-09-01 & 746  &  5  &  262.7  &   0.4  &  CHARIS & $JHK$ \\
    \enddata
        \tablecomments{These are the same data first used in \cite{2020ApJ_Currie_Thayne_HD33632}.}
\end{deluxetable*}

\begin{deluxetable*}{ccccccc}
\tablewidth{0pt}
    \tablecaption{Relative astrometry derived from \gaia EDR3 for wide stellar companions.\label{tab:relative_astrometry_stellar_companions}}
    \tablehead{ 
    \colhead{Date} & \colhead{Identifiers} & \colhead{Sep} & \colhead{$\sigma_{\rm Sep}$} & \colhead{PA} & \colhead{$\sigma_{\rm PA}$} & \colhead{Sep--PA correlation} \\ 
    \colhead{(UT)} & \colhead{} & \colhead{(mas)} & \colhead{(mas)} &
    \colhead{($\deg$)} & \colhead{($\deg$)} & \colhead{}
    }
    \startdata
    2016 Jan 1 & HD 33632 A/B & 33990.86 & 0.03 & \phantom{2}20.34068 & 0.00011 & $-$0.28 \\
    2016 Jan 1 & HD 72946/ HD 72945  & 10044.18 & 0.09 & 204.76056 & 0.00061 & \phantom{$-$}0.74 \\
    \enddata
        \tablecomments{All data are derived from \cite{2020GaiaEDR3_catalog_summary}. The raw positions and correlations were fetched from the \Gaia archive (\url{https://gea.esac.esa.int/archive/}).} 
\end{deluxetable*}

\subsection{HD~72946}
We use the two epochs of relative astrometry taken with VLT/SPHERE \citep{SPHERE_INST_PAPER} presented in \cite{2020AA_Maire_HD19467}.
HD~72946 has a co-moving stellar companion, HD~72945, at a separation of $\approx$10$^{\prime\prime}$ ($\approx$250 au) \citep{2020GaiaEDR3_catalog_summary}. 
We convert the \gaia EDR3 absolute astrometry of HD~72946 and HD~72945 into relative astrometry following the same procedure as for HD~33632~A/B. The resulting separation and PA for HD~72945 about HD~72946 are in Table~\ref{tab:relative_astrometry_stellar_companions}.

The RVs for HD~72946 come from the ELODIE \citep{1996A&AS..119..373B} and SOPHIE \citep{2006tafp.conf..319B} instruments as published by \cite{2016AA_Bouchy2016_HD72946}. These RVs span roughly sixteen years --- a full orbital period of HD~72946~B. 

\section{Host Star Astrometry}\label{sec:gaia_astrometry}

Absolute astrometry from \hipparcos and \gaia give powerful constraints on the masses of giant long-period companions. We use absolute astrometry of the host stars to measure the dynamical properties of the six systems with high precision. We follow the procedures described in \cite{brandt_cross_cal_gaia_2018, Dupuy+Brandt+Kratter+etal_2019, TimOrbitFitTemporary, Brandt2020betapicbc}, which are similar to those adopted by, e.g., \cite{2019MNRAS.490.5002F, Lagrange_beta_pic_c, AMLagrange2020betapicc_direct_detection}. In brief, we use the proper motion anomalies between \hipparcos, \gaia EDR3 and the \hipparcos-\gaia long-term proper motion to measure the acceleration vector of the host star in the plane of the sky. The acceleration offers additional constraints on the dynamical properties of the companion (and particularly its mass). 

We use calibrated \gaia EDR3 and \hipparcos astrometry from the \hipparcos-\gaia v.EDR3 catalog of accelerations, originally produced for \gaia DR2 by \citet{brandt_cross_cal_gaia_2018}. We use the HGCA because it rotates the \hipparcos, \gaia, and \Hipparcos-\Gaia proper motions into the same reference frame in order to make them suitable for orbit fitting. The HGCA also calibrates all uncertainties to produce Gaussian residuals with the expected variance.

HD~33632~A and HD~72456 have outer third bodies: HD~33632~B and HD~72945, respectively. We analyse both two-body (ignoring the outer stellar companion) and three-body orbital fits. \orbitcodename uses their proper motions and proper motion correlations to help constrain their orbit, both of which are available in the \gaia archive.\footnote{\url{https://gea.esac.esa.int/archive/}} However, neither HD~33632~B nor HD~72945 are in the HGCA. We apply a proper motion error inflation of a factor of 2 for HD~33632~B \citep{Cantat-Gaudin_Brandt_2021} and 1.37 for HD~72945 \citep{BrandtGaiaEDR3HGCA} to account for any low-level systematics. We correct for projection effects in the proper motion and apply the \citet{Cantat-Gaudin_Brandt_2021} magnitude-dependent correction, which aligns the proper motion of the Gaia EDR3 sources brighter than G=13 with the International Celestial Reference Frame. Both corrections are negligible compared to the inflated proper motion errors, but we include them for completeness. The final proper motions and errors for HD~33632~B are $-144.58 \pm 0.15\,\masyr$ in right-ascension and $-139.53 \pm 0.11\,\masyr$ in declination For HD~\addition{72495}, we use $-130.31 \pm 0.19\,\masyr$ in right-ascension and $-133.12 \pm 0.15\,\masyr$ in declination

\section{Orbit Fitting}\label{sec:orbitfitting}

We use \orbitcodename to fit for the orbital parameters of each system. The code employs MCMC with {\tt ptemcee} \citep{Foreman-Mackey+Hogg+Lang+etal_2013,Vousden+Farr+Mandel_2016}. Absolute astrometry is processed and fit for the five astrometric parameters by \htofcodename \citep{MirekHTOFtemporary, htof_zenodo} at each MCMC step. We use a parallel-tempered MCMC with 20 temperatures; for each temperature we use 100 walkers with at least 400,000 steps per walker, thinned at the end by at least a factor of 50. Our MCMC chains converge typically between 20,000 and 80,000 steps; we conservatively discard the first 75\% of each chain as burn in and use the remainder for inference. The chains for Gl~229 and the three-body fits to HD~33632 \addition{and HD~72946} were run for two million steps to ensure convergence and that the full parameter space was explored. We use the same criteria presented in \cite{Brandt2020betapicbc} to verify the convergence of our chains. The convergence criteria include the Gelman-Rubin Diagnostic \citep{Gelman+Rubin_1992_MCMCconvergence, Vivekananda_2019_MCMCconvergence}.

These aforementioned methods and analysis tools are nearly identical to those presented in \cite{brandt_gliese_229b_mass_htof}, \cite{2020ApJ_Currie_Thayne_HD33632}, \cite{Brandt2020betapicbc}, \cite{TimOrbitFitTemporary}, and \citet{2021_Li_Yiting_nine_masses_RV_planets}. We fit either 9 or 16 parameters for either 2 or 3 bodies total, respectively. These are the six Keplerian orbital elements\footnote{\orbitcodename fits for $\sqrt{e} \cos \omega$, $\sqrt{e} \sin \omega$ instead of the eccentricity $e$ and the argument of periastron $\omega$ directly.} for each companion plus its mass and an RV jitter to be added to the RV uncertainties. We use a single RV jitter per star rather than per instrument, attributing the jitter to stellar activity.  Our results are consistent \addition{if we adopt a} different jitter for each instrument. \orbitcodename marginalizes out each instrument's RV zero-point, parallax, and barycenter proper motion. We perform fits to HD~72946 and HD~33632 that include and exclude their widely-separated, co-moving stellar companions (i.e., we do both two and three-body fits to these two systems). \orbitcodename's three-body approach was shown to be accurate in \cite{Brandt2020betapicbc} via a set of \texttt{REBOUND} validation tests. We refer the reader to Section 2.3 of \cite{Brandt2020betapicbc} for the discussion of the three-body approach.

\subsection{\addition{Priors on Orbital Elements}}

We assume uninformative priors for all the orbital elements: uniform except for inclination $i$, where we assume the standard geometric prior, and with semi-major axis and companion mass where we assume log-flat priors. We adopt a log-flat prior on each RV jitter.

\subsection{Priors on Stellar Masses}\label{sec:stellar_mass_priors}
We assume uniform priors on the masses of HD~13724~A, Gl~758~A, and Gl~229~A. We adopt stellar evolution masses as priors on the primary mass of the other three systems (HD~33632, HD~19467, and HD~72946), for which the RV baseline is short.
For these three systems, the constraints on the mass of the primary from any orbital fit are many factors worse than those known (even loosely) from stellar evolution. For instance, a completely uninformative fit to HD~19467 yields a posterior on the primary mass of $1.6 \pm 2 \Msun$. This is a G3 dwarf, and so we know that the mass is near $1\Msun$ with much higher confidence than $\pm 2 \Msun$. Adopting a prior informed by stellar evolution theory is appropriate. For similar reasons, we could adopt a prior on the primary mass of Gl~758, however, adopting a tight prior on the primary mass adds a negligible improvement to the inferred secondary mass.

We adopt the same \addition{Gaussian} priors on the primary masses as were used in the most recent dynamical analyses of the systems. These are $1.1 \pm 0.1 \Msun$ for HD~33632~A \citep{2020ApJ_Currie_Thayne_HD33632}, $0.953 \pm 0.022 \Msun$ for HD~19467~A \citep{2020AA_Maire_HD19467}, and $0.986 \pm 0.027 \Msun$ for HD~72946~A \citep{2020AA_Maire_HD72946}. 
These choices enable direct comparisons of our results to the preceding orbital analyses. Moreover, they are consistent with isochronal mass estimates in the literature (compare and see \citealp{2011AA...530A.138C, 2012ApJ...756...46R, 2017AA...604A.108M}).
%
%
Our use of an informative prior ultimately has a negligible effect on our mass constraints for HD~19467~B and HD~33632~Ab. But it improves the precision of our inferred mass for HD~72946~B by a factor of $\approx$3.


We also adopt priors on the distant stellar companions in the three-body fits to the HD~72946 and HD~33632 systems in Section \ref{sec:results}. As we show in Section \ref{sec:results}, adding the third body (HD~72945 or HD~33632~B) does not change significantly the inferred parameters of the BD companion. However, in both cases adding the tertiary stellar body without placing a prior on its mass degrades the convergence of the chains because the stellar companion's mass is unconstrained by the data. We adopt stellar-evolution based priors on the masses of the stellar companions for these two systems. For the M dwarf \citep{2020ApJ_Currie_Thayne_HD33632} companion HD~33632~B, we adopt a $0.22 \pm 0.03 \Msun$ prior consistent with the mass-magnitude relation from \citet{2019ApJ...871...63M}, and with stars of similar spectral type (roughly M4; \citealt[]{2016A&A...587A..51S}), e.g., V1352 Ori; $0.23 \Msun$; GJ 3709 B, $\approx 0.27 \Msun$ or HD~239960, $\approx 0.21 \Msun$; \citealt[]{2014ApJ...791...54G}. For the F8V-type companion HD~72945 \citep{2012AstL...38..331A_extended_hipcompilation}, we use a mass prior of $1.1 \pm 0.1 \Msun$ prior (consistent with e.g., $1.15 - 1.25\,\Msun$ from \citealt[]{Luck_2017}; or $1.21 \pmoffs{0.02}{0.03}$\,\Msun\ found by \citealt[]{2012ApJ...756...46R}).





\section{Orbit \& Dynamical Mass Results}\label{sec:results}
In this Section we discuss the inferred orbital elements and masses for each star from our MCMC orbit fits. We improve the secondary mass constraints for all systems, and obtain a large (factor of 7) improvement on the mass precision of Gl~229~B. 

We use the reduced chi-squared statistic to assess the goodness of fit: $\chi^2/n_{\rm dof} = \sum ({\rm data - model})^2/n_{\rm dof} \sigma^2$, where $n_{\rm dof}$ is the number of degrees of freedom. Our use of an RV jitter term enforces a reduced chi-squared near unity for the RV data, but there is no such condition for the relative or absolute astrometry. Table \ref{tab:chisquared_statistics} gives the chi-squared statistics from the best-fit orbit (for every source) for relative and absolute astrometry.

\begin{figure*}
    \centering
    \includegraphics[width=\textwidth]{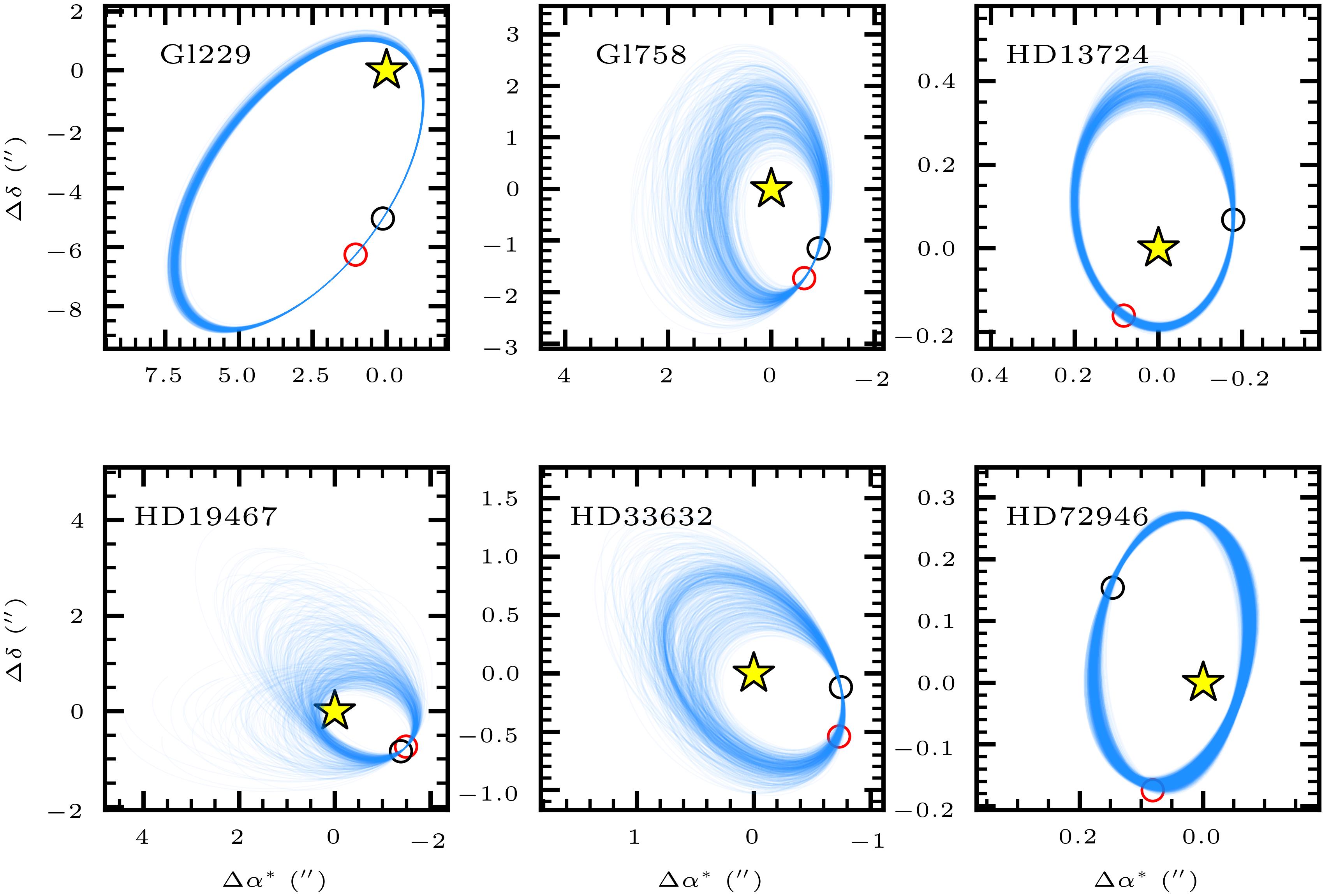}
    \caption{Relative orbits (in arc seconds) from fitting RVs, relative astrometry, and absolute astrometry from the v.EDR3 HGCA. Five hundred random orbital draws are shown. Positions at 2010 (red) and 2020 (black) are marked by circles. The host-star of each system is marked with the star symbol at the origin.}
    \label{fig:astrometric_orbits}
\end{figure*}

\begin{figure*}
    \centering
    \includegraphics[width=0.32\textwidth]{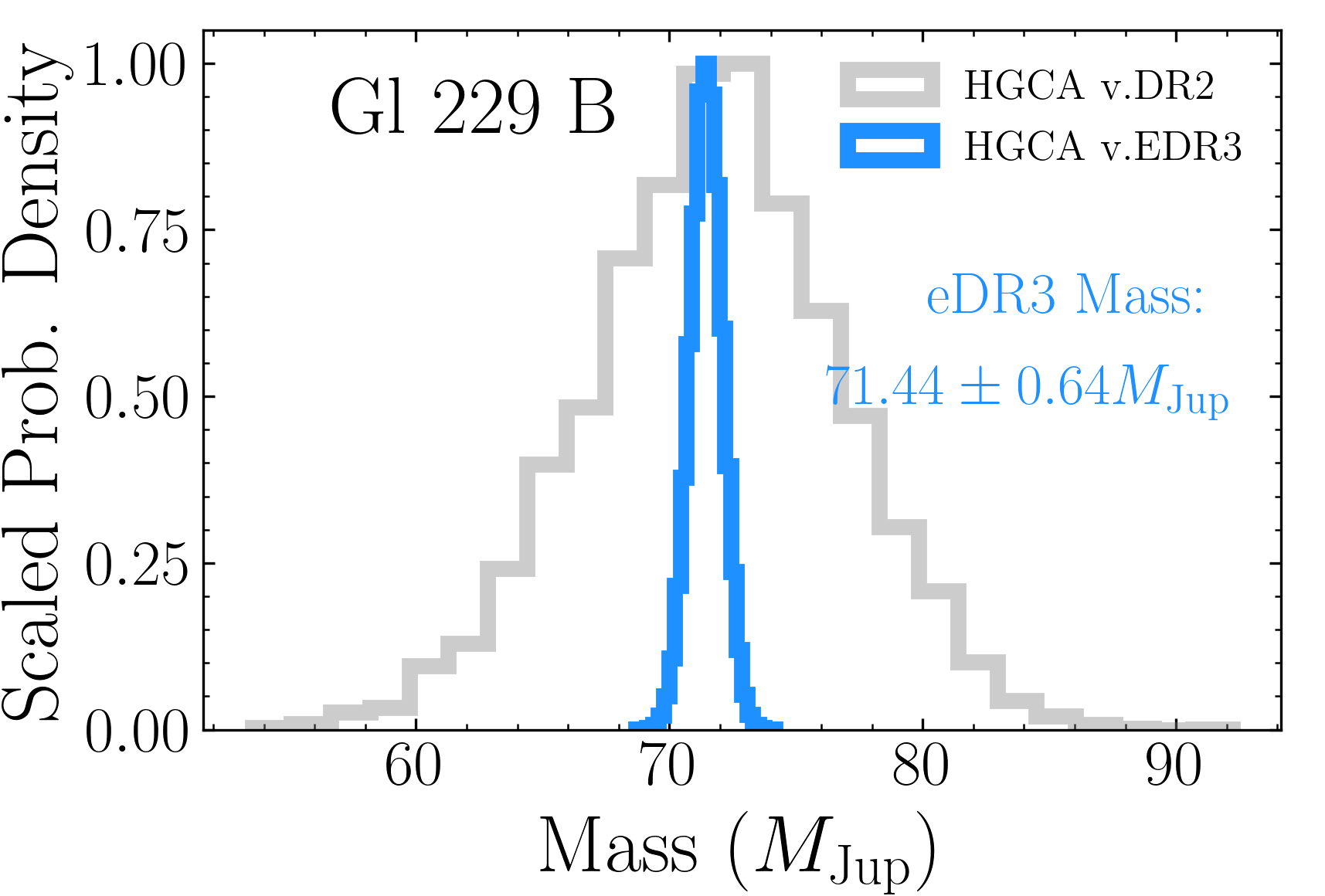}
    \includegraphics[width=0.3\textwidth]{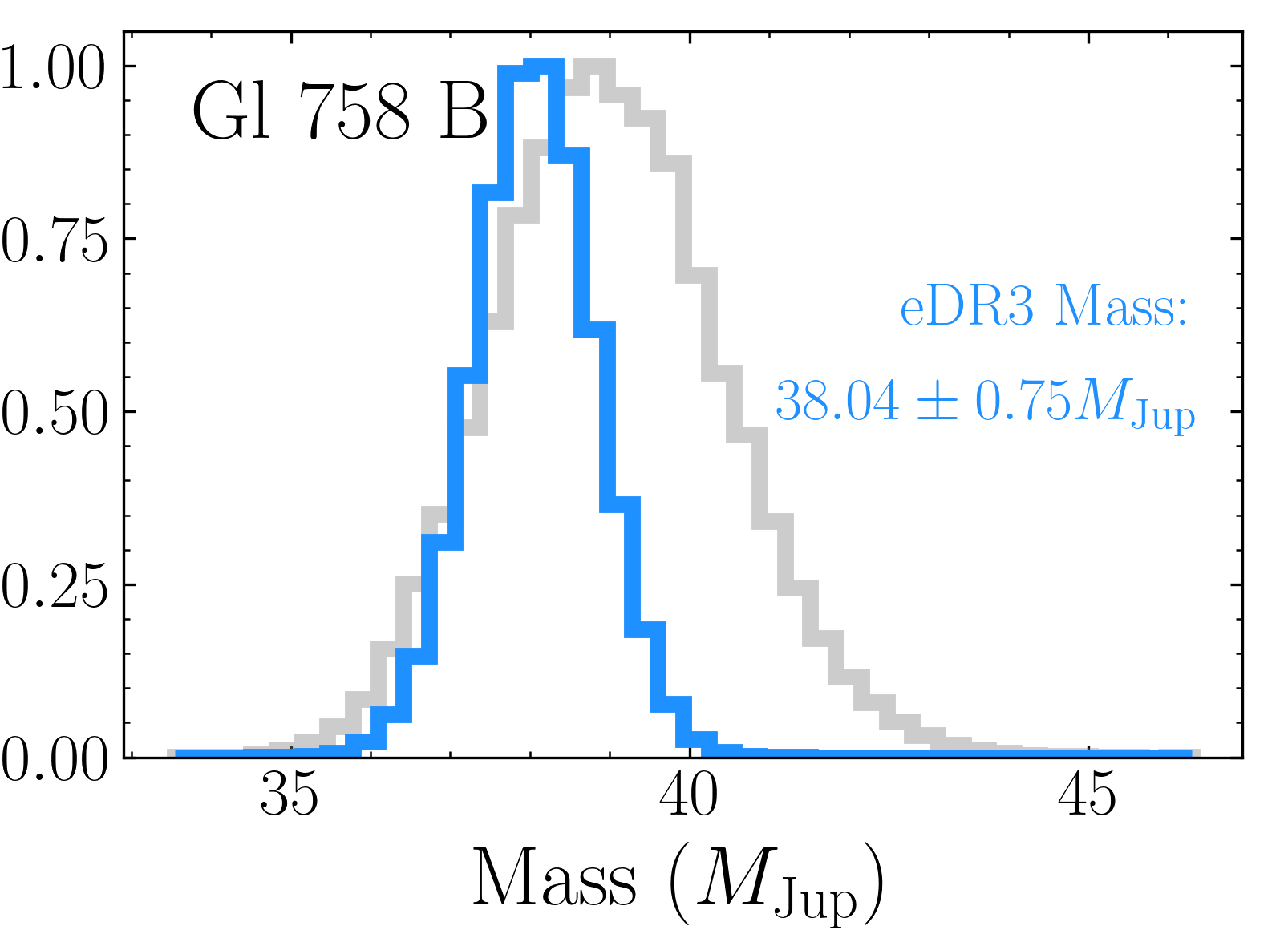}
    \includegraphics[width=0.3\textwidth]{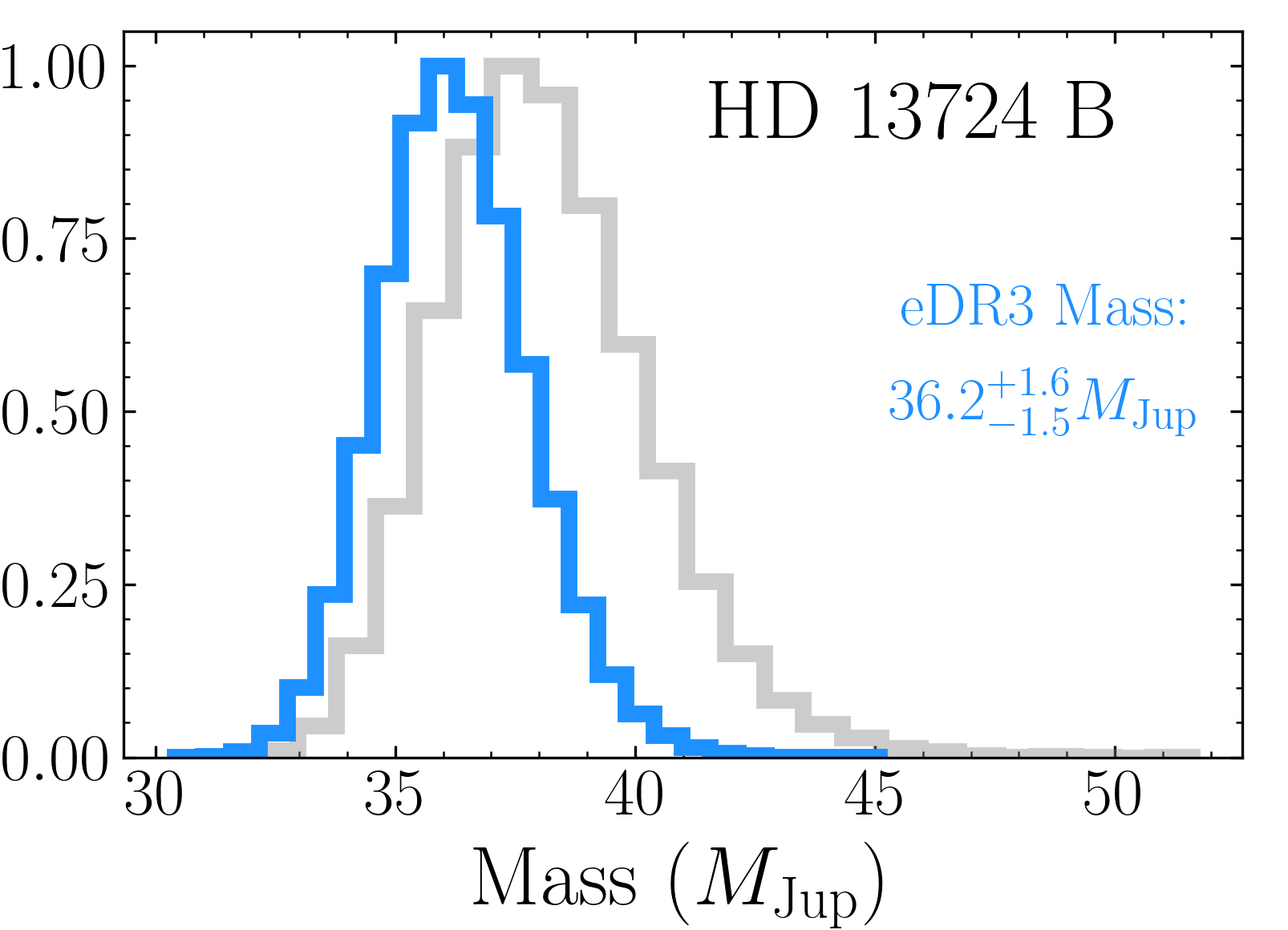}
    \includegraphics[width=0.32\textwidth]{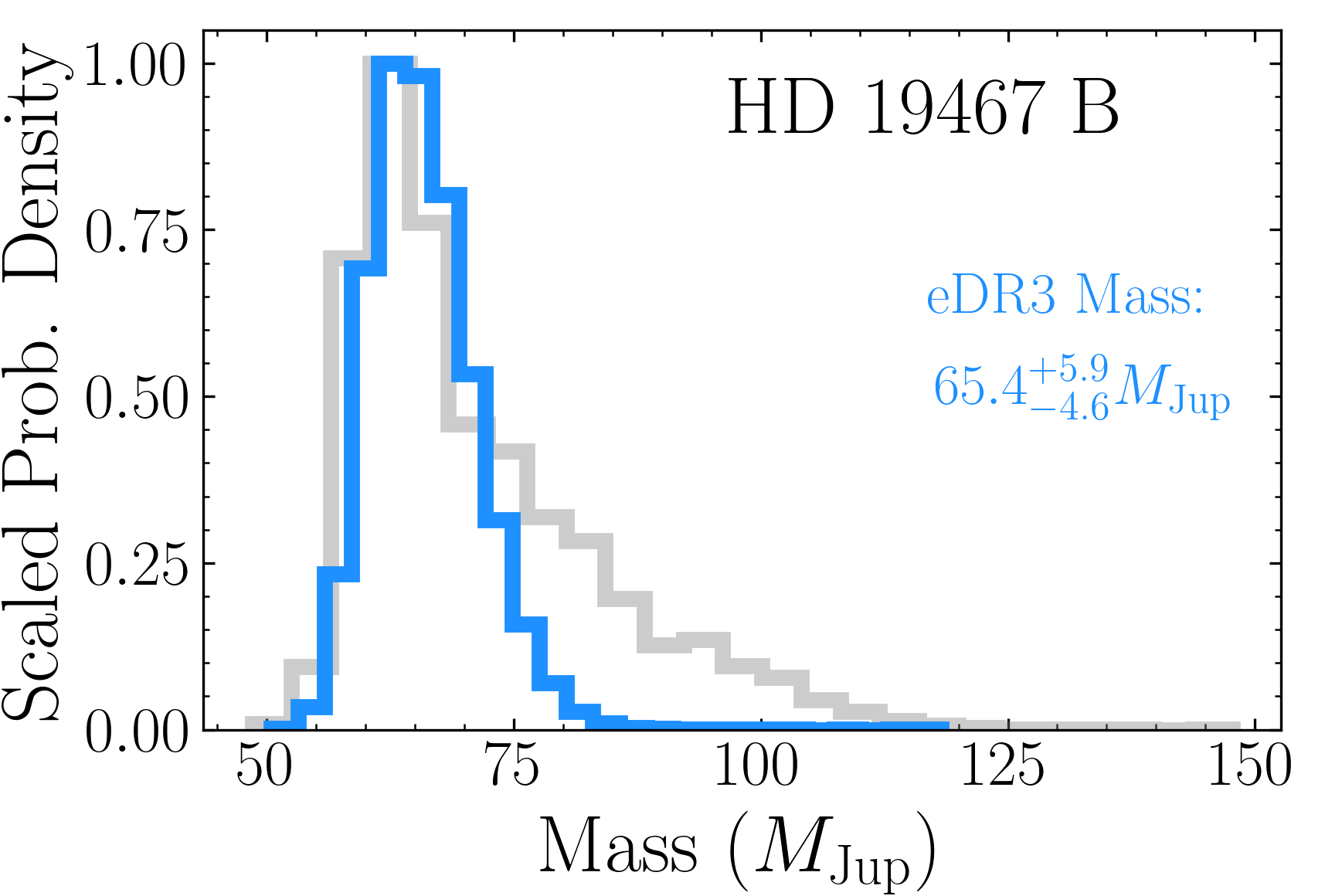}
    \includegraphics[width=0.3\textwidth]{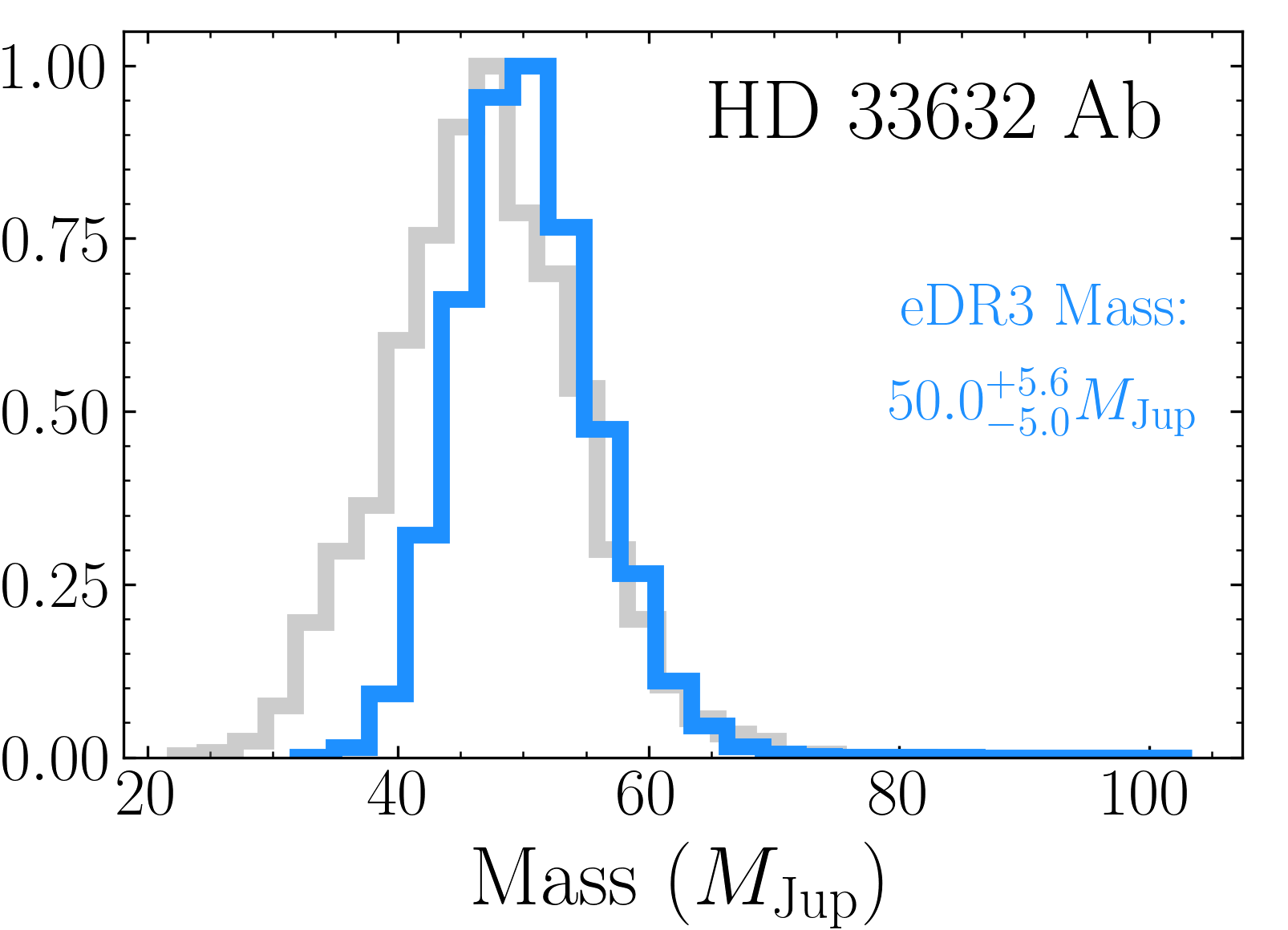}
    \includegraphics[width=0.3\textwidth]{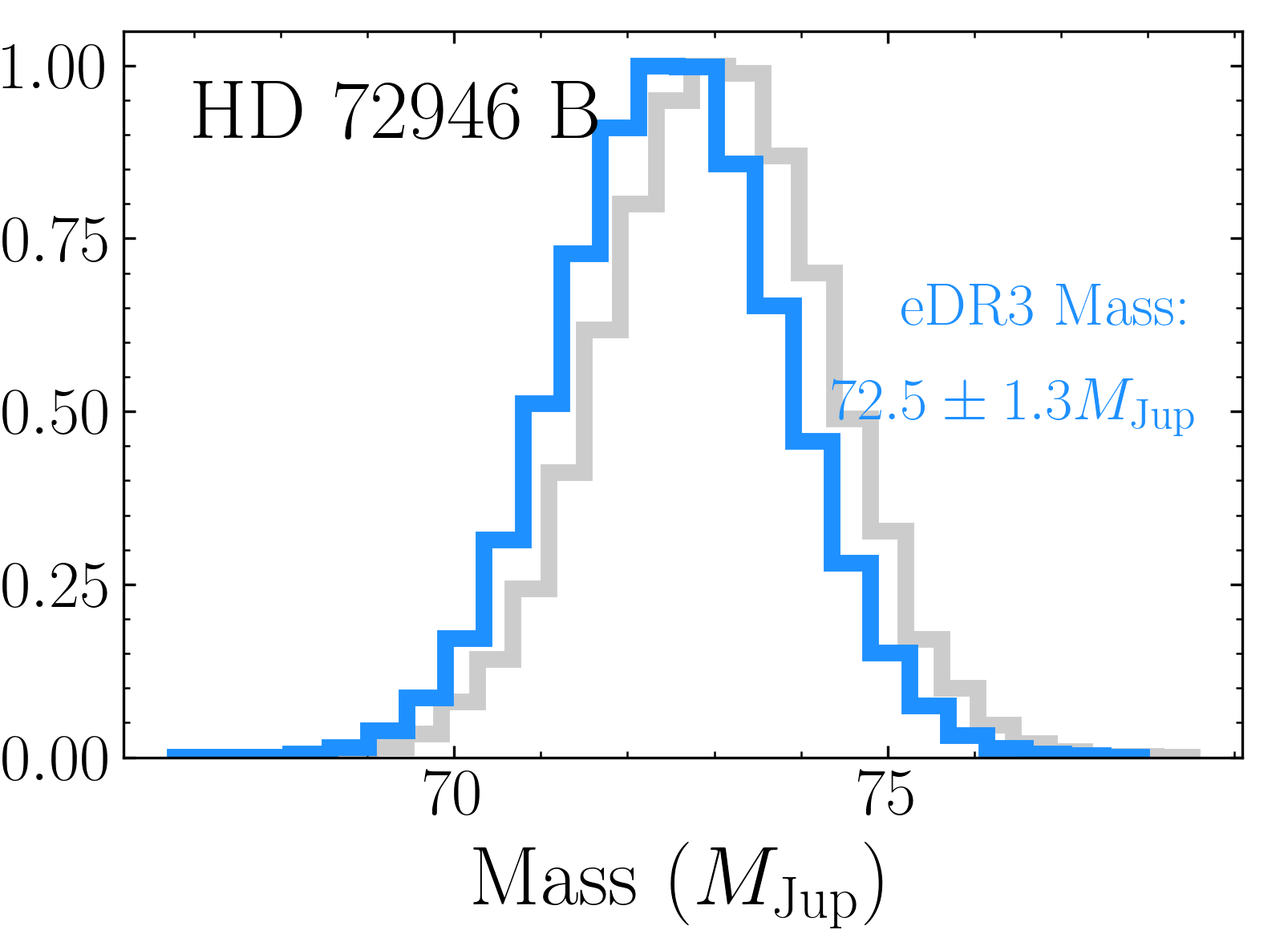}
    \caption{Marginalized mass posteriors for the six companions using the HGCA v.EDR3 (blue; \citealt{BrandtGaiaEDR3HGCA}) and HGCA v.DR2 \citep[gray;][]{brandt_cross_cal_gaia_2018}. Each posterior has been scaled to a peak value of 1. The median and 1$\sigma$ confidence intervals of our new \gaia EDR3 mass posteriors are listed in each sub panel. The HGCA v.EDR3 yields masses that are more precise by factors of 2--5 over HGCA v.DR2 for four systems. The other input data (RVs and relative astrometry) are identical between the fits shown here.}
    \label{fig:marginalized_mass_pdfs_all}
\end{figure*}

Figure \ref{fig:astrometric_orbits} shows the relative orbits of the six systems studied here. Figure \ref{fig:marginalized_mass_pdfs_all} summarizes our improvements to the BD masses, displaying the marginalized mass posteriors using the HGCA v.EDR3 in comparison to an otherwise identical analysis using the HGCA v.DR2. We improve the mass precision of four of the BDs by factors of two to five after adopting the \gaia EDR3 astrometry. Predicted positions (separation, PA etc.) are available at any epoch via \url{http://www.whereistheplanet.com/} \citep{2021ascl.soft01003W}. \addition{The chains used for the predicted positions on \url{http://www.whereistheplanet.com/} are included in the supplemental data.}

Corner plots for every fit, which show the orbital parameter covariances, are contained within Figure set \ref{fig:Corner_Gl229}. Figure sets \ref{fig:fit_propermotion_Gl229}, \ref{fig:fit_relative_astrometry_Gl229}, and \ref{fig:fit_RVs_Gl229} show the fits to the proper motions, relative astrometry, and RVs, respectively.

\begin{deluxetable*}{cccccc}
\tablewidth{0pt}
    \tablecaption{The goodness-of-fit to the relative and absolute astrometry of each orbital fit. \label{tab:chisquared_statistics}}
    \tablehead{
    \colhead{System} &
    \colhead{$\chi^2_{\rm PA}$ + $\chi^2_{\rm Sep}$} &
    \colhead{$N_{\rm PA}$ + $N_{\rm Sep}$} & 
    \colhead{$\chi^2_{\Hipparcos \, \mu}$} &
    \colhead{$\chi^2_{\Gaia \, \mu}$} &
    \colhead{$\chi^2_{{\rm HGCA long-baseline} \, \mu}$}
    }
    \startdata
    Gl~229  &  15.82  &  18  &  5.77  &  0.45  &  0.49 \\
    Gl~758  &  5.06  &  8  &  2.64  &  0.61  &  0.36 \\
    HD~13724  &  6.22  &  18  &  2.65  &  0.07  &  0.08 \\
    HD~19467  &  11.73  &  14  &  0.26  &  2.91  &  0.69 \\
    HD~33632  &  1.41  &  8  &  5.87  &  0.01  &  0.01 \\
    HD~72946  &  0.01  &  4  &  4.59  &  0.27  &  1.84 \\
    \enddata
        \tablecomments{The $\chi^2$ quoted here are for the maximum likelihood orbits. $\chi^2_{\rm PA}$ + $\chi^2_{\rm Sep}$ is the total $\chi^2$ of the fit to the relative astrometry. The reduced chi-squared of the RVs are near one by construction, and so are unlisted. The $\chi^2$ for each proper motion ($\mu$) includes both $\mu_{\delta}$ and $\mu_{\alpha}$, and so it is composed of 2 data points. $N_{\rm PA}$ + $N_{\rm Sep}$ is the combined number of PA and separation measurements (twice the number of relative astrometry measurements).}
\end{deluxetable*}

\subsection{Gl~229}

\begin{figure*}
    \centering
    \includegraphics[width=\linewidth]{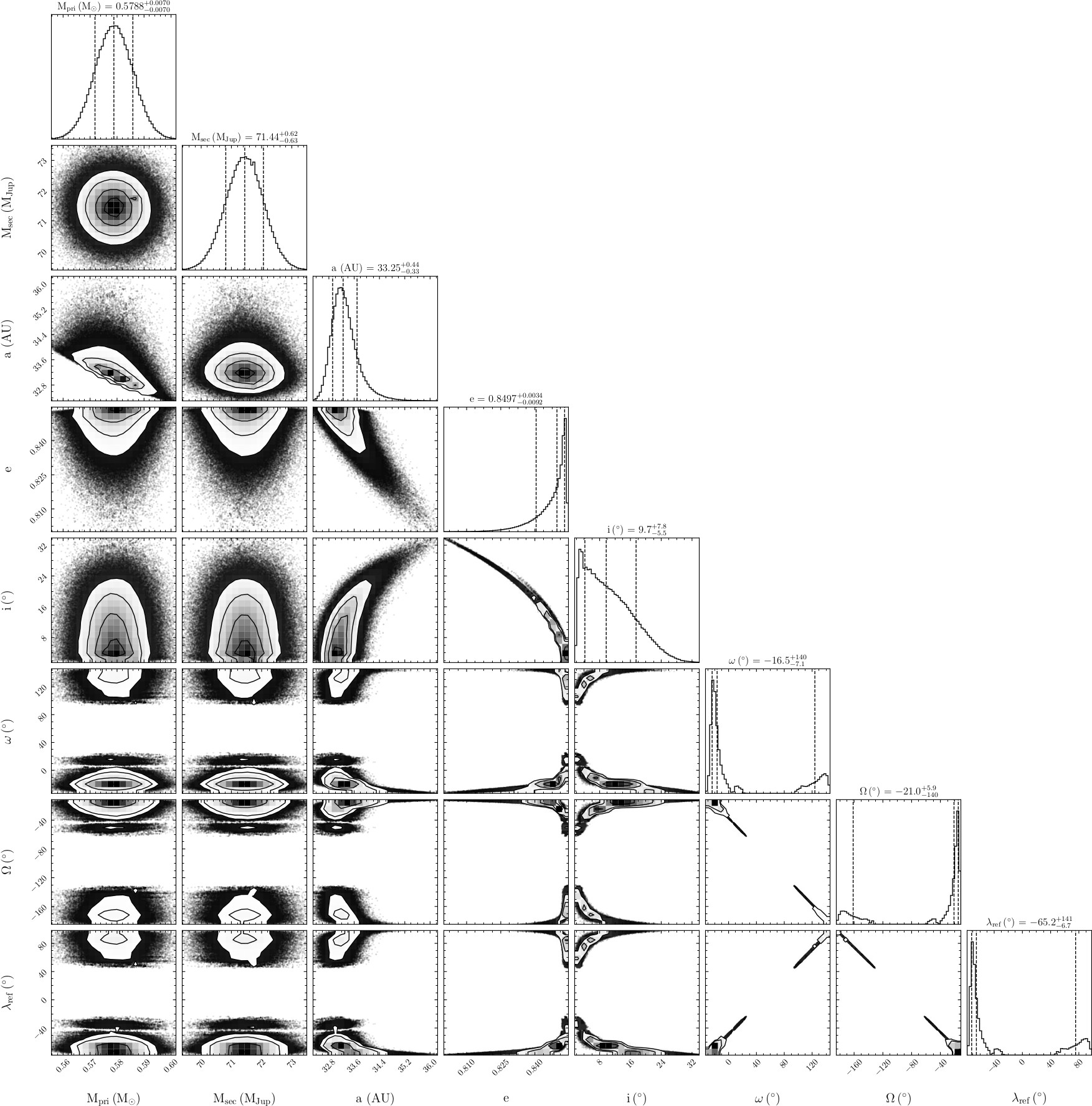}
    \caption{Orbital elements, with respect to the star, for Gl 229 B. In the 1D histograms, the vertical-dashed lines about the center dashed lines give the 16\% and 84\% quantiles around the median. In the 2d histograms, the contours give the 1-$\sigma$, 2-$\sigma$, and 3-$\sigma$ levels. The Figure set contains the corner plots for every fit. The complete Figure set (10 images) for this version is available as a compressed file (figset\_cornerplots.zip) with the supplemental data.}
    \label{fig:Corner_Gl229}
\end{figure*}

\begin{figure*}[!ht]
    \centering
    \includegraphics[width=\linewidth]{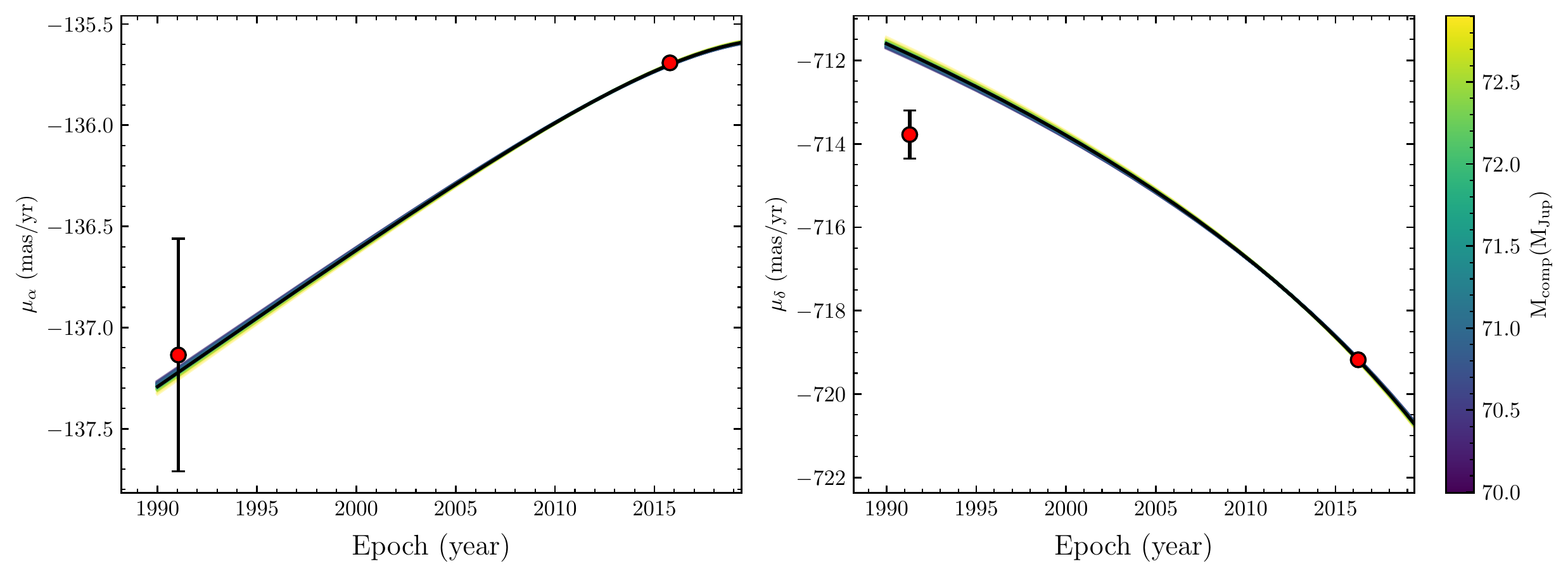}
    \caption{Model proper motions compared to the calibrated \hipparcos (dot at 1991.25) and \gaia EDR3 proper motions (dot near 2016) from the HGCA. The best fit orbit is shown in black. A random sampling of other orbits from the MCMC chain are shown and are color coded by the mass of Gl~229~B. The formal $\chi^2$ of the fit to each proper motion are listed in Table \ref{tab:chisquared_statistics}. The complete Figure set (6 images) for this version is available as a compressed file (figset\_PM.zip) with the supplemental data.}
    \label{fig:fit_propermotion_Gl229}
\end{figure*}

\begin{figure*}[!ht]
    \centering
    \includegraphics[height=0.42\linewidth]{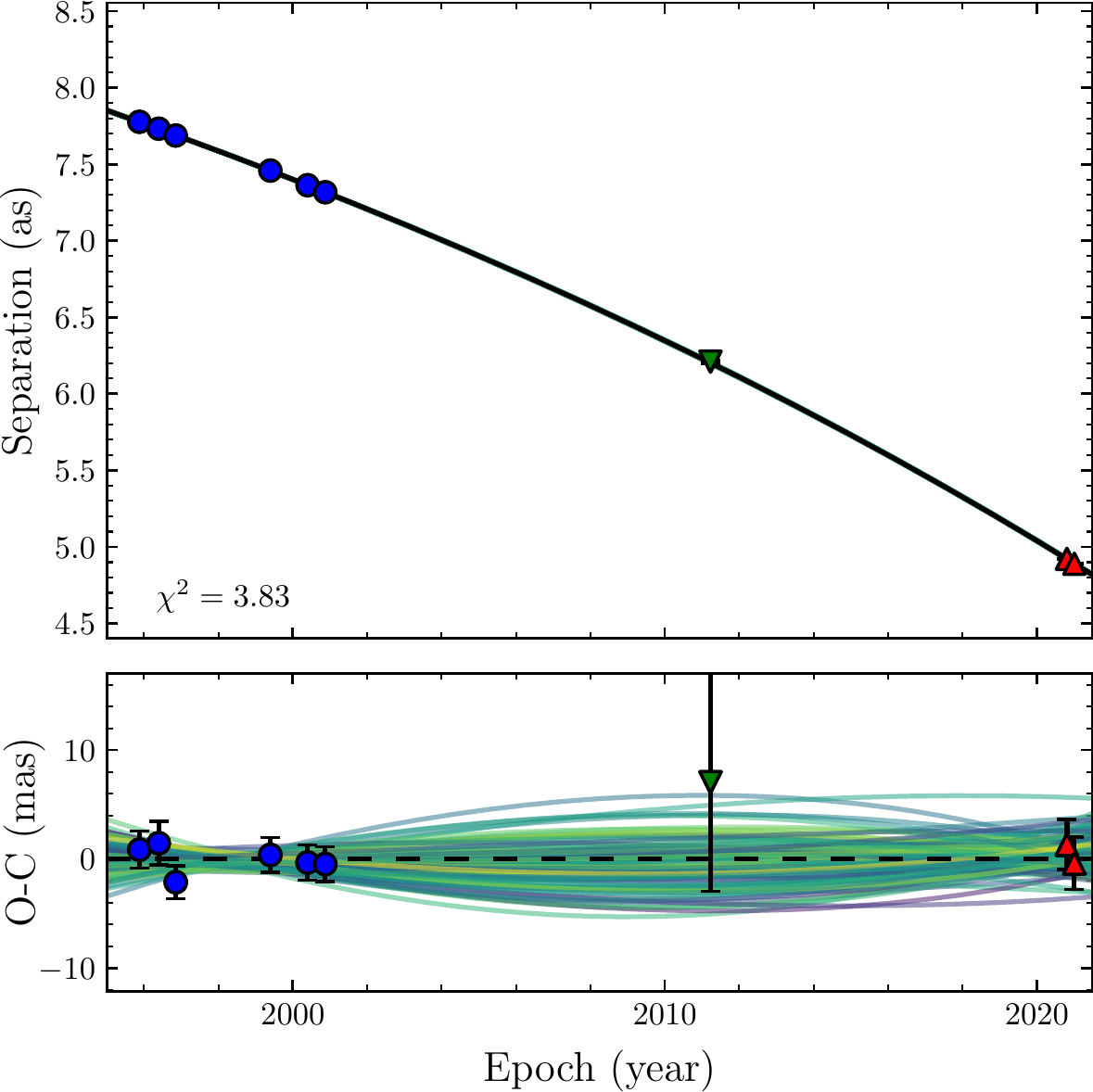} \hskip 0.1 truein
    \includegraphics[height=0.42\linewidth]{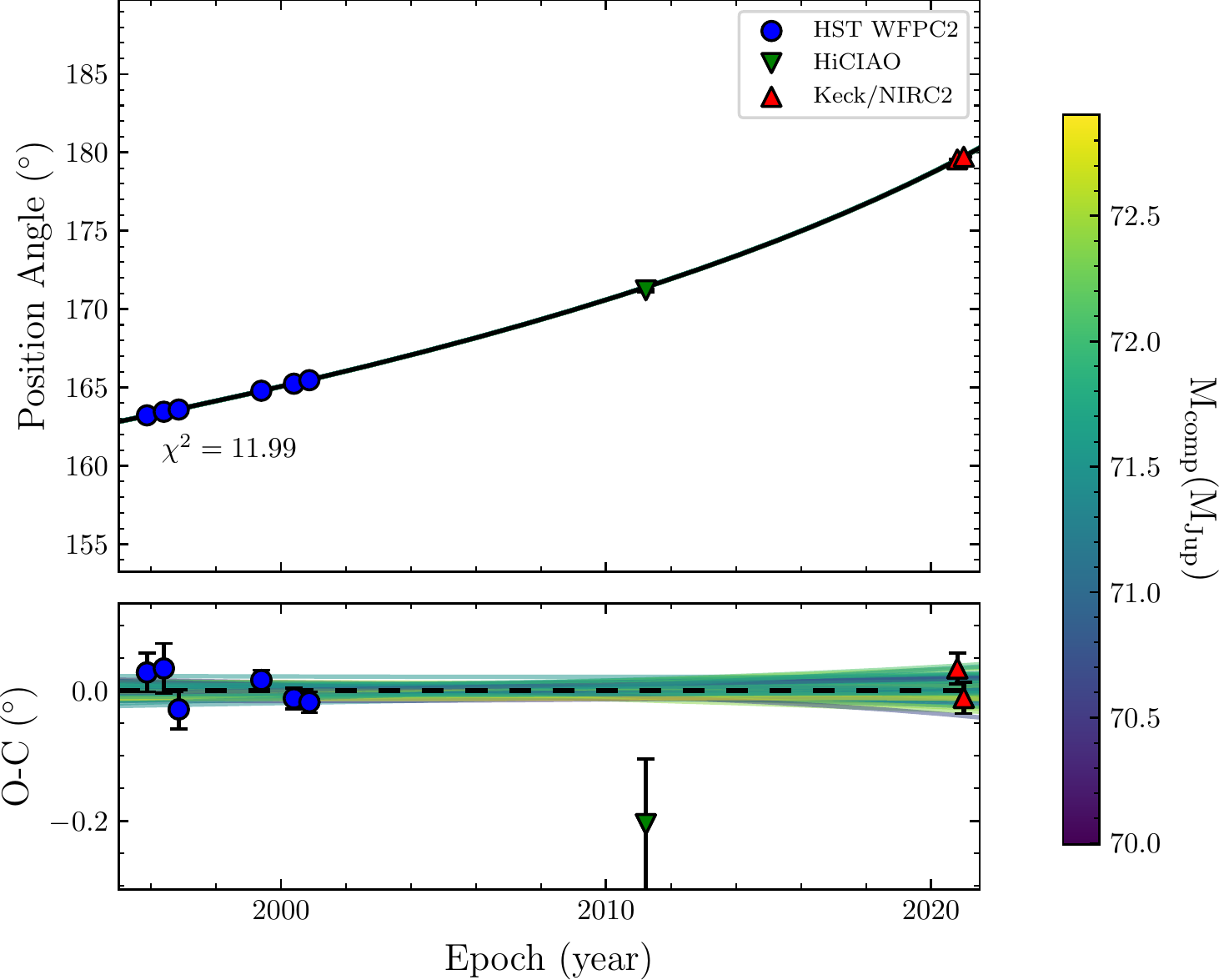}
    \caption{Left: relative separation of Gl~229~B.  Right: PA of Gl~229~B. A random sampling of orbits from other MCMC steps are shown and are color coded by the mass of Gl~229~B. The best fit orbit is shown in black. The formal $\chi^2$ of the fit to the data are inset and listed in Table \ref{tab:chisquared_statistics}. The complete Figure set (12 images) for this version is available as a compressed file (figset\_relast.zip) with the supplemental data.}
    \label{fig:fit_relative_astrometry_Gl229}
\end{figure*}

\begin{figure}[!ht]
    \centering
    \includegraphics[width=\linewidth]{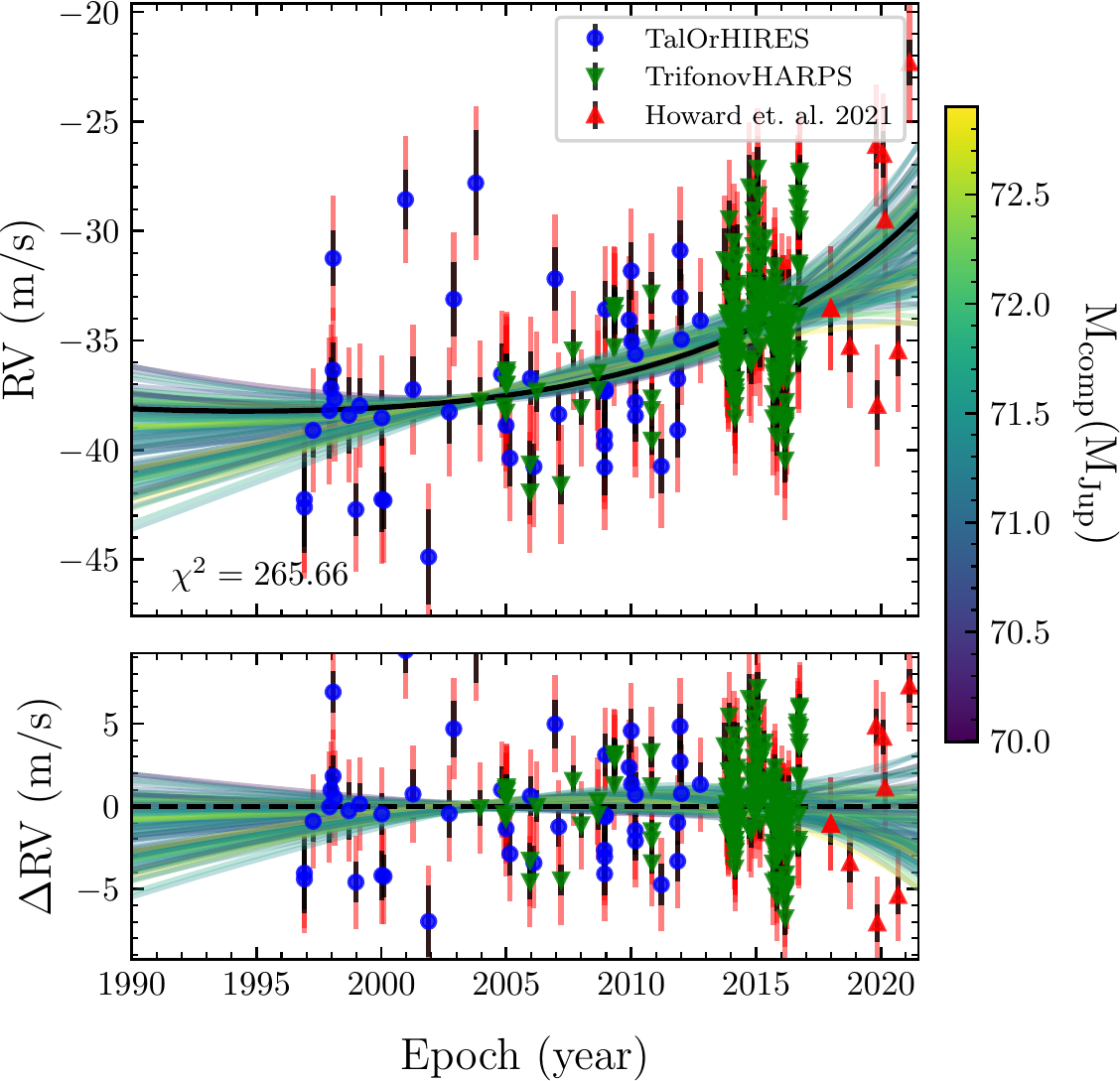}
    \caption{Top panel: The observed RVs of Gl~229~A overplot with the best fit orbit (in black) and a random sampling of other orbits from the MCMC chain. Bottom panel: The RV residuals with respect to the best fit orbit. Both panels: The random sampling of other orbits from the MCMC chain are color coded by the mass of Gl~229~B. The black error bars give the observed errors. The red error bars include the best fit jitter added in quadrature to the observed errors. The formal $\chi^2$ of the fit to the data is inset in the top panel. The complete Figure set (6 images) for this version is available as a compressed file (figset\_RV.zip) with the supplemental data.}
    \label{fig:fit_RVs_Gl229}
\end{figure}

Our orbital posteriors are summarized in Table \ref{tab:posteriors_Gl229}, \addition{which is} presented in the appendix. The corner plot and covariances of select orbital parameters are shown in Figure \ref{fig:Corner_Gl229}. We infer a mass of $71.4 \pm0.6 \,\Mjup$ for Gl~229~B, and an eccentricity of $0.851 \pmoffs{0.002}{0.008}$, the highest precision of both to-date. Our mass agrees with the previously published value of $70.4 \pm 4.8 \, \Mjup$ \citep{brandt_gliese_229b_mass_htof} yet is a factor of seven more precise. The $\chi^2$ on the \hipparcos-\gaia long term proper motion is just 0.5 (Table \ref{tab:chisquared_statistics}); the observed proper motion anomaly of Gl~229~A is in almost exactly the same direction predicted by the best-fit orbit. The fit to the RVs is summarized in Figure \ref{fig:fit_RVs_Gl229}. Because of the long period and significant RV jitter, the \hipparcos-\gaia absolute astrometry plays a crucial role in constraining the mass of the secondary. Figure \ref{fig:fit_relative_astrometry_Gl229} showcases the fit to the relative astrometry, and Figure \ref{fig:fit_propermotion_Gl229} shows the fit to the \gaia and \hipparcos proper motions.
The goodness-of-fit statistics are good for the relative astrometry ($\chi^2 = 16$ for 18 data points), but the proper motion in declination from \hipparcos is discrepant and leads to a poor $\chi^2$ of 5.5 (with 2 data points). Excluding both \hipparcos proper motions from the fit changes our best-fit mass and errors by $\leq$0.1\,$\Mjup$ ($\leq$0.2$\sigma$). Likewise, omitting the new relative astrometry and/or the new HIRES RVs changes the mass by $\lesssim$0.2\,$\Mjup$ and has a negligible effect on the precision of our mass measurement.  Nearly all of the improvement in our mass constraint comes from the \gaia EDR3 proper motions.


The new NIRC2 relative astrometry improves the mass constraint on the primary star Gl~229~A by a factor of $\approx$5, removes almost all of the covariance between primary and secondary masses (upper left panel of Figure \ref{fig:Corner_Gl229}), and reduces the semi-major axis uncertainty by a factor of 3. The other parameters are not affected by the new relative astrometry.

For Gl~229, neither the RVs nor relative astrometry cover a significant fraction of the $\approx$240-year period (Table \ref{tab:posteriors_Gl229}), as both data sets have baselines of $\approx$25 years. These facts, in addition to the nearly face-on orbit, result in a degeneracy between $\omega$ and $\Omega$. There are four local maxima for $\omega$ and $\Omega$ in the posterior distribution, with one mode significantly higher than the others 
(Figure \ref{fig:Corner_Gl229}). 
Revealing this multi-modality in an MCMC analysis requires exhaustively exploring the parameter space. 
An analysis that begins its MCMC chains from previously published orbital parameters could miss such additional modes in the posteriors. 


The true value of $\Omega$ and $\omega$ could be identified using high-precision relative astrometry. A few VLT/GRAVITY observations, taken now, would serve the same purpose (from the perspective of constraining power) as several $\sigma\approx3$\,mas measurements over the next decade from more `classic' direct-imaging instruments such as NIRC2 or HiCIAO. A single, $100~\mu$as precise GRAVITY observation \citep{Nowak_2020_beta_pic_c_direct_detection, AMLagrange2020betapicc_direct_detection} would improve the orbital period, inclination, eccentricity, $\Omega$, and $\omega$ constraints by 20\% to 80\%.  As we discuss in Section \ref{sec:bdmodels}, Gl~229~B is surprisingly massive, and could be a BD-BD binary.  Ultra-precise GRAVITY astrometry might detect the astrometric signature of such an unseen companion.



\citet{Tuomi+Jones+Barnes+etal_2014} found evidence for a new super-Earth-sized planet in the Gl~229 system, Gl~229~b. \cite{2020ApJS..246...11F} found that Gl~229~b was still yet to be confirmed but reported the discovery of an additional planet, Gl~229~c. Both are at least super-Earth-sized; their minimum masses are $8.5 \pm 2.0\,M_{\oplus}$ (Gl~229~b) and $7.3 \pm 1.3\,M_{\oplus}$ (Gl~229~c) with RV semi-amplitudes between 1 and 2 m/s \citep{2020ApJS..246...11F}. We perform an identical analysis including these two planets in our orbital fit by subtracting off their RV signals. We use the \textit{maximum a posteriori} (MAP) orbital elements provided in Table~2 of \cite{2020ApJS..246...11F}. We infer a mass for Gl~229~B (the BD) that is just 0.24~$\Mjup$ (0.4$\sigma$) higher than that from the case that ignored the candidate inner planets. The other orbital elements are nearly identical as well. 
These two inner planets combined contribute $<$3\,m/s of RV perturbations, with short orbital periods compared to the total RV baseline. This is only a factor of $\approx$2 larger than the median RV error (including jitter). The reported planets have too little mass and are too close to the star (both have semi-major axes $<1$ au; \citealt[]{2020ApJS..246...11F}) to impact the inferred mass of Gl~229~B. 


We infer a mass of $0.579 \pm 0.007 \,\Msun$ for the primary, Gl~229A. This agrees with the v.DR2 fit by \cite{brandt_gliese_229b_mass_htof}, who found $0.54 \pmoffs{0.04}{0.03} \,\Msun$. Our improved primary mass precision is entirely due to the two additional epochs of relative astrometry from NIRC2. In a fit that ignores that new relative astrometry, we find a primary mass with a factor of $\approx$5 worse precision: $0.545 \pmoffs{0.033}{0.030} \,\Msun$. This is expected because Gl~229~B has a long orbital period ($237.9 \pmoffs{5.1}{4.6}$ years) and so the \gaia proper motion is quasi-contemporaneous (i.e., all scans occurred approximately at the same orbital phase of Gl~229~B). In the single-epoch approximation (\citealp{Brandt_Dupuy_Bowler_2018}), the astrometric acceleration of Gl~229~A on the sky, combined with the parallax and angular separation, constrains only the mass of the companion; it yields no constraint on the mass of the primary. These two facts are partly why we obtain such a better secondary mass constraint after adopting \gaia EDR3 astrometry. The additional relative astrometry drives most of the precision increase on the orbital elements (including both period and primary mass, related by Kepler's third law), while the improved \gaia EDR3 absolute astrometry drives the improved precision on the secondary mass.

\subsubsection{Potential mass systematics below the 1\% level}

We achieve our highest mass precision for Gl~229~B (0.9\%), so we consider here potential systematics below the $\sim$1\% level. Unknown systematics within \gaia or \hipparcos proper motions are unlikely to be a concern; the HGCA dealt with these systematics and corrected them far below this level (Figure~6 of \citealt{BrandtGaiaEDR3HGCA}). The RV star reference set in that work (all non-accelerators according to RV trends) is nicely calibrated into a Gaussian core, with minimal evidence for outliers. 

However, a potential source of systematics is the fact that we do not have the \gaia EDR3 intermediate astrometric data (the individual positions and uncertainties per transit).
This systematic is rooted in the fitting, per MCMC step, of the five-parameter astrometric model to \gaia transits. We use the resulting positions and proper motions to compute a likelihood given the measured HGCA proper motions. A \gaia transit consists of four components: the transit time, the scan angle of the transit, the along-scan formal error, and whether this particular transit was used in the final solution. 
\htofcodename uses scan angles and epochs from the \gaia GOST\footnote{\url{https://gaia.esac.esa.int/gost/}} tool. \htofcodename automatically rejects GOST observations that fall into the documented satellite dead times. \htofcodename assumes uniform along-scan errors for all observations of one source. Deviations from these assumptions, whether from varying precision or additional rejected observations, will change the relative weighting of different transits in the astrometric fit.  As a result, the time of minimal positional uncertainty---the central epochs in the HGCA---may differ between the forward-modeled and catalog values.  Using the incorrect central epochs would lead to inferring an incorrect astrometric acceleration.  




For Gl~229, we find that \htofcodename's computed central epochs from the \gaia EDR3 GOST scanning law are only 0.036 yr and 0.017 yr different from the true \gaia EDR3 values in right ascension and declination, respectively. The acceleration that we measure is primarily between the midpoint of \hipparcos and \gaia, around 2004, and \gaia in 2016.  A discrepancy of 0.036 years is about 0.3\% of this baseline, and would lead to a $\approx$0.3\% error in the astrometric acceleration. This is a factor of $\approx$3 smaller than the $\approx$1\% precision of the HGCA acceleration measurement, though the acceleration of Gl~229~A is increasing as its companion approaches periastron.


The following test shows quantitatively the impact of the GOST approximation on Gl~229~B. By disabling \htofcodename in \orbitcodename, \orbitcodename employs a different approximation (\citealp{Brandt_Dupuy_Bowler_2018}) that forces the central epoch to be equal exactly to the catalog values. In this case, we find the best fit companion mass grows by $0.5\,\Mjup$ --- slightly less than $1\sigma$. We do not need access to the full \gaia intermediate astrometric data for Gl~229, but it will become essential in the future to push mass precisions well below 1\%. 


\subsection{Gl~758}

Gl~758~B \citep{2009_Thalmann_Gl758_discovery}, a late-T dwarf, has a rich history of dynamical mass measurements. \cite{Bowler+Dupuy+Endl+etal_2018} measured $42 \pmoffs{19}{7} \,\Mjup$ using RVs and relative astrometry. \cite{Calissendorff+Janson_2018} and \cite{Brandt_Dupuy_Bowler_2018} improved this estimate with \hipparcos-\gaia DR2 accelerations; \cite{Brandt_Dupuy_Bowler_2018} inferred $38.1 \pmoffs{1.7}{1.5} \,\Mjup$. 

We add newly calibrated RVs from \cite{TalOrHIRES} and update the absolute astrometry using 
the HGCA v.EDR3.  
Table~\ref{tab:posteriors_Gl758} (presented in the appendix) summarize our posteriors and priors. We infer a mass for Gl~758~B of $38.0 \pm 0.8 \,\Mjup$, twice as precise as the previous estimate. We infer an eccentricity of $0.24 \pm 0.11$; a circular orbit remains allowed at $\approx$2$\sigma$. The secondary mass posterior is nearly Gaussian. 
Our priors are all uninformative, but adopting a stellar-evolution informed prior on the mass of Gl~758~A has a negligible effect on Gl~758~B's mass measurement. Using a primary mass prior of $0.96 \pm 0.03\,\Msun$ (consistent with \citealt[]{Takeda_2007, Luck_2017}) yields a secondary mass that is shifted by only 0.1\Mjup. 

Our inferred eccentricity is more modest than that found by \cite{Bowler+Dupuy+Endl+etal_2018} ($0.58\pmoffs{0.07}{0.11}$), but consistent with the most recent estimate of $0.26 \pm 0.11$ by \cite{Brandt_Dupuy_Bowler_2018}. The latter work included absolute astrometric accelerations. Astrometric accelerations favor lower eccentricities than what one would infer from RVs alone. The RV baseline is short compared to the orbital period and so the RV constraint on the eccentricity is relatively weak. The mild eccentricity that we confirm for Gl~758~B cements its place in like company with HD~33632~Ab. 

\subsection{HD~13724}
We derive a dynamical mass of $36.2 \pmoffs{1.6}{1.5} \,\Mjup$ for HD~13724~B with uninformative priors on all system parameters. The posteriors are summarized in Table \ref{tab:posteriors_HD13724}. All posteriors (except for the argument of periastron, $\omega$) are nearly Gaussian.

\addition{The mass of this primary star is weakly constrained with RVs, relative astrometry, and \gaia DR2 astrometric accelerations, and poorly constrained if one excludes astrometric accelerations.} Thus, previous studies assumed an informative prior on the mass of HD~13724~A, allowing better constraints on the secondary mass. \cite{Rickman2020_HD13724} placed a Gaussian prior of $1.14 \pm 0.06 \,\Msun$ on HD~13724~A, a range inferred from the \citet{Ekstrom_etal_2012} and \citet{Georgy_etal_2013} grids of Geneva stellar models. Adopting this prior on the primary has a small ($\approx$1-2$\Mjup$, or roughly 1$\sigma$) effect on our inferred mass for the secondary.

\addition{Combining the new, higher precision \gaia EDR3 accelerations with RVs and relative astrometry yields a useful dynamical constraint on the primary mass. We find} a dynamical mass constraint 
of $0.95 \pmoffs{0.076}{0.067} \,\Msun$ for HD~13724~A, \addition{using an uninformative prior on the primary mass.} 
Our dynamical mass precision is comparable to the precision of predictions from stellar evolution: e.g.,~$1.08 \pmoffs{0.04}{0.03} \,\Msun$ reported by \cite{2018AA...614A..55A}, and the \cite{Rickman2020_HD13724} prior. We discuss the discrepancy between our dynamical mass and those from stellar evolution in Section \ref{ssec:hd13724_mesa_mass}.

Our dynamical mass for HD~13724~B, $36.2 \pmoffs{1.6}{1.5} \,\Mjup$, is in tension with the first dynamical mass measurement of $50.5 \pmoffs{3.3}{3.5} \,\Mjup$ found by \cite{Rickman2020_HD13724}, but agrees well with the minimum mass ($m \sin(i)= 26.77 \pmoffs{4.4}{2.2} \Mjup$, determined from RVs alone in the initial discovery by \cite{2019AA_RickmanHD13724_discovery}. Using our inferred inclination of $45.1 \pmoffs{2.1}{1.8}$ degrees and the minimum mass from \cite{2019AA_RickmanHD13724_discovery}, we calculate $m = 37.8 \pmoffs{7.2}{4.3} \,\Mjup$.
Our inferred eccentricity of $0.335 \pm 0.026$ is significantly more modest than $e=0.64 \pm 0.07$ reported by \cite{Rickman2020_HD13724}. Our parameters are consistent with $e=0.34 \pmoffs{0.09}{0.05}$ found by \cite{2019AA_RickmanHD13724_discovery}.


The two salient differences between our analysis and that of \cite{Rickman2020_HD13724} are that we include \hipparcos-\gaia accelerations and that we do not adopt a prior on the mass of HD~13724~A. 
If we instead adopt the same stellar mass prior of $1.14 \pm 0.06 \,\Msun$, we find a secondary mass of $38.6\pmoffs{1.2}{1.1} \,\Mjup$ and an eccentricity of $0.346 \pm 0.026$, consistent with our results using an uninformative prior on the primary star's mass. Entirely excluding the \hipparcos-\gaia accelerations does not resolve the tension either, although it does weaken it. Such an analysis yields an eccentricity of $0.39\pm 0.15$ and mass of $41.1\pmoffs{9.9}{6.2} \, \Mjup$. 

\subsection{HD~19467}

Using the same primary mass prior as \citet{2020AA_Maire_HD19467}, we infer a mass of $65.4 \pmoffs{5.9}{4.6} \,\Mjup$ for HD~19467~B. All posteriors are summarized in Table~\ref{tab:posteriors_HD19467}.
Our BD mass is consistent with $74 \pmoffs{12}{9} \,\Mjup$ as found by \cite{2020AA_Maire_HD19467} using \gaia DR2, and is roughly twice as precise. We infer an eccentricity of $0.54 \pm 0.11$, in good agreement with the $0.56 \pm 0.09$ found by \cite{2020AA_Maire_HD19467}, as well as the earlier measurement of $0.39\pmoffs{0.26}{0.18}$ by \cite{Bowler_2020_Blunt_Nielsen}. Our inferred period of $320 \pmoffs{200}{80} \,\rm{years}$ is shorter than (but fully consistent with) both $420 \pmoffs{170}{250}$ years inferred by \cite{Bowler_2020_Blunt_Nielsen} and $398\pmoffs{95}{93}$ years from \cite{2020AA_Maire_HD19467}. 

Removing the primary mass prior results in a secondary mass that is fully consistent with the measurement with a prior. However, as mentioned in Section \ref{sec:stellar_mass_priors}, this yields a posterior for the mass of the G3 dwarf star that is much broader than constraints from stellar evolution. 


HD~19467~B's high eccentricity is unlikely to be due to the interactions between it and an undiscovered, inner and massive companion; the RV curve has no signatures of residual few-year signals with semi amplitudes $\gtrsim$10\,m/s.
HD~19467~B's high eccentricity places it in like-company (among HR~7672~B and 1RXS2351+3127~B) with the BD population-level peak of eccentricities near $e=0.6-0.9$ studied by \cite{Bowler_2020_Blunt_Nielsen}. Without \gaia EDR3, \cite{Bowler_2020_Blunt_Nielsen} could only exclude zero eccentricity at $\approx$2$\sigma$.  

\subsection{HD~33632~A, Ab, \& B}


For the L/T transition object HD~33632~Ab 
we derive a dynamical mass of $50 \pmoffs{5.6}{5} \,\Mjup$, which is consistent with and 1.5 times more precise than the mass of $46.4 \pmoffs{8.1}{7.5}$ derived by \cite{2020ApJ_Currie_Thayne_HD33632}. This precision improvement is due to \gaia EDR3 astrometry. 
The other orbital parameters are modestly improved and are summarized in Table~\ref{tab:posteriors_HD33632}.
Our preferred fit uses a prior of $1.1 \pm 0.1 \, \Msun$ on the host star (Section \ref{sec:stellar_mass_priors}). An uninformative prior slightly degrades our inferred secondary mass to $50.5\pmoffs{6.2}{5.1} \Mjup$, with a posterior of $1.05\pmoffs{0.26}{0.21}\, \Msun$ for the primary. 

\cite{2020ApJ_Currie_Thayne_HD33632} found bimodalities in the posteriors for both the eccentricity and semi-major axis for HD~33632~B. 
Our new analysis (using the informative primary mass prior) and more precise \gaia EDR3 astrometry breaks both degeneracies: 
we find $a = 23.6 \pmoffs{3.2}{4.5} \,\rm{au}$ 
and $e = 0.12 \pmoffs{0.18}{0.09}$. 
Our inclination constraint is modestly improved, $45.2 \pmoffs{4.7}{11} \,\rm{degrees}$, compared to the \cite{2020ApJ_Currie_Thayne_HD33632} result.


HD~33632~A has a widely-separated co-moving M dwarf companion at a common parallax. The analysis of \cite{2020ApJ_Currie_Thayne_HD33632} noted this companion but did not include it in a dynamical fit. We perform a three-body fit to the system, adopting the priors discussed in Section \ref{sec:stellar_mass_priors}.

\gaia EDR3 provides a $\approx$20 $\mu$as constraint (a fractional separation error of $10^{-6}$) on the relative position between HD~33632~A and B. This results in sharp likelihood peaks and ridges across parameter space and slows convergence of our MCMC chain. We reduce the precision of the \gaia EDR3 relative astrometry (Table \ref{tab:relative_astrometry_stellar_companions}) by a factor of 100 (resulting in separation error of 3~mas), this results in poorer constraints on HD~33632 B but converged posteriors. Additionally, we run the chains for two million steps. 

In Figure set \ref{fig:Corner_Gl229}, we show orbital elements for HD~33632~Ab and HD~33632~B (the stellar companion), respectively, from the three-body fit. We summarize all posteriors and priors in Table~\ref{tab:posteriors_HD33632_3body}. The RV, relative astrometry, and proper motion fits look identical to the two-body fits and so those are not present in the corresponding Figure sets.


The key conclusion is that including the stellar companion does not appreciably shift the MAP values for the BD. The constraints on most of the orbital parameters of the stellar-companion are weak. However, we find a modest constraint on the inclination of HD~33632~B of $74.6\pmoffs{4.1}{11}$ degrees. We adopt the results of the two-body fit with HD~33632~A and Ab as our preferreed orbital elements for the BD.




Unlike the other BDs considered in this study, HD~33632~Ab appears to have a definitively low eccentricity. The MAP value is near 0.06 and circular orbits are allowed. But, like the other BD's studied herein, HD~33632~Ab is massive, confidently weighing between 40 and 60 $\Mjup$.

\subsection{HD~72946}
We infer a secondary mass of $72.5 \pm 1.3 \, \Mjup$ using the primary mass prior from Section \ref{sec:stellar_mass_priors} (the same prior adopted by \citealt[]{2020AA_Maire_HD72946}). This agrees well with the mass of $72.4 \pm 1.6 \,\Msun$ found by \cite{2020AA_Maire_HD72946}. We derive an eccentricity, period, and inclination of $0.489 \pm 0.007$, $15.92 \pm 0.10 \,\rm{years}$, and $59.5 \pmoffs{1.2}{1.1}$ degrees, respectively. These agree with the values reported by \cite{2020AA_Maire_HD72946}, but we improve the precision in period by a factor of $\approx$1.5 and inclination by a factor of $\approx$2.
Our eccentricity, period, and $M \sin i$ constraints all agree with the initial RV discovery work by \cite{2016AA_Bouchy2016_HD72946}.

The posteriors from this two-body fit are summarized in Table \ref{tab:posteriors_HD72946}.
Our analysis adopts an informative prior on the primary star's mass.  With an uninformative primary mass prior, we infer a much less precise BD mass of $76.2 \pmoffs{4.6}{4.2} \,\Mjup$ together with a primary mass of $1.11 \pmoffs{0.14}{0.12} \,\Msun$. 


As noted by \cite{2020AA_Maire_HD72946} and \cite{2016AA_Bouchy2016_HD72946}, HD~72946 has a wide stellar companion, HD~72945, separated by $\approx$10$^{\prime\prime}$ ($\approx$250 au) \citep{2020GaiaEDR3_catalog_summary}. Neither of the latter authors included this companion in their fit. We perform a three-body fit to the system, including this companion. We adopt priors as discussed in Section \ref{sec:stellar_mass_priors}.

In Figure set \ref{fig:Corner_Gl229}, we show orbital elements (with covariances) for HD~72946~B and HD~72945 from the three-body fit. The orbital elements are tabulated in Table~\ref{tab:posteriors_HD72945_3body}. The RV, relative astrometry, and proper motion fits look identical to the two-body fits and so those are not present in the corresponding Figure sets. As in the three-body fit to HD~33632~A/Ab/B, convergence is slowed due to the $\approx$100\,$\mu$as precision on the separation of the stellar companion.  However, the effect here is much smaller than the case with HD~33632 (where we had to inflate the \gaia EDR3 errors). We therefore quote results from fits using the relative astrometry from Table \ref{tab:relative_astrometry_stellar_companions} without inflating those errors.

The exceptional precision of EDR3 provides a good measurement of $200 \pmoffs{52}{41} \,\rm{A.U.}$ for the semi-major axis of the orbit of HD~72945 about HD~72946 (see the marginalized posterior in the corner plot from Figure set \ref{fig:Corner_Gl229}). We obtain a good measurement because we have four constraints for the six phase space components of the stellar companion: separation, PA, and both proper motions.

Comparing the three-body and two-body fits, The inferred BD mass is shifted by less than 0.5$\sigma$, and the eccentricity is nearly identical. Like with HD~33632, including the outer stellar companion, HD~79245, does not appear to influence significantly the inferred properties of the BD companion. As with HD~33632, we adopt the two-body parameters for the BD due to the exceptional quality of those chains.



HD~72946~B, like Gl~229~B, is a BD whose mass is near the hydrogen-burning limit.  Continued orbital monitoring and better measurements of the RV trend will establish the orbit of the outer, stellar companion and determine whether the HD~72946~AB / HD~72945 system is unstable to Kozai-Lidov oscillations \citep{Kozai_1962}.

\section{Primary masses and stellar evolution}\label{sec:stellar_masses_from_mesa}

We highlight here our model-independent primary masses for the three systems where we have useful constraints. These are HD~13724, Gl~229, and HD~72946. We focus our discussion on the dynamical masses of HD~13724~A and Gl~229~A.
Our measurement of the high-mass M-dwarf Gl~229~A is especially precise (1.2\%; $\pm 0.007 \Msun$). The high-mass end of M dwarfs is particularly interesting due to the well-known $\approx$10\% tension between models and observations in radius-mass space \citep[see Figure 21 of][]{2016ApJ...823..102C}. A relatively small number of precise individual masses constrain the $\approx$0.4--0.6\,\Msun\ regime of the mass-magnitude relation \citep[Figure 21 of][]{2016AJ....152..141B}. 


We calculate bolometric luminosities for HD~13724~A, Gl~229~A, and HD~72946~A by combining Tycho \citep{Hog_2000} $V_T$ and $B_T$ magnitudes with a parallax-distance from \gaia EDR3 and a bolometric correction. This procedure is described in detail in \citet{2021_Li_Yiting_nine_masses_RV_planets}. In brief, we convert the Tycho $B_T - V_T$ index into the Johnson $B-V$ index with the transformations provided by \cite{HIP_TYCHO_ESA_1997} (Eqn. 1.3.26 therein). We adopt the bolometric corrections from Table 5 of \cite{Pecaut_2013}. We use the \gaia EDR3 parallax to obtain a distance posterior, and thereafter bolometric magnitude and luminosity posteriors. Table \ref{tab:stellar_luminosities} shows the resulting bolometric luminosities and 1-$\sigma$ confidence intervals together with our dynamical mass measurements.

\begin{deluxetable}{cCc}
\tablewidth{0pt}
    \tablecaption{Bolometric luminosities for the three sources with dynamical primary mass constraints.\label{tab:stellar_luminosities}}
    \tablehead{
    \colhead{Identifier} & \colhead{Bol. Luminosity} & \colhead{Mass} \\
    \colhead{} & \colhead{$\Lsun$} & \colhead{$\Msun$}
    }
    \startdata
    HD 13724 A & 1.199\pm0.014 & $0.95 \pmoffs{0.08}{0.07}$\\
    Gl 229 A & 0.0430\pm0.0005 & $0.579 \pm 0.007$\\
    HD 72946 A & 0.871\pm0.009 & $1.11 \pmoffs{0.14}{0.12}$\tablenotemark{a} \\
    \enddata
        \tablenotetext{a}{This posterior comes from the orbital fit to HD~72946~A/B that did not include a prior on the mass of A.}
\end{deluxetable}

\subsection{Gl~229~A \& the Mass-Luminosity-Relation}\label{ssec:MLR_relation_gl229A}

With our new 1.2\% precise mass for Gl~229~A, the star becomes one of the few early-M dwarfs with an individually measured dynamical mass.  
We now compare it to calibrations of mass-magnitude relations from \citet{2016AJ....152..141B} and \citet{2019ApJ...871...63M}, sometimes also referred to as the Mass-Luminosity Relation (MLR). \citet{2016AJ....152..141B} used mass measurements of individual stars within binaries; \citet{2019ApJ...871...63M} used the larger sample of binaries with measured total system masses. 


We use $V=8.129$\,mag from \cite{2015A&A...580A..23P}, adopting an error of $0.010$\,mag, and the \gaia EDR3 parallax of $173.574 \pm 0.017$\,mas to calculate an absolute magnitude of $M_V = 9.326 \pm 0.010$\,mag. That agrees with $M_V = 9.33 \pm 0.01$\,mag reported by \cite{2009A&A...501..941H}. 
Gl~229~A is so bright in the infrared that 2MASS gives a poor photometric measurement, so we use $K_s = 4.15\pm0.05$\,mag from \citet{Leggett1992_ApJS_82_351}, assuming that the conversion between 2MASS and CIT photometric systems is negligible within the errors.

\begin{figure}
    \centering
    \includegraphics[width=\linewidth]{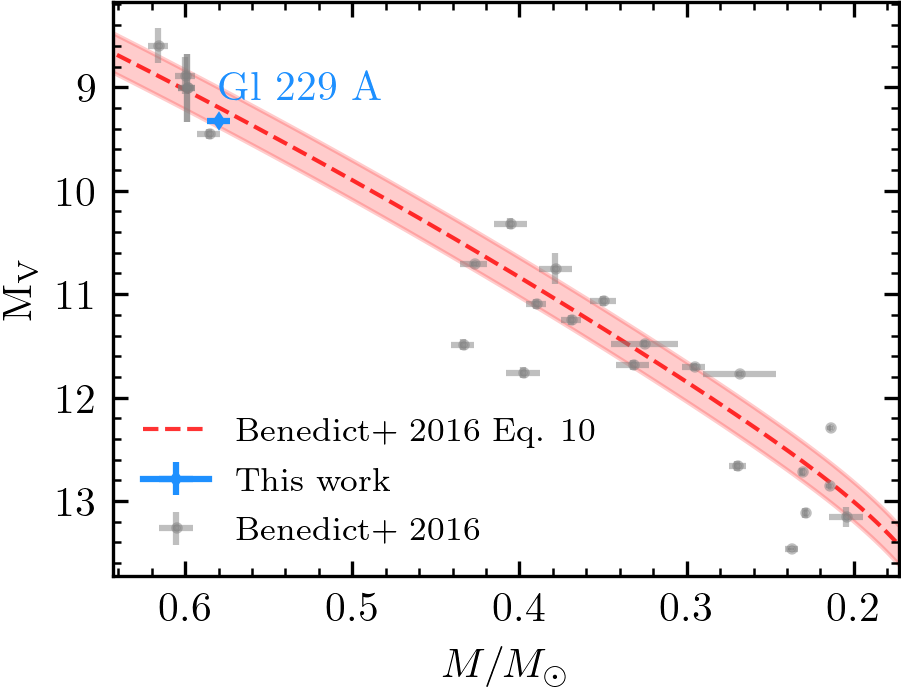}
    \includegraphics[width=\linewidth]{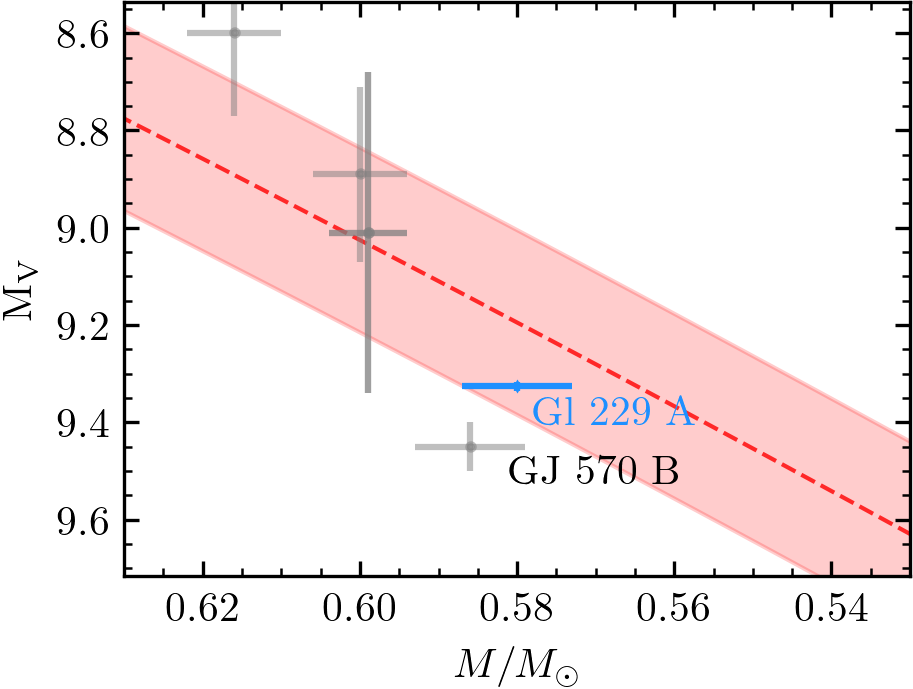}
    \caption{Absolute magnitude as a function of mass for Gl~229~A (shown in blue using the mass from this work) and binaries from \cite{2016AJ....152..141B} (shown in gray). The best-fit mass-magnitude relation from \cite{2016AJ....152..141B} is the dashed red line, and the red region gives the root-mean-square scatter about the best fit relation (0.19 in $M_{\rm V}$; \citealp{2016AJ....152..141B}). The error bar on $M_{\rm V}$ for Gl~229~A is smaller than the diamond plotting symbol in both panels. Gl~229~A borders a gap at $\approx$0.45--0.55\,\Msun\ in the \cite{2016AJ....152..141B} calibration sample. The lower panel shows a zoom-in highlighting Gl~229~A and the five other stars on the high-mass end of the \cite{2016AJ....152..141B} sample (GJ~570~B, GJ~278~C/D, GU~Boo~A/B). Note that GJ~278~C/D overlap.}
    \label{fig:Benedict_MLR_2016}
\end{figure}

Figure~\ref{fig:Benedict_MLR_2016} displays Gl~229~A and the binaries from \cite{2016AJ....152..141B} in mass-$M_{V}$ space. Their best-fit double-exponent empirical mass-magnitude relation is overplotted as a dashed red line. 
Our mass for Gl~229~A is as precise as the other five stars above $0.5 \,\Msun$ from the \cite{2016AJ....152..141B} sample. However, binary stellar evolution has played an unknown but potentially important role in some of the other stars that constrain the mass-magnitude relation at this high-mass end. GJ~278~C/D, for example, has a sub-1~day orbital period \citep{2013ApJ_GJ278CD}, implying that stellar tides may have influenced their evolution. Gl~229~A, with its distant BD companion, is free from such concerns. 
As the bottom panel of Figure~\ref{fig:Benedict_MLR_2016} shows, our dynamical mass for Gl~229~A agrees within the uncertainties of the mass-$M_V$ relation from \cite{2016AJ....152..141B}.

We use the code\footnote{\url{https://github.com/awmann/M_-M_K-}} provided by \cite{2019ApJ...871...63M} to estimate a mass of $0.549\pm0.015\,\Msun$ for Gl~229~A from its $K$-band photometry and parallax. This is lower than our measured mass but consistent within 1.8$\sigma$. Thus, Gl~229~A provides a remarkable corroboration of both the empirical mass-magnitude relation of \citet{2016AJ....152..141B} and \citet{2019ApJ...871...63M} at their quoted uncertainties.


\subsection{Comparing the dynamical and stellar-evolution masses for HD~13724~A}\label{ssec:hd13724_mesa_mass}

Our dynamical mass for HD~13724~A is $0.95 \pmoffs{0.08}{0.07} \,\Msun$, about half as precise as predictions from stellar evolution. However, stellar evolution predictions favor systematically higher masses: $1.101 \pm 0.022 \,\Msun$ \citep{2019AA...624A..78D_HD13724}, $1.120 \pm 0.010 \,\Msun$ \citep{2021AA...646A..77G}, and $1.08 \pmoffs{0.04}{0.03}\,\Msun$ \citep{2018AA...614A..55A}. 
To confirm whether our dynamical mass is discrepant with stellar evolution models, we build solar-like, non-rotating models using Modules for Experiments in Stellar Astrophysics (MESA) \citep[version 15140;][]{2011ApJS..192....3P,2013ApJS..208....4P,2015ApJS..220...15P,2018ApJS..234...34P,2019ApJS..243...10P}. 

We start by calibrating a solar model using the simplex test suite in MESA, which adopts \cite{GS98} abundances and includes the effects of diffusion \citep{Thoul1994} and exponential overshoot mixing \citep{Herwig2000}. 
The key parameters are the initial mass fractions of helium ($Y$) and metals ($Z$), the mixing length parameter ($\alpha$), and the overshoot parameter ($f_{\mathrm{ov}}$). The solar-calibrated values of $\alpha$ and $f_{\mathrm{ov}}$ are then used to generate a set of models with masses ranging from $0.95 \, \Msun$ to $1.15 \, \Msun$. We use the tracks in this mass range to extract a range of masses that agree with the observed luminosity and $T_{\rm eff}$. 

We adopt a linear enrichment law, $Y = Y_{\mathrm{p}} + ( \mathrm{d} Y / \mathrm{d} Z ) Z$ where $ Y_{\mathrm{p}} = 0.249 $ and $ \mathrm{d} Y / \mathrm{d} Z  = 1.4$. We use the effective temperature (\Teff) of $5824 \pm 19$\,K from \cite{2013A&A...555A.150T}. We inflate the errors to $\pm$50K to reflect the spread of the other \Teff\, measurements in the literature from high resolution spectra \citep{2014A&A...562A..92D, 2015A&A...574A.124D, Soubiran+Campion+Brouillet+etal_2016}. 

\begin{figure}
    \centering
    \includegraphics[width=\linewidth]{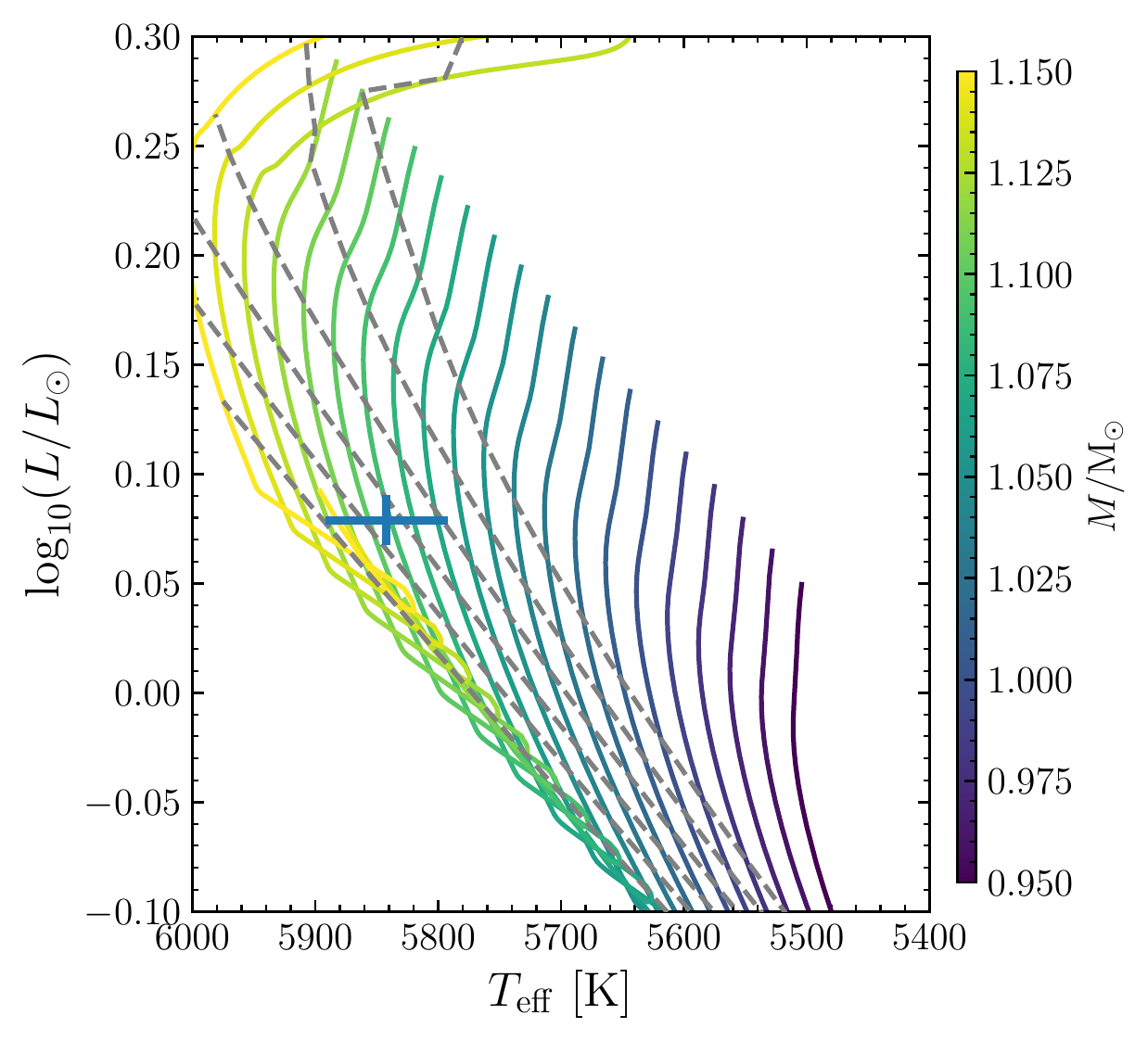}
    \caption{Newly computed MESA evolution tracks at a variety of masses, tuned to HD~13724~A's surface composition and calibrated with the simplex test suite. 
    The blue point and error bars give the luminosity from Table \ref{tab:stellar_luminosities} and the (error inflated) $T_{\rm eff}$ from \cite{2013A&A...555A.150T}.
    Each model is run until core hydrogen is exhausted. The dashed gray lines are isochrones at 1,2,3,4,5 and 6 Gyr; 1 Gyr is the bottom leftmost and 6 Gyr is the top rightmost. Models with masses of $1.10 \pm 0.02 \Msun$ and ages 1--4 Gyr are consistent with the observed $T_{\rm eff}$ and luminosity.}
    \label{fig:HD13724_logg_L_MESA}
\end{figure}

Figure \ref{fig:HD13724_logg_L_MESA} plots our MESA models in a Hertzsprung-Russell diagram, along with isochrones at 1 Gyr steps. Models with an initial $[Z/X] = 0.26$--$0.27$, ages of 1--4\,Gyr, and masses $1.10 \pm 0.02\,\Msun$ simultaneously match the observed luminosity ($1.199\pm0.014$\,\Lsun), \Teff, and surface $[Z/X]$ ($0.241$; \citealp{2020AA_Nissen_HD13724_metallicity}). Our 0.95-\Msun\ model is a factor of 1.5 too low in luminosity at the measured effective temperature, firmly ruling it out. 
However, HD~13724's dynamical mass has a sufficiently large uncertainty that the tension with our MESA-derived mass of $1.10 \pm 0.02\,\Msun$ is $\approx$2$\sigma$.  Future data will improve the mass precision and determine whether this system is meaningfully discrepant with the predictions of stellar evolutionary models.

Evolutionary models can also constrain the star's age given a luminosity and either an effective temperature or a dynamical mass. Figure \ref{fig:HD13724_logg_L_MESA} shows that MESA currently provides only a weak constraint of 1--4\,Gyr. Even this constraint depends on modeling details in the convective zone that set the effective temperature. A precise dynamical mass would remove this dependence and enable a better age estimate from stellar models.
%



\section{Benchmark Tests of Substellar Evolutionary Models}\label{sec:bdmodels}
Substellar cooling models predict an object's luminosity given its age, mass, and composition.  Benchmark BDs with known physical parameters provide the strongest tests of these models.
Here we combine our masses and ages with measured BD luminosities to assess substellar evolutionary models.

\subsection{Overview of Evolutionary Models}

We consider three sets of evolutionary models.  Each makes different assumptions about the most influential unknown: the atmospheric boundary condition. The earliest-developed models in our set are from \cite{Burrows+Marley+Hubbard+etal_1997}.  These make a number of different assumptions than the more recent models we consider, the most important being their lower atmospheric opacities. 
This is partly due to knowledge of opacity sources improving and expanding as more complete molecular line lists have been developed with time. 
The \cite{Burrows+Marley+Hubbard+etal_1997} models also use lower-opacity ``gray'' atmospheres at higher temperatures, as their main focus 
was to explore cooler BDs and giant planets ($\Teff<1300$\,K). 

The second set of substellar models is from \cite{Saumon+Marley_2008}.  The ``hybrid'' calculations of these evolutionary models assume cloudy atmospheres at warmer effective temperatures ($\Teff>1400$\,K), no clouds at $\Teff<1200$\,K, and a combination of cloudy and cloud-free atmospheres at intermediate temperatures. 

The third set are the ATMO~2020 evolutionary models from \cite{phillips_2020_atmo_models}.  These are the latest cloud-free evolutionary models in the same lineage as ``Cond'' \citep{Baraffe+Chabrier+Barman+etal_2003} and BHAC15 \citep{bhac15}. Unlike the hybrid models from \cite{Saumon+Marley_2008} that are applicable over the whole \Teff\ range of the companions we examine here, ATMO~2020 is only intended to apply to cloud-free, later-type T~dwarfs like Gl~229~B, Gl~758~B, and (marginally) HD~19467~B. For completeness, we still compare all companions to all models, even though the ATMO~2020 and \cite{Burrows+Marley+Hubbard+etal_1997} models are only intended for cooler BDs.

\begin{figure}
    \centering
    \includegraphics[width=3.25in]{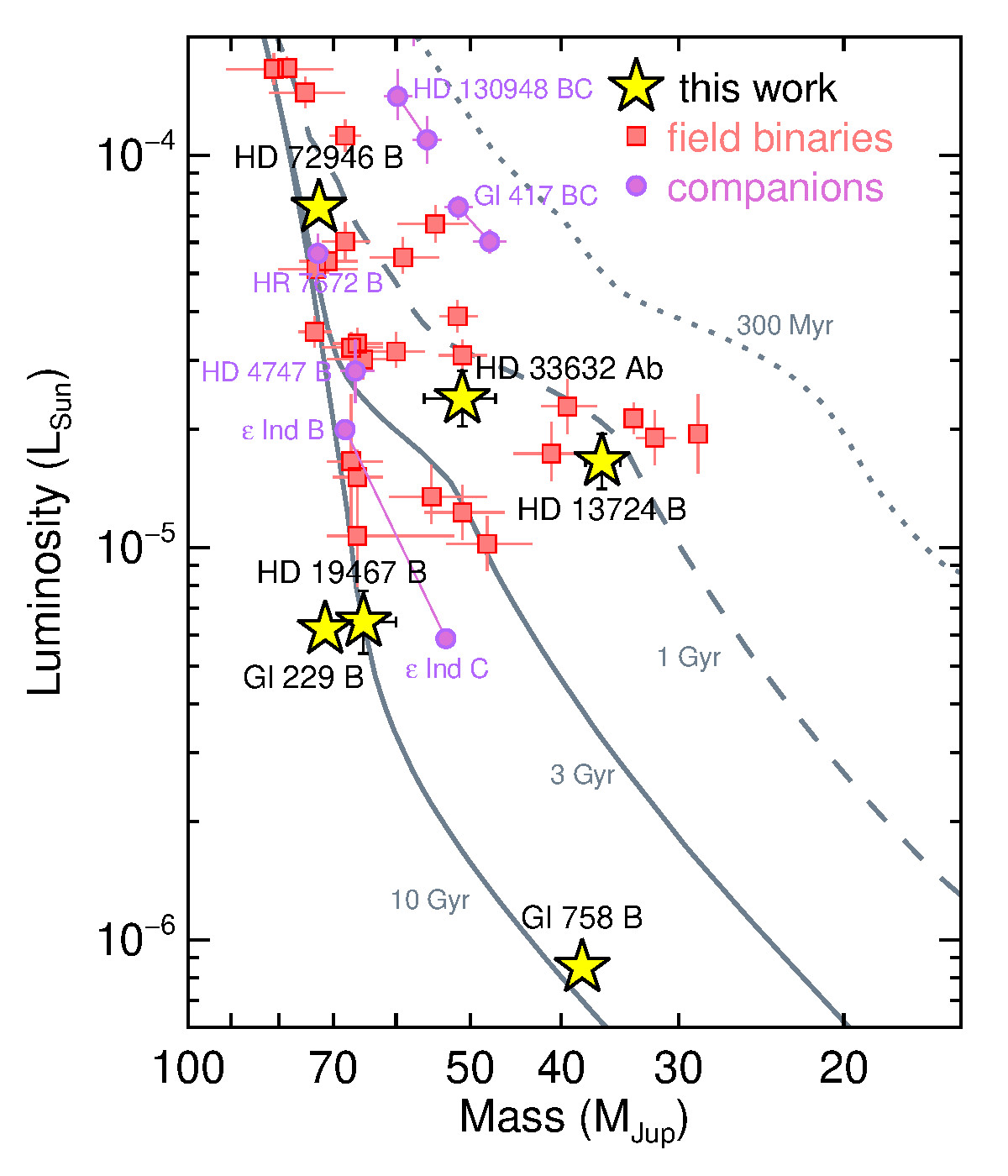}
    \caption{Luminosity as a function of mass for our sample (yellow stars) compared to the field dynamical mass sample (red squares), other companions (purple circles), and hybrid \citet{Saumon+Marley_2008} evolutionary models (gray lines). The ensemble of 41 objects plotted here generally overlap with evolutionary models at the ages expected for field dwarfs; only Gl~229~B is more massive (less luminous) than the oldest model isochrone at its luminosity (mass). Most of the 35 literature masses here are from \citet{Dupuy+Liu_2017}, with additional masses from \citet{cardoso2012}, \citet{lazorenko+sahlmann2018}, \citet{Brandt_Dupuy_Bowler_2018}, \citet{Dupuy2019_AJ_158_174}, and \citet{2020MNRAS.495.1136S,2021MNRAS.500.5453S}. \addition{There are dynamical mass measurements for $\epsilon$~Ind~BC from both \citet{cardoso2012} and \citet{2018ApJ_epsi_indi_bc_Dieterich}. We opt to plot the former because they are in better agreement with a new analysis of the orbit by Chen et~al.\ (2021, in prep.).}}
    \label{fig:m-l}
\end{figure}

\subsection{Description of Benchmark Tests}

We perform two types of benchmark tests.  For one, we use our determinations of age and \Lbol\ to derive a model mass that we compare to our dynamical masses.  For the other, we derive BD ages from models given \Lbol\ and mass measurements and compare to our inferred host star ages. Not all models are computed beyond an age of 10\,Gyr, so in the following analysis we restrict all age distributions to $\leq$10\,Gyr. Figure~\ref{fig:m-l} shows our sample in comparison to evolutionary model isochrones, as well as other previous mass measurements for ultracool dwarfs.

Our method for computing posterior distributions is based on the Monte Carlo rejection sampling approaches described in \cite{Dupuy+Liu_2017} and \cite{Dupuy2018_AJ_156_57}. We begin by drawing random masses and ages.  When inferring mass, we use our age posterior and a distribution uniform in $\log{M}$; when inferring age we use a distribution uniform in time and our dynamical mass posterior.  We then bi-linearly interpolate the evolutionary model grid at each mass and age to compute a test $L_{\rm bol}^{\prime}$ along with effective temperature, radius, and surface gravity. For each luminosity test value, we compute $\chi^2 = (\Lbol-L_{\rm bol}^{\prime})^2/\sigma_{L_{\rm bol}}^2$ and then determine the global minimum of all trials ($\chi^2_{\rm min}$). We accept each pair of mass and age into our output posterior with probability $\exp(-(\chi^2-\chi^2_{\rm min})/2)$. This produces not only an output distribution of the parameter of interest (mass or age) but also any other properties interpolated from the evolutionary models. 

To perform quantitative benchmark tests, we then compare the model-derived posterior distributions to the observed ones.  We use a one-tailed test of the null hypothesis (that the two distributions are consistent), following \cite{Bowler+Dupuy+Endl+etal_2018}. Given independent draws from the two distributions, we compute the probability that a draw from the model-derived posterior is larger or smaller than the draw from the observed posterior.  We convert this probability into a Gaussian sigma. 

\subsection{Summary of Luminosity Measurements}

Not all of our benchmark sample have published \Lbol\ values, so we derive those that are needed using empirical relations. For Gl~229~B and Gl~758~B we use the same \Lbol\ values from \citet{2015ApJ...810..158F} and \citet{Bowler+Dupuy+Endl+etal_2018}, respectively. 

For HD~19467~B and HD~33632~Ab, we use the $K$-band absolute magnitude--\Lbol\ relation of \citet{Dupuy+Liu_2017} and the photometry reported by \citet{2014Crepp_Johnson_TrendsV} and \citet{2020ApJ_Currie_Thayne_HD33632} to compute luminosities of $\log(\Lbol/\Lsun) = -5.19\pm0.08$\,dex and $-4.62\pm0.07$\,dex, respectively. Our derived luminosity for HD~19467~B is consistent with those by \citet{2020AA_Maire_HD19467} who found $\log(\Lbol/\Lsun) = -5.17\pmoffs{0.10}{0.08}$\,dex and $-5.31\pm0.12$\,dex from $J$ and $K$ bands, respectively. Our derived luminosity for HD~33632~Ab agrees with \citet{2020ApJ_Currie_Thayne_HD33632}, who found $\log(\Lbol/\Lsun) = -4.62\pmoffs{0.04}{0.08}$\,dex. 

For HD~13724~B, we first convert the \citet{Rickman2020_HD13724} SPHERE medium-band photometry to standard systems. We compute synthetic photometry from their best-matching template spectrum \citep[2MASS~J10595185+3042059;][]{2009AJSheppard_Cushing}. For the MKO system, we find $J_{\rm MKO}=17.41\pm0.05$\,mag, $H_{\rm MKO}=17.61\pm0.14$\,mag, $K_{\rm MKO}=17.16\pm0.17$\,mag. For the 2MASS system we find $J_{\rm 2MASS}=17.62\pm0.05$\,mag, $H_{\rm 2MASS}=17.55\pm0.14$\,mag, $K_{\rm S, 2MASS}=17.03\pm0.17$\,mag. Using the $K_{\rm MKO}$-band photometry and \citet{Dupuy+Liu_2017} \Lbol\ relation we find $\log(\Lbol/\Lsun) = -4.78\pm0.07$\,dex. 

\addition{For HD~4113~C, we adopt a bolometric correction of BC$_J = 2.0\pm0.5$\,mag based on Figure~12 of \citet{2015ApJ...810..158F} that, combined with the photometry reported in \citet{Cheetham+Segransan+Peretti+etal_2018}, gives $\log(\Lbol/\Lsun) = -6.30\pm0.22$\,dex.}

While HD~72946~B has a luminosity of $-4.11\pm0.10$\,dex from \citet{2020AA_Maire_HD72946}, for consistency and improved precision, we compute one here using the \citet{Dupuy+Liu_2017} relation. As with HD~13724~B, we first convert the SPHERE photometry from \citet{2020AA_Maire_HD72946} to the MKO and 2MASS systems using their best-matching template spectrum \citep[2MASS~J03552337+1133437;][]{2014ApJ...794..143B}. We find $H_{\rm MKO} = 14.55\pm0.05$\,mag and $H_{\rm 2MASS} = 14.50\pm0.05$\,mag, resulting in $\log(\Lbol/\Lsun) = -4.133\pm0.023$\,dex.

\begin{figure*}
    \centering
    \includegraphics[width=0.32\textwidth]{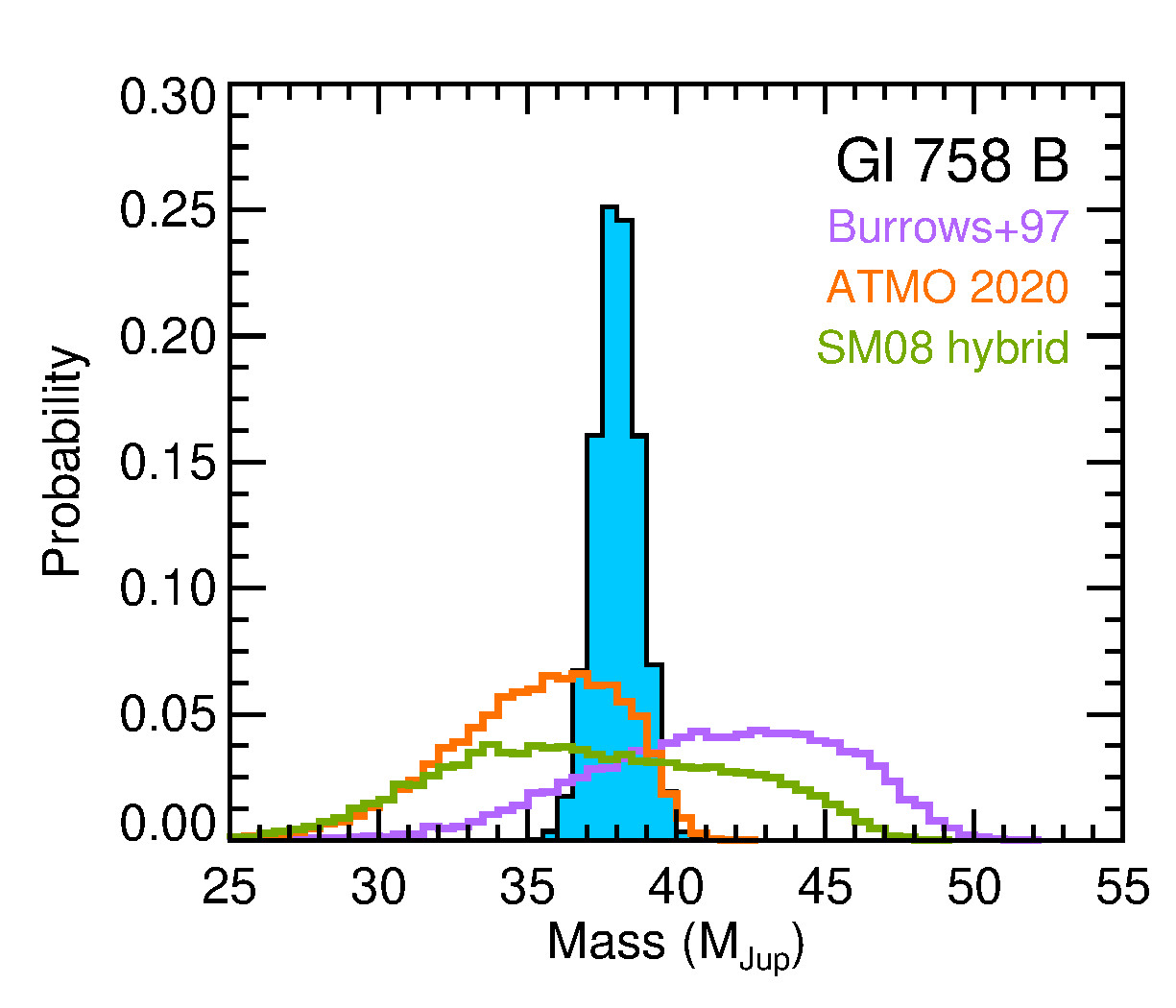}
    \includegraphics[width=0.32\textwidth]{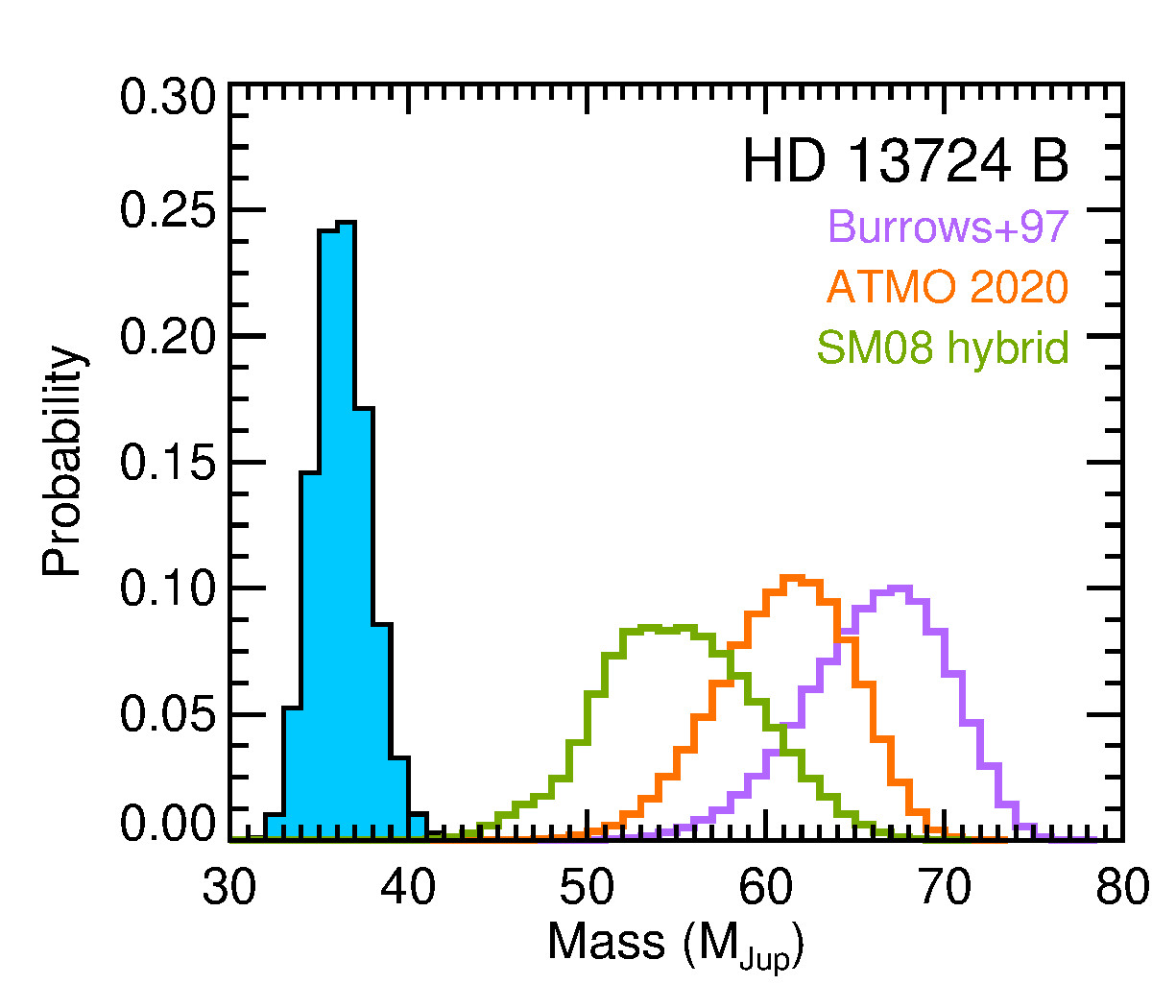}
    \includegraphics[width=0.32\textwidth]{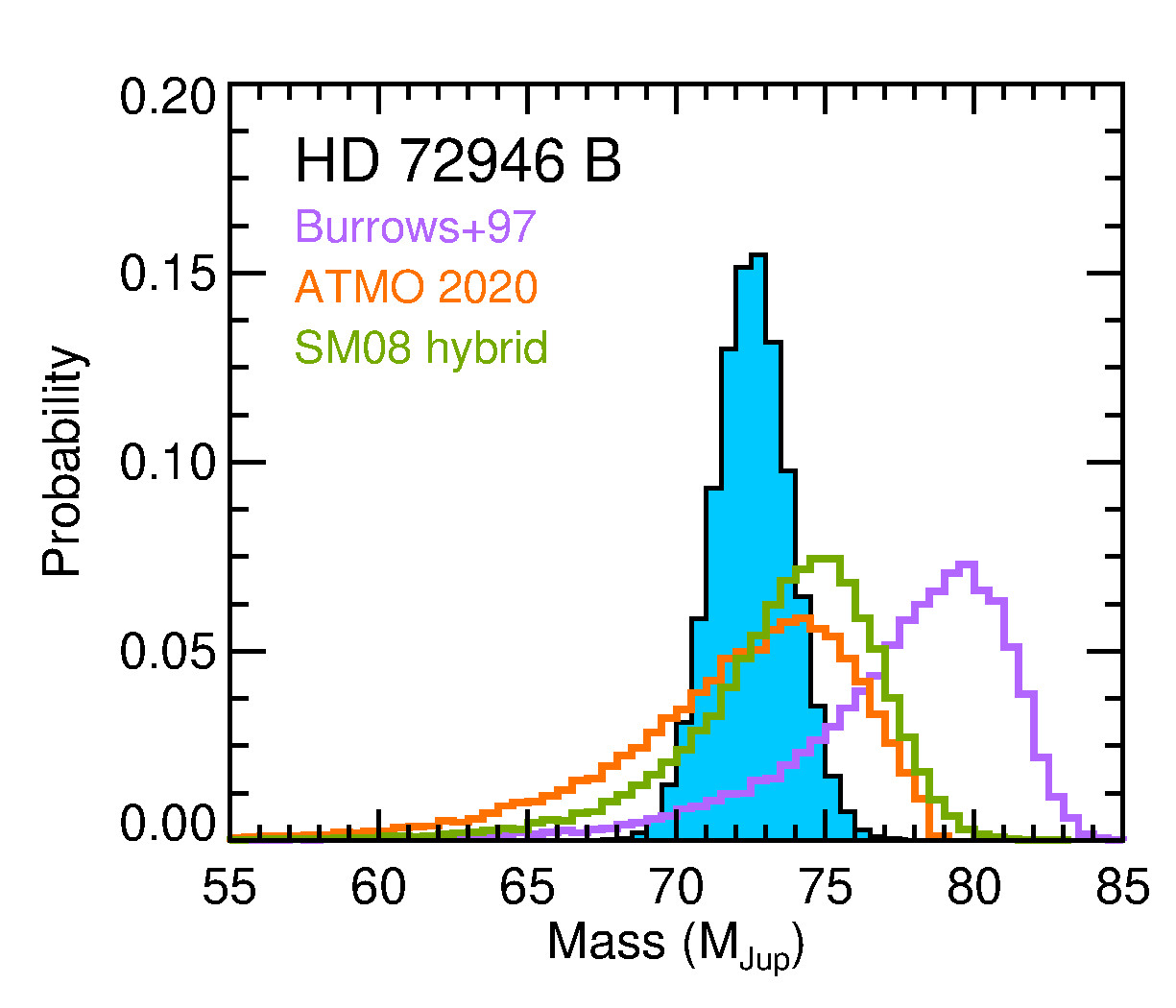} \\
    \includegraphics[width=0.32\textwidth]{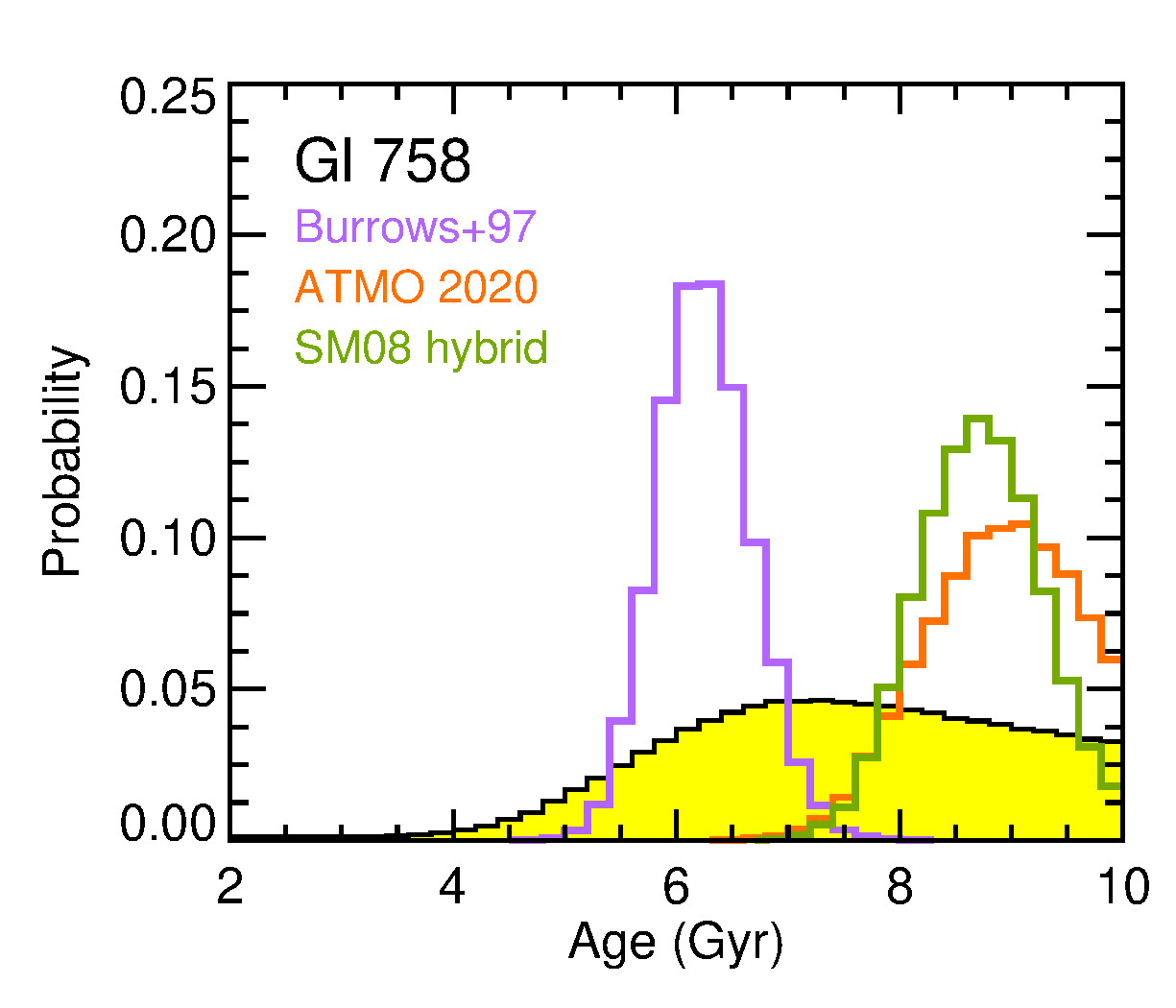}
    \includegraphics[width=0.32\textwidth]{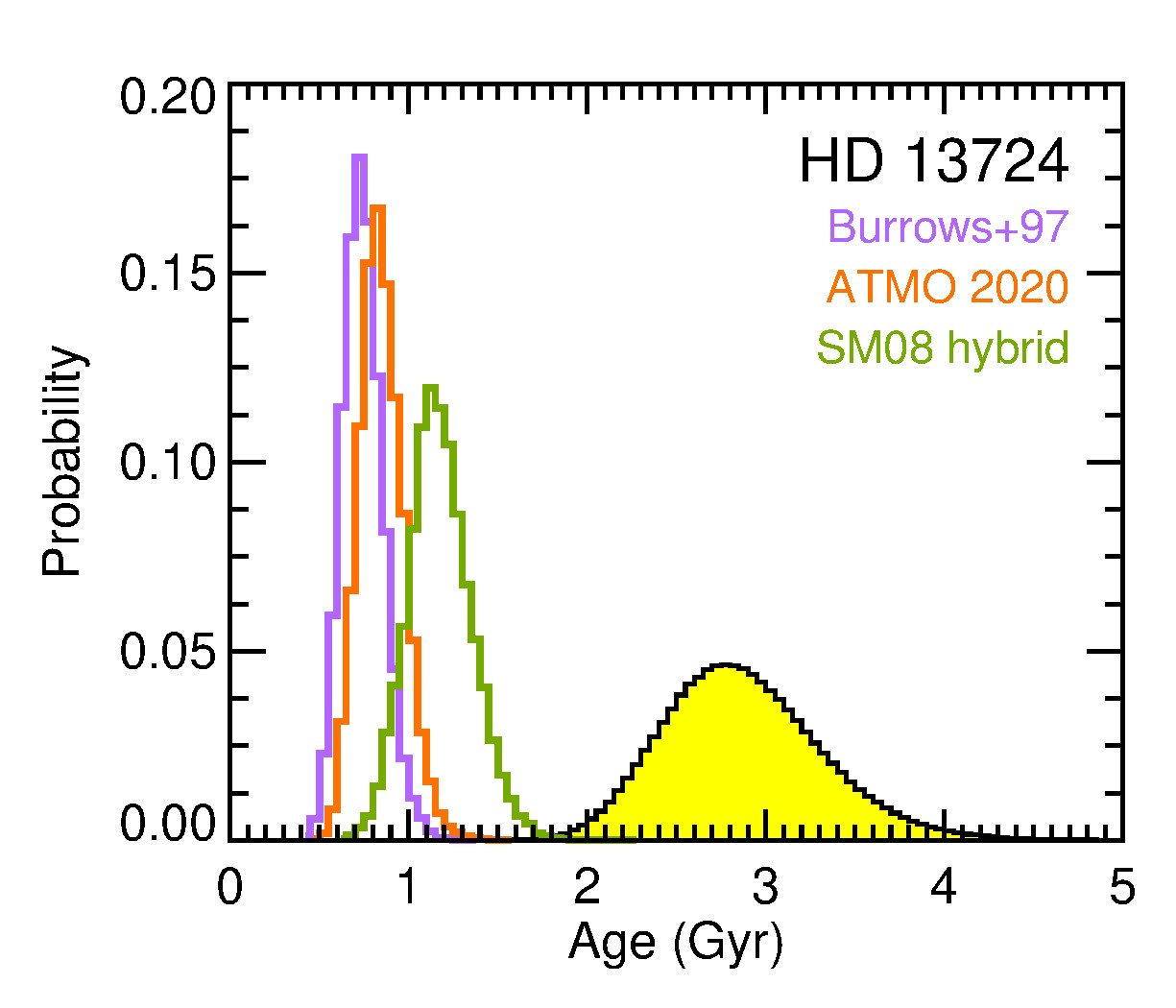}
    \includegraphics[width=0.32\textwidth]{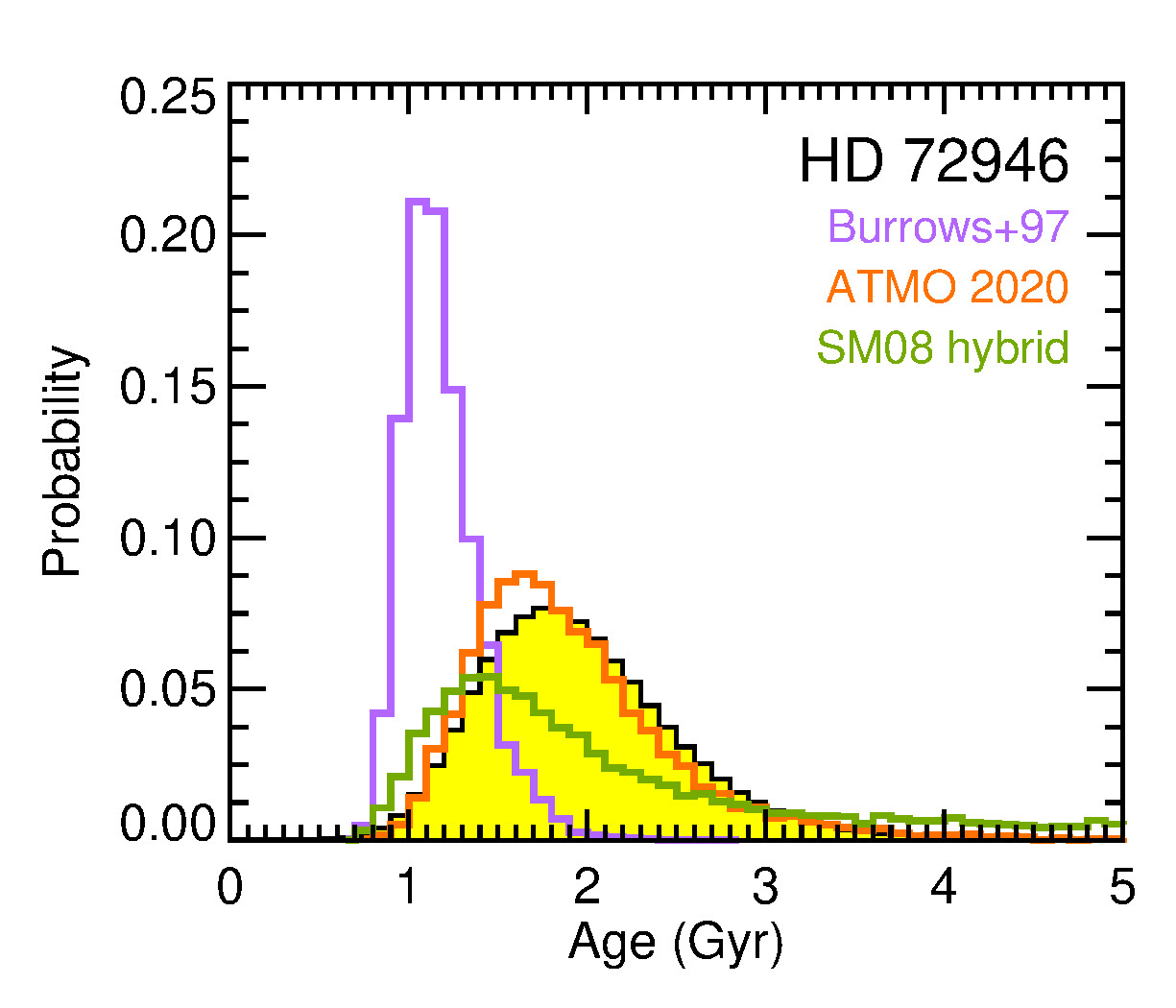}
    \caption{Top row: dynamical mass posterior distributions (filled blue histograms) compared to the mass posteriors derived from substellar evolutionary models given \Lbol\ and host star ages (unfilled histograms). Bottom row: host star activity-based age probability distributions (filled yellow histograms) compared to the age posteriors derived from evolutionary models given \Lbol\ and mass (i.e., substellar cooling ages). Objects shown here have masses measured so precisely that the limiting factor is the uncertainty in the host star's age. Gl~229~B would be shown here, but as discussed in Section~\ref{sec:bdmodels}, its luminosity is too low to be consistent with the ATMO~2020 or SM08 models even at 10~Gyr. HD~13724~B is the only object here that appears significantly discrepant with any models. As a mid-T dwarf, clouds are likely to be important in modeling its evolution, thus only the hybrid models (3.8$\sigma$ disagreement) are relevant for it.}
    \label{fig:benchmark1}
\end{figure*}

\begin{figure*}
    \centering
    \includegraphics[width=0.32\textwidth]{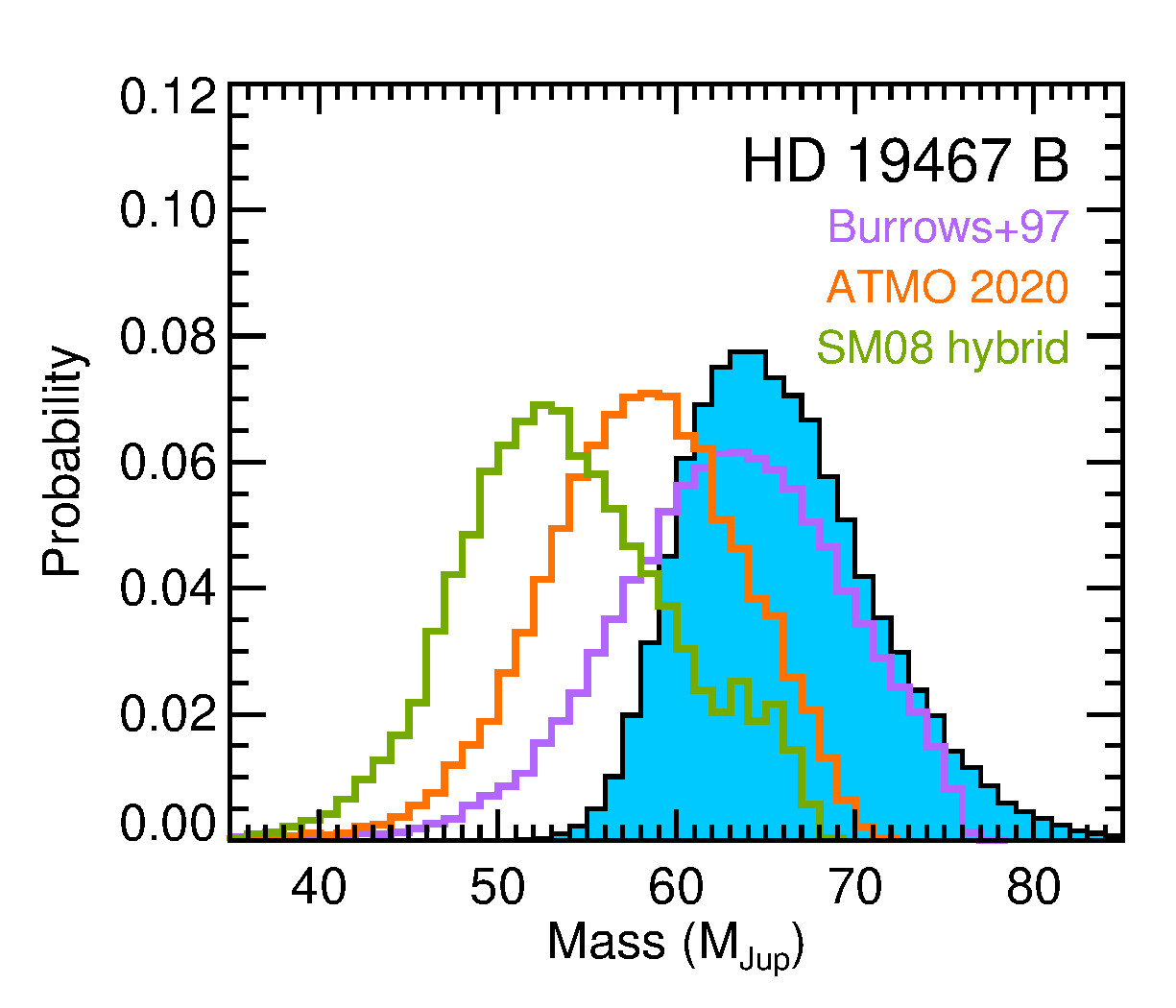}
    \includegraphics[width=0.32\textwidth]{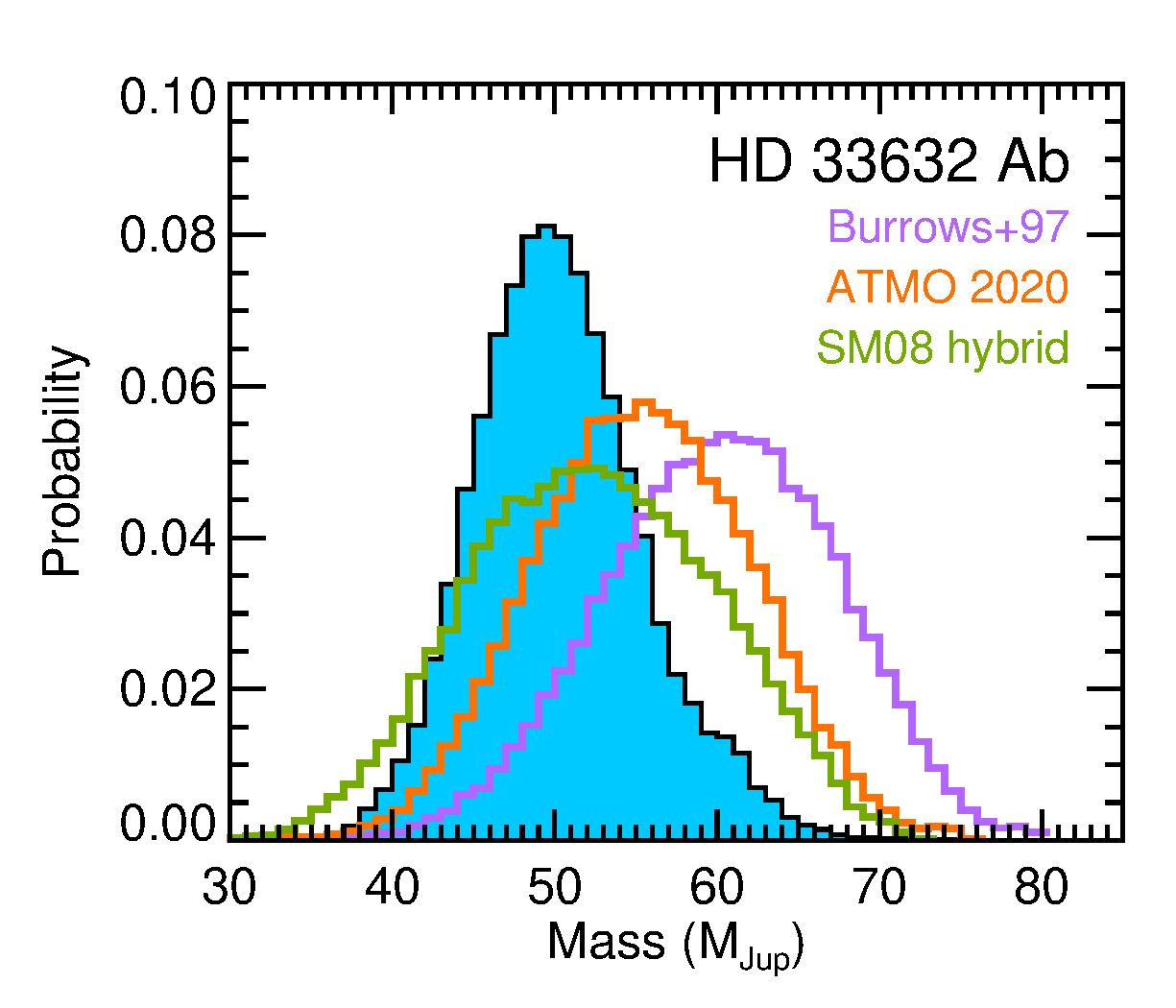} \\
    \includegraphics[width=0.32\textwidth]{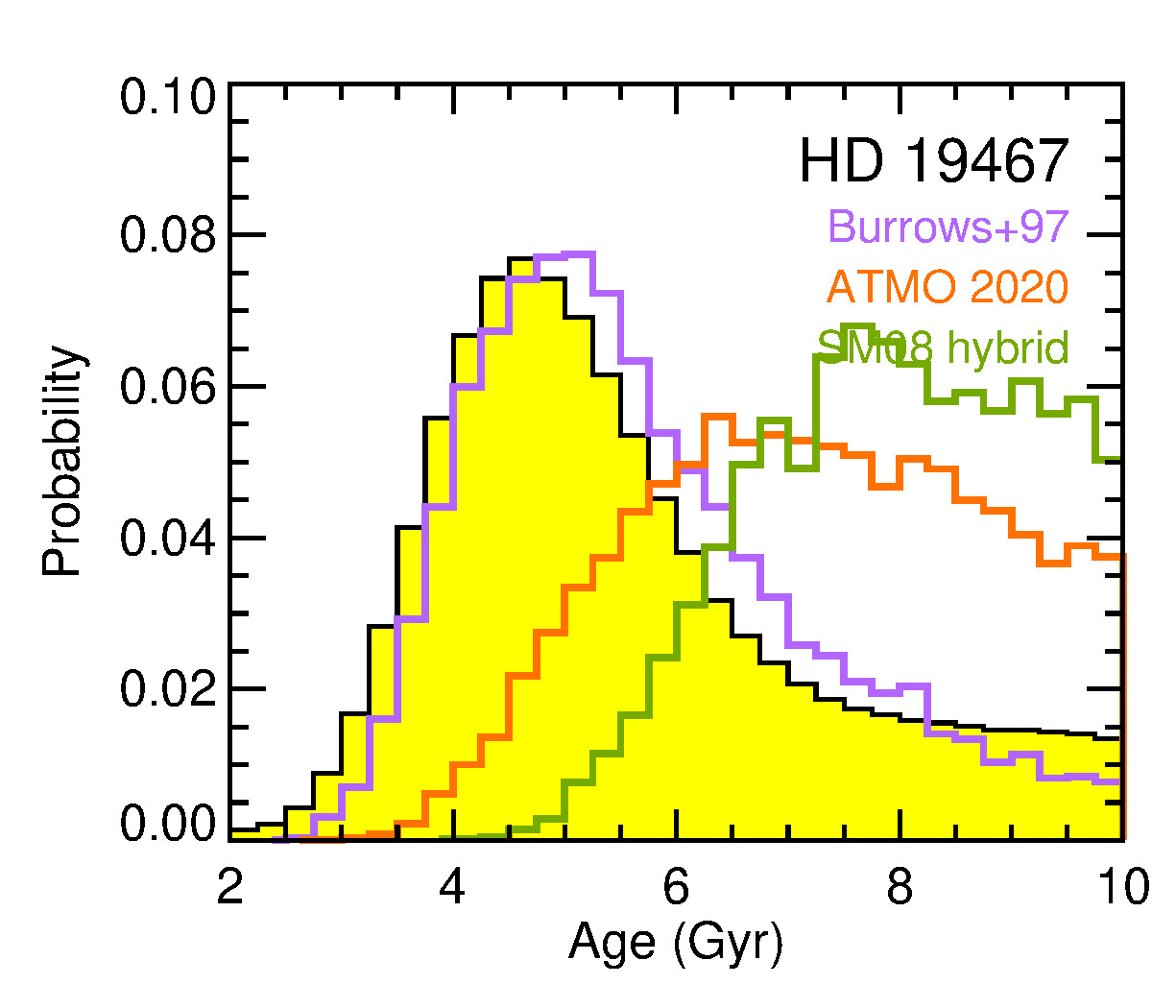}
    \includegraphics[width=0.32\textwidth]{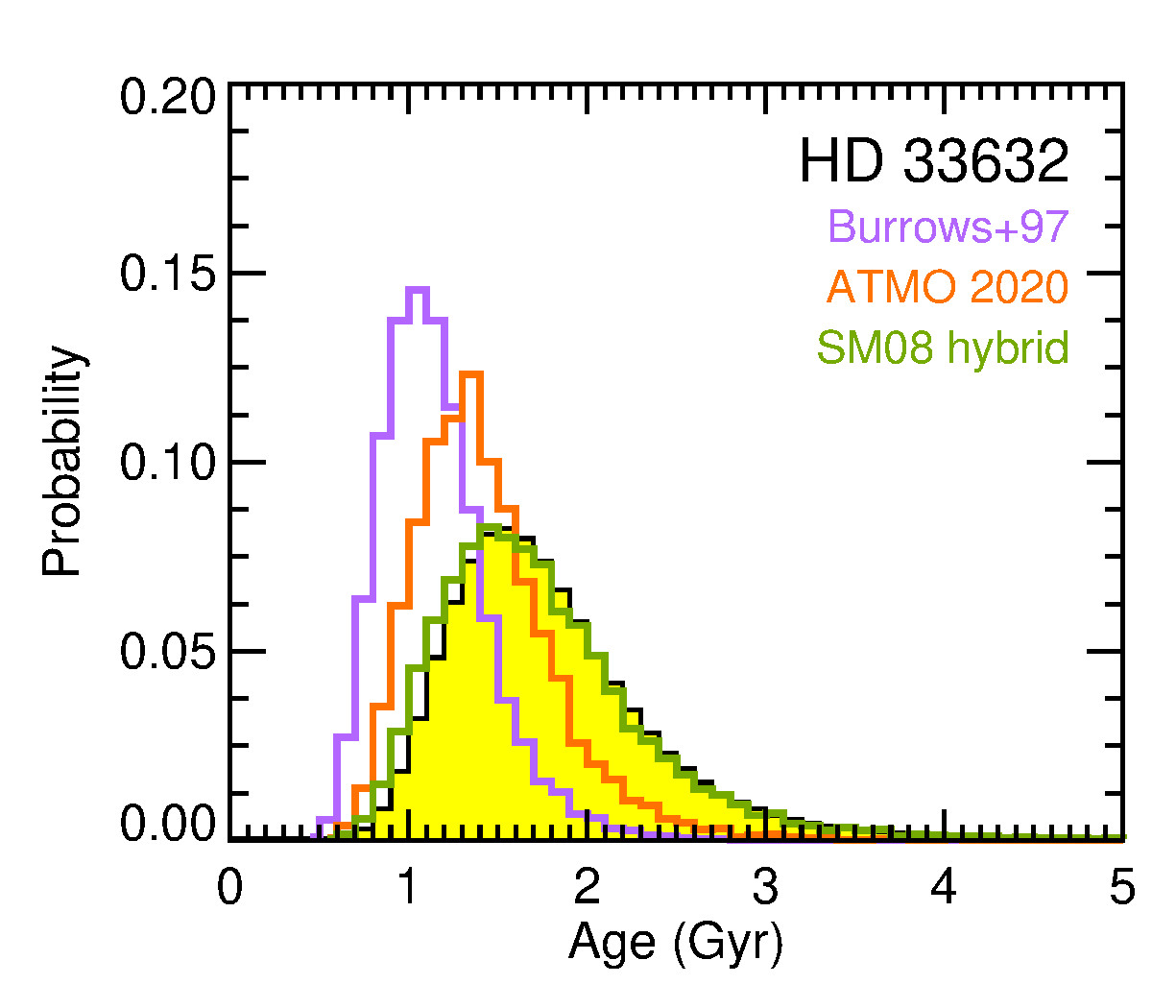}
    \caption{Figure~\ref{fig:benchmark1} gives a description of the posteriors. The objects displayed here have comparable uncertainties in mass and age. The only significant discrepancy with models is the mid-T dwarf HD~19467~B where the measured mass is higher than expected from the \cite{Saumon+Marley_2008} hybrid models.}
    \label{fig:benchmark2}
\end{figure*}

\subsection{Discussion of Individual Objects}

Our benchmark tests fall into two categories: cases where our measured mass is significantly more precise than the model-derived mass (Figure~\ref{fig:benchmark1}) or where the precision of the two are comparable (Figure~\ref{fig:benchmark2}). 
For a substellar object of fixed \Lbol, mass $M$ and age $t$ scale approximately as $M \propto t^{0.49}$ \citep{Burrows_2001RvMP...73..719B}.  
Thus, as \cite{Liu2008_ApJ_689_436} noted while discussing the first T-dwarf mass benchmark,
a 5\% mass uncertainty propagates to a 10\% uncertainty in the model-derived age.  
The limiting factor in the benchmark test will be the independently determined age, unless the fractional error in the age is no more than twice the fractional error in mass. Our age determinations range in precision from 15\%--30\%, while our measured masses range in precision from 1\%--10\%. Our masses for HD~19467~B and HD~33632~Ab are within a factor of two of the precision of their stellar ages, but in most cases our benchmarks are dominated by the precision of the age determination and not the mass precision (Gl~229~B, Gl~758~B, HD~13724~B, and HD~72946~B).

\begin{figure}
    \centering
    \includegraphics[width=\linewidth]{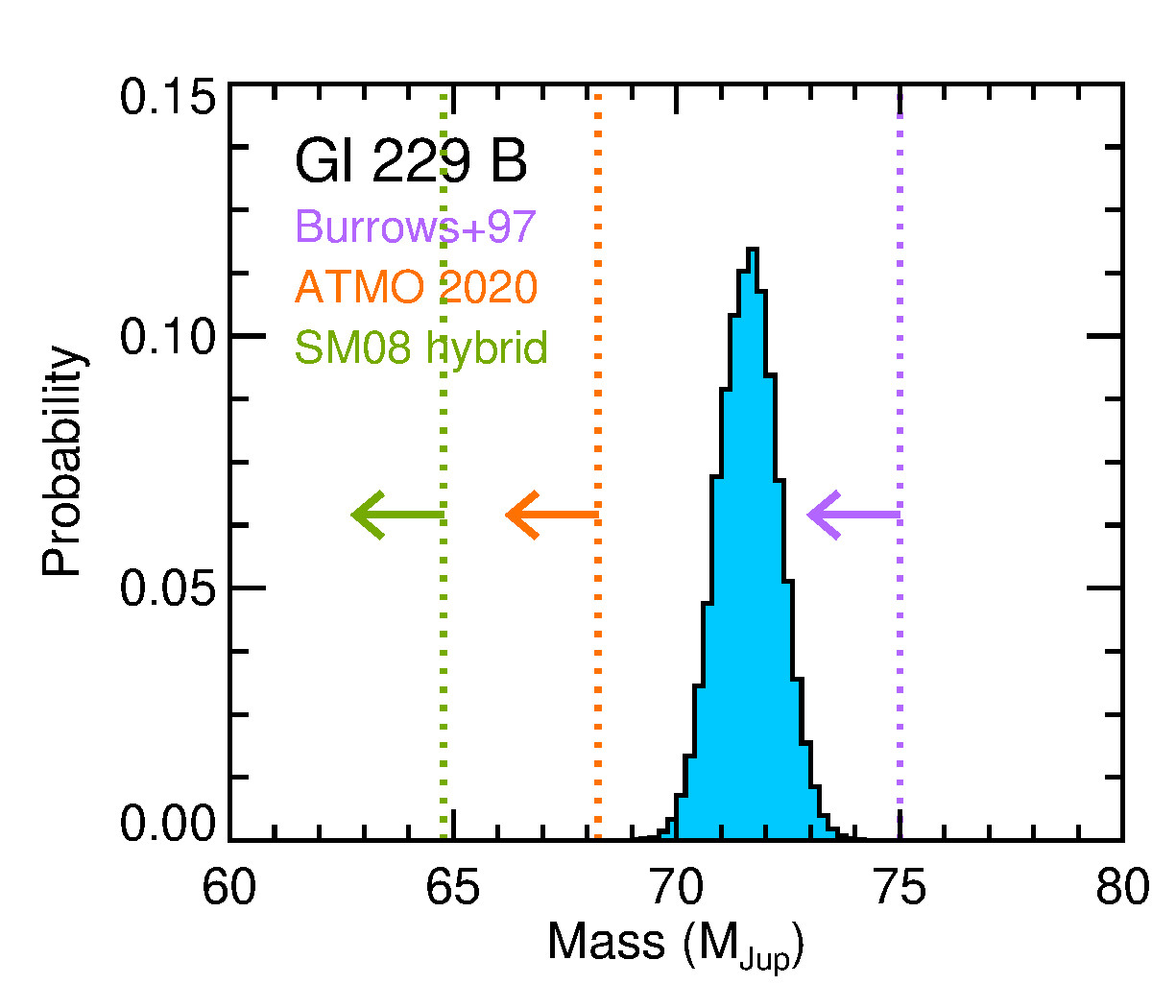}
    \caption{Same plot as in Figures~\ref{fig:benchmark1} and \ref{fig:benchmark2} except showing the 3$\sigma$ upper limits in mass predicted by models corresponding to an age of $<$10\,Gyr for Gl~229~B. Only the evolutionary models of \citet{Burrows+Marley+Hubbard+etal_1997} are consistent with our measurement. No other models, to our knowledge, predict that such a massive object can achieve such a low luminosity within a Hubble time.}
    \label{fig:benchmark3}
\end{figure}

\paragraph{Gl~229~B} 
\citet{brandt_gliese_229b_mass_htof} found an unexpectedly high mass of $70\pm5$\,\Mjup\ for Gl~229~B. Our new mass of $71.4\pm0.6$\,\Mjup\ greatly increases the significance of the tension with evolutionary models (Figure~\ref{fig:benchmark3}). Gl~229~B's luminosity is 11$\sigma$ lower than predicted by the hybrid \citet{Saumon+Marley_2008} models for an object of this mass even at 10\,Gyr. 

The \citet{Burrows+Marley+Hubbard+etal_1997} models overlap with our mass measurement well and give a cooling age of $7.5\pmoffs{0.3}{0.4}$\,Gyr. As discussed in detail by \citet{Saumon+Marley_2008}, the primary reason that the \citet{Burrows+Marley+Hubbard+etal_1997} models predict lower luminosities at a given mass and age is their lower global opacity ($\Lbol\propto\kappa_R^{0.35}$, where $\kappa_R$ is the Rosseland mean atmospheric opacity; \citealp{burrows+liebert_1993}). The differences in predicted luminosity are 0.3--0.6\,dex (Figure~6 of \citealp{Saumon+Marley_2008}), especially around the H-fusion mass boundary at old ages. 

There are two ways Gl~229~B could have such a low global opacity. First, it and its host star could inherit a low metallicity. However, \citet{brandt_gliese_229b_mass_htof} concluded that a sub-solar metallicity for Gl~229~A is implausible given an assortment of measurements that are consistent with solar metallicity \citep{neves_etal_2013,2014ApJ...791...54G,2014MNRAS.443.2561G}. 
Secondly, Gl~229~B could have acquired a sub-solar metallicity during its formation. However, companions formed by disk fragmentation are expected to be at least as metal rich as the host star \citep[e.g., ][]{2010ApJ...724..618B}. The only processes that alter the companion metallicity, such as concentration of solids at the site of fragmentation \citep[e.g., ][]{Haghighipour_boss_2013,2006MNRAS.372L...9R} or planetesimal capture \citep[e.g., ][]{helled+bodenheimer+2010}, only increase metallicity.


There are two possibilities by which Gl~229~B's mass could be reconciled with models without needing an unusually low opacity: either it is a tight binary, or there is another massive companion in orbit around Gl~229~A, which would muddle our interpretation of the astrometric acceleration. The latter scenario was ruled out in \citet{brandt_gliese_229b_mass_htof} and is even more unlikely with the more precise \gaia EDR3 proper motions and the additional RVs. The HGCA acceleration is significant at $\approx$115$\sigma$ \addition{(a difference of 13,000 in $\chi^2$)}, and it points in exactly the direction expected. An additional massive companion would be very unlikely to preserve the low $\chi^2$ value for the proper motion anomalies of just 0.49 ($n_{\rm dof}=2$).


Gl~229~B itself being an unresolved binary remains a plausible explanation that would not require radical changes to substellar evolutionary models. \citet{brandt_gliese_229b_mass_htof} noted that it is not unusually luminous for its spectral type, making a nearly equal-flux companion unlikely. A faint companion would need a sufficiently small orbit to elude detection by astrometric perturbations in the relative astrometry, and we discuss this possibility in more detail below in Section~\ref{ssec:emerging_trends}.


\paragraph{Gl~758~B}
This late-T-type companion has one of the most precise masses in our sample, with a fractional error of 2.0\%. Our benchmark test is dominated by the uncertainty in the host star's age, which we discussed in detail in Section \ref{sec:stellar_ages}. 
All three substellar models' cooling ages are consistent with our broad host star age distribution (4.7--10\,Gyr at 2$\sigma$). The
\citet{Burrows+Marley+Hubbard+etal_1997} models give a much younger age ($6.2\pm0.4$\,Gyr) than hybrid ($8.7\pm0.6$\,Gyr) or ATMO ($8.9\pmoffs{0.8}{0.6}$\,Gyr) models. This companion remains the sole test of models at the cold temperatures (ATMO~2020-derived $\Teff=603\pm9$\,K) of older, lower-mass BDs.

\paragraph{HD~13724~B}
As discussed in Section~\ref{sec:stellar_ages}, we find a host star age from gyrochronology that is significantly older than other determinations in the literature \citep[e.g., $1.0\pm0.9$\,Gyr;][]{Rickman2020_HD13724}. Here we conservatively adopt the youngest of our age distributions ($2.8\pmoffs{0.5}{0.4}$\,Gyr) that uses the stellar rotation period of 21\,days from \citet{Arriagada_2011}. A very similar age posterior would result from using the $20.2 \pm 1.2$ day rotation period from \citet{2019AA_RickmanHD13724_discovery}. As a mid-T dwarf, the hybrid evolutionary models are the most appropriate for HD~13724~B, and indeed they provide the best agreement in our benchmark test. Still, they give a substellar cooling age that is highly discrepant (3.8$\sigma$) with our host star age distribution (Figure \ref{fig:benchmark1}). 

Given the disagreement in the literature about the age of this host star, it is possible that this is a case where the host star itself is atypical (e.g., rotating slowly at a young age). However, it is also possible that substellar evolutionary models are to blame for the young derived age, as this phenomenon has been observed before for moderately young ($\lesssim$1\,Gyr) BDs \citep{Dupuy+Liu+Ireland_2009a,dupuy14_gl417}. We discuss both of these possibilities in more detail in Section \ref{ssec:emerging_trends}.

\paragraph{HD~19467~B} 
\citet{2020AA_Maire_HD19467} concluded that both hybrid and cloud-free models predict a luminosity $\approx$1$\sigma$ higher than is observed given their measured mass. 
Our benchmark test yields qualitatively similar results: the \cite{Burrows+Marley+Hubbard+etal_1997} models best match the observed luminosity given the measured mass and host-star age, even though those models should not be appropriate for a mid-T dwarf. Quantitatively, there is only a 1.4$\sigma$ difference between our host star age distribution and the cooling age derived from the (most appropriate) \cite{Saumon+Marley_2008} hybrid models. Therefore, with a mass of $65\pmoffs{6}{5}$\,\Mjup\ and a spectral type of T$5.5\pm1.0$ \citep{Crepp+Rice+Veicht+etal_2015}, HD~19467~B joins the ranks of unexpectedly (but not anomalously) massive T~dwarfs near the substellar mass boundary, comparable to WISE~J0720$-$0846~B \citep[$66\pm4$\,\Mjup\ and T$5.5\pm0.5$;][]{Dupuy2019_AJ_158_174}.

\paragraph{HD~33632~Ab}
This is the other L/T transition object in our sample, and our mass measurement of $50\pmoffs{6}{5}$\,\Mjup\ is about twice as precise as that in \cite{2020ApJ_Currie_Thayne_HD33632}. Our host star age of $1.7\pmoffs{0.4}{0.6}$\,Gyr agrees very well with all model-derived cooling ages, especially the \citet{Saumon+Marley_2008} hybrid models that are appropriate for an object of this type.

\paragraph{HD~72946~B} 
Our mass agrees well with that found by \cite{2020AA_Maire_HD72946} who discussed in detail its host star's somewhat young age (0.8--3\,Gyr) and somewhat metal-rich composition (${\rm [Fe/H]} = 0.11\pm0.03\,$dex; \citealt[]{2016AA_Bouchy2016_HD72946}). All three substellar model-derived ages agree with our adopted host star age distribution ($1.9\pmoffs{0.5}{0.6}$\,Gyr), with the largest difference being 1.4$\sigma$ for the \cite{Burrows+Marley+Hubbard+etal_1997} models. Such cloud-free, low-opacity models should not be appropriate for this companion's spectral type (L$5.0\pm1.5$), and they should be especially ill-suited given the metallicity implied by the host star. HD~72946~B is therefore a case of a warm companion (hybrid-derived $\Teff=1700\pm90$\,K) with a mass ($72.5\pm1.3$\,\Mjup) that could be on either side of the hydrogen-fusion boundary, and a host-star age that is consistent with its substellar cooling age.

\begin{figure*}
    \centering
    \includegraphics[height=3.0in]{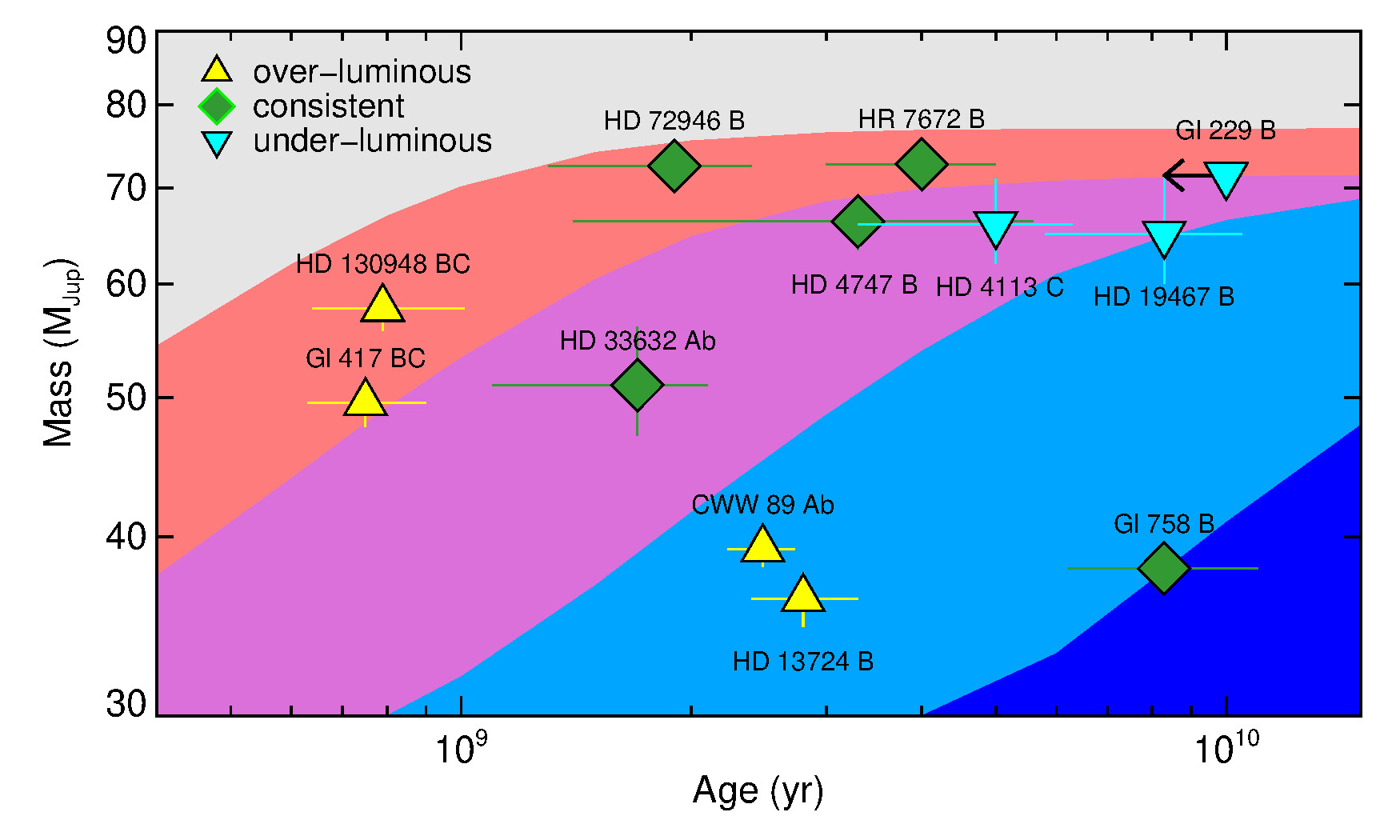}
    \caption{All BDs that have dynamically-measured masses, directly-determined luminosities, and independently determined ages. Each system is classified as being either consistent with models ($<$1$\sigma$ discrepant; green diamonds) or over- or under-luminous as compared to models given the mass and host star age (up-pointing and down-pointing triangles, respectively). In fact, of the discordant objects, all but HD~19467~B (1.1$\sigma$) are $\geq$2.5$\sigma$ discrepant with models. The background shading indicates effective temperature as predicted by hybrid \citet{Saumon+Marley_2008} evolutionary models ranging from $>$1800\,K (gray), 1800--1400\,K (red; $\approx$L4--L9), 1400--1100\,K (purple; $\approx$T0--T4), 1100--600\,K (light blue; $\approx$T5--T7), and $<$600\,K (dark blue). While no trend is apparent with temperature, the ensemble of measurements is consistent with observations favoring higher luminosities than models for young and low-mass BDs, lower luminosities for old and high-mass BDs, and agreement in between. Note that the average masses of HD~130948~BC \citep{Dupuy+Liu+Ireland_2009a} and Gl~417~BC \citep{dupuy14_gl417} are plotted because those benchmark tests are based on total, not individual, masses.}
    \label{fig:discrep}
\end{figure*}

\begin{deluxetable*}{lcccccccccccc}
    \tablewidth{0pt}
    \setlength{\tabcolsep}{3pt}
    \tablecaption{Comparison of observed and BD evolutionary model-derived fundamental properties of benchmark systems.\label{tab:benchmarks}}
    \tablehead{\colhead{} & \colhead{} & \multicolumn{3}{c}{System age} & \colhead{} & \multicolumn{3}{c}{Companion mass} & \colhead{} & \multicolumn{3}{c}{Companion $\log(\Lbol/\Lsun)$} \\\cline{3-5}\cline{7-9}\cline{11-13}
               \colhead{Object} & \colhead{} & \colhead{Host star} & \colhead{BD model} & \colhead{$\Delta$} & \colhead{} & \colhead{Dynamical} & \colhead{BD model} & \colhead{$\Delta$} & \colhead{} & \colhead{Observed} & \colhead{BD model} & \colhead{$\Delta$} \\[-4pt]
               \colhead{} & \colhead{} & \colhead{(Gyr)} & \colhead{(Gyr)} & \colhead{} & \colhead{} & \colhead{(\Mjup)} & \colhead{(\Mjup)} & \colhead{} & \colhead{} & \colhead{(dex)} & \colhead{(dex)} & \colhead{}}
    \startdata
    Gl~417~BC\tablenotemark{a}    && $0.74_{-0.15}^{+0.12}$ & $0.62\pm0.03$          & $+0.9\sigma$ && $110.1_{-1.5}^{+1.4}$ & $118\pm6$             & $-1.2\sigma$ && $-4.23\pm0.03$    & $-4.45\pm0.09$          & $+1.6\sigma$ \\
    HD~130948~BC\tablenotemark{a} && $0.79_{-0.15}^{+0.14}$ & $0.44_{-0.03}^{+0.04}$ & $+3.0\sigma$ && $116.0\pm0.4$         & $141_{-5}^{+6}$       & $-4.5\sigma$ && $-3.93\pm0.06$    & $-4.27\pm0.11$          & $+3.0\sigma$ \\
    HD~33632~Ab                   && $1.7_{-0.6}^{+0.4}$    & $1.7_{-0.6}^{+0.4}$    & $+0.1\sigma$ && $50\pmoffs{5.6}{5}$              & $52\pm8$              & $-0.2\sigma$ && $-4.69\pm0.07$    & $-4.65_{-0.11}^{+0.14}$ & $-0.2\sigma$ \\
    HR~7672~B\tablenotemark{b}    && $1.9\pm0.3$            & $4_{-2}^{+3}$          & $-1.3\sigma$ && $72.7\pm0.8$          & $71.6_{-1.3}^{+1.8}$  & $+0.7\sigma$ && $-4.25\pm0.05$    & $-4.21_{-0.07}^{+0.06}$ & $-0.5\sigma$ \\
    HD~72946~B                    && $1.9_{-0.6}^{+0.5}$    & $1.8_{-0.6}^{+0.5}$    & $-0.0\sigma$ && $72.5\pm1.3$  & $73.7_{-1.5}^{+1.8}$  & $-0.4\sigma$ && $-4.12\pm0.02$    & $-4.21_{-0.10}^{+0.13}$ & $+0.7\sigma$ \\
    CWW~89~Ab\tablenotemark{c}    && $2.48\pm0.25$          & $0.35_{-0.12}^{+0.09}$ & $+7.3\sigma$ && $39.2\pm1.1$          & $74\pm3$              & $-3.9\sigma$ && $-4.19\pm0.14$    & $-5.22\pm0.09$          & $+6.1\sigma$ \\
    HD~13724~B                    && $2.8_{-0.5}^{+0.4}$    & $1.18\pm0.17$          & $+3.8\sigma$ && $36.2_{-1.5}^{+1.6}$  & $55_{-5}^{+4}$        & $-3.6\sigma$ && $-4.78\pm0.07$    & $-5.44_{-0.13}^{+0.12}$ & $+3.9\sigma$ \\
    HD~4747~B\tablenotemark{b}    && $2.9_{-0.5}^{+0.4}$    & $3.6_{-2.0}^{+1.4}$    & $-0.4\sigma$ && $66\pm3$              & $66_{-3}^{+5}$        & $+0.1\sigma$ && $-4.55\pm0.08$    & $-4.59_{-0.08}^{+0.06}$ & $+0.3\sigma$ \\
    HD~4113~C\tablenotemark{d}    && $5.0_{-0.8}^{+0.7}$    & $14.0_{-0.7}^{+1.0}$   & $-4.4\sigma$ && $66_{-4}^{+5}$        & $23\pm5$              & $+6.1\sigma$ && $-6.30\pm0.22$    & $-4.72_{-0.17}^{+0.18}$ & $-5.2\sigma$ \\
    HD~19467~B                    && $5.1_{-1.7}^{+1.3}$    & $9_{-3}^{+2}$          & $-1.4\sigma$ && $65.4_{-4.6}^{+5.9}$        & $53_{-5}^{+6}$        & $+1.8\sigma$ && $-5.16\pm0.08$    & $-4.7_{-0.3}^{+0.2}$    & $-1.3\sigma$ \\
    Gl~758~B                      && $7.5_{-1.4}^{+1.8}$    & $8.8\pm0.6$            & $-0.8\sigma$ && $38.0\pm0.75$  & $35\pm4$              & $+0.8\sigma$ && $-6.07\pm0.03$    & $-5.98_{-0.16}^{+0.09}$ & $-0.7\sigma$ \\
    Gl~229~B                      && $<10$                  & \nodata                & \nodata      && $71.4\pm0.6$          & $64.78\pm0.10$       &  $+10\sigma$ &&$-5.208\pm0.007$   & $-4.52_{-0.07}^{+0.06}$& $ -11\sigma$ \\
    \enddata
    \tablecomments{The source of the data tabulated here is from this work unless otherwise noted. For Gl~229~B, model results are given at a fixed age of 10\,Gyr.}
    \tablenotetext{a}{For Gl~417~BC and HD~130948~BC, results in the system age and companion mass columns are based on their total dynamical mass rather than individual masses. In the companion $\log(\Lbol/\Lsun)$ column, we quote results for the fainter component assuming the model-derived mass ratio for the system. We use the most recent orbital fits from \citet{Dupuy+Liu_2017}, replacing the \hipparcos\ parallaxes used in that work with \Gaia~EDR3 parallaxes.}
    \tablenotetext{b}{The mass and luminosity measurements of HD~4747~B and HR~7672~B are from \citet{Brandt_Dupuy_Bowler_2018}, and we have updated the stellar activity ages here using the same methods as in \citet{Brandt_Dupuy_Bowler_2018}.}
    \tablenotetext{c}{The mass, cluster age, and luminosity measurements of CWW~89~Ab are from \citet{2018AJ....156..168B}.}
    \tablenotetext{d}{The dynamical mass of HD~4113~C is from \citet{Cheetham+Segransan+Peretti+etal_2018}, the luminosity is from this work, and we have computed a stellar activity age posterior for consistency with other systems presented here.}
\end{deluxetable*}

\subsection{Emerging Trends in Benchmark Tests}\label{ssec:emerging_trends}

We have derived masses and ages for six benchmark BDs.  We now place these six systems in context, combining them with other benchmark BD systems with measured masses, ages, and luminosities.

Two of these are HD~130948~BC \citep{Dupuy+Liu+Ireland_2009a} and Gl~417~BC \citep{dupuy14_gl417}, BD+BD binaries orbiting young, solar-type host stars. In both cases their total masses have been measured dynamically. \addition{Previous results were based on \hipparcos parallactic distances, so we have checked whether newer \gaia~EDR3 parallaxes would significantly change the results. For HD~130948, the \gaia parallax is consistent within 0.1\%, implying negligible changes. But for Gl~417, the \gaia~EDR3 parallax is 3.4\% (3.5$\sigma$) smaller than the \hipparcos parallax used by \citet{dupuy14_gl417}. By Kepler's Third Law, $M_{\rm sys} \propto a^3$, the larger semimajor axis implied by the larger \gaia distance results in an 11\% higher system mass of $110.1\pmoffs{1.4}{1.5}$\,\Mjup\ as compared to previous work. This reduces the tension with models, corresponding to a 1.6$\sigma$ discrepancy in luminosity (models too faint) for the companion Gl~417~C.}

Three more objects, HR~7672~B and HD~4747~B \citep{Crepp+Johnson+Fischer+etal_2012,Crepp+Gonzales+Bechter+etal_2016,Brandt_Dupuy_Bowler_2018} and HD~4113~C (\citealp{Cheetham+Segransan+Peretti+etal_2018}; Rickman et. al., in prep), are directly imaged companions that have measured masses using \Gaia~DR2, and where the RV phase coverage is sufficient that \Gaia~EDR3 measurements do not significantly change published results. 

The only BD mass benchmark with an age determined by association to a stellar cluster, rather than gyrochronology, is CWW~89~Ab, a transiting BD in Ruprecht~147 \citep[$2.48\pm0.25$\,Gyr; ][]{2018ApJ...866...67T}. We use the secondary-eclipse, irradiation-corrected luminosity from \citet{2018AJ....156..168B} and dynamical mass from \citet{2019AJ....158...38C}. We limit our discussion to the evolutionary parameter space circumscribed by these older ($\gtrsim$300\,Myr), massive ($>$30\,\Mjup) BDs, and do not consider the much younger and/or lower mass objects such as $\beta$~Pic~bc \citep{Nowak_2020_beta_pic_c_direct_detection, AMLagrange2020betapicc_direct_detection,Brandt2020betapicbc} and 2MASS~J15104786$-$2818174~Aab \citep{2020NatAs...4..650T}.

There are other examples of BD benchmarks in the literature that we do not include in the following due to large uncertainties in their ages or masses. Gl~802~B is a likely old (probable thick-disk member) and over-luminous BD with a mass of $66\pm5$\,\Mjup\ \citep{2008ApJ...678..463I}. HD~47127~B is a massive ($>68$\,\Mjup) and old (7--10\,Gyr) BD companion to a white dwarf \citep{2021ApJ...913L..26B}. And finally, the BD binary $\epsilon$~Ind~BC has discrepant dynamical mass measurements \citep{cardoso2012,2018ApJ_epsi_indi_bc_Dieterich} and a poorly-constrained host-star age \citep{2010A&A...510A..99K}.

We place benchmark results into three broad categories: consistent with evolutionary models, over-luminous, and under-luminous. The over-luminous cases are BDs that are brighter than models predict given their independently measured mass and age. Equivalently, their measured masses are surprisingly low, or their host stars' ages are older than the substellar cooling ages. Such systems include HD~13724~B (cooling ages are 3.8$\sigma$ younger than the star's activity age) as well as literature systems HD~130948~BC (3.0$\sigma$), Gl~417~BC (0.9$\sigma$), and CWW~89~Ab (7.3$\sigma$). The under-luminous cases are observed to be fainter than model predictions given their mass and age; their dynamical masses are higher than expected.  These include Gl~229~B ($\approx$10$\sigma$) and HD~19467~B (though at just 1.4$\sigma$), as well as HD~4113~C ($\approx$5$\sigma$). 

Here we are considering only the benchmark results where models are appropriate to each object's corresponding spectral type. We use hybrid \citet{Saumon+Marley_2008} models for all objects and ATMO~2020 for late-T objects. While the \cite{Burrows+Marley+Hubbard+etal_1997} models provide the best match the observed mass, age, and luminosity for some BDs, they lack state-of-the-art cloud and opacity treatments, and have also been shown to overpredict the lithium-depletion mass boundary \citep{Dupuy+Liu_2017}. 


Figure~\ref{fig:discrep} displays the benchmark test results as a function of mass and age, with model-predicted \Teff\ ranges highlighted with background shading. \addition{Table~\ref{tab:benchmarks} lists the quantitative discrepancies from each benchmark test.} Each temperature band contains a mixture of benchmark results, suggesting that any significant problems with substellar cooling models are not restricted to a particular surface temperature.  
Instead, there is a trend of objects being over-luminous at low masses and young ages and under-luminous at high masses and old ages. This more closely resembles a correlation with surface gravity, which increases towards higher masses and older ages (as the radius decreases). 
This analogy is not precise, as $\log{g}$ does not actually map one-to-one on a mass-age diagram.

The over-luminosity problem at moderately young ages has been known since the earliest measurement of a substellar mass-age-\Lbol\, benchmark (\addition{HD 130948~BC;} \citealp{Dupuy+Liu+Ireland_2009a}).  Our addition of HD~13724~B stakes out the lowest mass at which this has now been observed, comparable to the unique mass-radius-age-\Lbol\, benchmark CWW~89~Ab. 
CWW~89~Ab's radius is consistent with evolutionary models for its mass and age, implying that its interior structure agrees with fundamental degeneracy physics.  This rules out deposition of excess energy in the deep interior (e.g., from tides) as the cause of the over-luminosity. 
\cite{2018AJ....156..168B} suggest that strong alkali absorption of flux from CWW~89~A could induce a thermal inversion if the BD's cooling is inhibited.  This could occur if a high C/O ratio in the BD makes CH$_4$ and H$_2$O chemically unfavorable relative to CO.
Such a mechanism would not be available to wide companions like HD~13724~B, HD~130948~BC, and Gl~417~BC and is regardless ruled out by the strong H$_2$O features in their spectra.

Magnetic fields have been unsuccessful at reproducing the over-luminosity problem \citep{2010ApJ...713.1249M}, and other potential solutions remain elusive. One promising avenue may be to examine the temperature-pressure profile directly through retrieval methods. This has shown evidence for upper atmosphere heating in L~dwarfs \citep[e.g.,][]{burningham_2017}, and perhaps such surface processes could make some BDs appear over-luminous.


In contrast, the under-luminosity problem (a.k.a., the ``over-massive BD problem'') is fundamentally different; it can always be explained by unresolved multiplicity. While we place HD~19467~B in this category, its dynamical mass is only 1.8$\sigma$ different from the model-derived mass. Gl~229~B and HD~4113~C are the true touchstones for the BDs that are much too faint for their mass. Both Gl~229~B and HD~4113~C could be unresolved binaries,
\addition{although the multi-decade collection of high-quality observations of Gl~229~B greatly restrict what binary scenarios are plausible with a total mass of 71.4\,$\Mjup$. AO imaging like in Figure~\ref{fig:Gl229 PSF fit} rules out wide, massive companions ($\gtrsim$0.3\,au), such as a 41+30\,\Mjup\ system. The lack of any perturbations $\gtrsim$2\,mas in relative astrometry implies the photocenter orbit of a putative binary must be $\lesssim$0.012\,au. For instance, a 10-\Mjup\ companion (contributing negligible flux) could only be orbiting B with $a\lesssim0.04$\,au. A 10-\Mjup\ companion on a 0.04\,au circular orbit would cause Gl~229~B to display significant RV variations with a semi-amplitude of 5.4\,km\,s$^{-1} \times (M_2\sin{i}/10\,\Mjup)(a/0.04\,{\rm au})^{-1/2}$. A brighter, more massive secondary would have a higher upper limit on $a$ and impart a larger RV semi-amplitude on Gl~229~B (as long as the binary orbit is not near face-on).
Such RV monitoring has not been carried out, but} if future observations rule out a massive companion, then something more fundamental must be amiss with evolutionary models. This would likely require extreme changes to how models treat properties like the equation of state, heat transport in the interior, or the influence of rotation or magnetism.


In between the two extremes outlined above, BDs with masses of 40--70\,\Mjup\ at intermediate ages of 1--4\,Gyr seem to agree very well with evolutionary models. Unfortunately, age constraints for these stars are relatively weak: magnetic fields weaken and magnetic braking becomes less efficient, limiting the precision of gyrochronology \citep[e.g.,][]{vanSaders+Ceillier+Metcalfe+etal_2016}.  
BDs with more precise age measurements at a few to several Gyr, whether from BDs in more distant open clusters or from asteroseismic ages of their host stars, would provide stronger tests of BD cooling models in this region of parameter space.


\section{Conclusions}\label{sec:conclusions}

We have derived masses and orbits for six systems containing BD companions: Gl~229, Gl~758, HD~13724, HD~19467, HD~33632, and HD~72946. Our analysis utilizes long-term RV monitoring, relative astrometry (including new Keck/NIRC2 measurements we report for Gl~229~B), and \hipparcos-\gaia proper motion accelerations. 
We summarize our main results below.

\begin{itemize}[noitemsep]
    \item [1.] We measure the most precise masses to-date for the late-T dwarfs Gl~229~B ($71.4 \pm 0.6\,\Mjup$) and Gl~758~B ($38.0 \pm 0.8 \,\Mjup$) and the M-dwarf Gl~229~A ($0.579 \pm 0.007 \,\Msun$). Notably, our masses for Gl~229~A and B have uncertainties $\lesssim$1\%, despite the fact that the system has been observed for only 10\% of the orbital period. We find good constraints on the masses of the other four BD companions, with a typical improvement in precision of a factor of two compared to previously published results. For the mid-T dwarfs HD~13724~B and HD~19467~B we infer masses of $36.2 \pmoffs{1.6}{1.5} \,\Mjup$ and $65 \pmoffs{6}{5} \,\Mjup$, respectively, while for the L/T transition companion HD~33632~Ab we find $50 \pmoffs{6}{5} \,\Mjup$. The mass of $72.5 \pm 1.3 \, \Mjup$ that we measure for the mid-L dwarf HD~72946~B places the object on the boundary between stars and BDs. 
    
    \item [2.] We perform mass--\Lbol--age benchmark tests of substellar evolutionary models and compare the results with six other such systems from the literature. We identify a pattern of BDs being over-luminous at younger ages and lower masses and under-luminous (or over-massive) at older ages and higher masses.  
    
    \item [3.] The mass and luminosity of Gl~229~B is highly discrepant with modern substellar evolutionary models. We reaffirm that neither an unusually low metallicity for Gl~229~B nor a massive, interior companion are likely to reconcile this discrepancy. It seems more likely that Gl~229~B itself may be an unresolved binary. 

    \item [4.] While companion mass is most directly constrained by astrometric accelerations measured for the host stars, our joint orbital analysis also results in well-measured orbital parameters for most systems. Companion eccentricities range widely from near-circular, like $0.12\pmoffs{0.18}{0.09}$ for HD~33632~Ab, to moderately eccentric, like $0.335 \pm 0.026$ for HD~13724~B; and even as high as $0.851\pmoffs{0.002}{0.008}$ for Gl~229~B.

    \item [5.] Our 1.2\% mass measurement of Gl~229~A provides a strong validation of mass--magnitude relations for low-mass stars at the few-percent level. \addition{On the other hand, our dynamical mass for HD~13724~A is $2 \sigma$ lower than expected from stellar evolution models given its \Teff\ and \Lbol.}
    \addition{This illustrates how exoplanet dynamical analyses are seeping into the domain where they provide meaningful constraints on the astrophysics of their host stars.}
\end{itemize}

\gaia EDR3 accelerations and the wealth of relative astrometry and RVs have allowed us to reach mass precisions on these six BDs where the model-testing error budget is dominated by the host star's age. In other words, substellar evolutionary models predict cooling ages (given our high-precision masses) that are comparable to or much better than the age constraints on the host star.

\addition{Sub-1\% uncertainties in BD masses, like we have demonstrated for Gl~229~B, will be especially valuable for certain applications. High precision masses will aid studies where $\log{g}$ is important, like atmosphere modeling and retrievals (e.g., combining the analyses of \citealp{2020A&A...640A.131M} and \citealp{2021ApJ...915L..16B}). High precision masses would also result in ultra-precise BD-cooling ages that would be especially useful for calibrating stellar age-dating methods at a wide range of ages, in between the usual benchmark stellar clusters. Sub-1\% mass precision is also crucial for precisely identifying the mass boundary between stars and BDs, as this is a relatively sharp transition often with small, few-\Mjup\ variation between model predictions (e.g., see Section~7.1 of \citealp{Dupuy+Liu_2017}).}

\gaia EDR3 accelerations continue to broaden the applications of direct imaging studies, now allowing for a 1\% dynamical mass for the host star, as well as the companion, in at least the case of the Gl~229 system. Dynamical primary masses may prove useful, in the coming decade, for constraining stellar evolutionary models of the host stars. As the baseline of the \gaia mission grows, and with the release of the individual observation epochs in \gaia DR4 (expected some years from now), dynamical constraints on BD's will continue to improve, test, and progress our understanding of brown dwarf and giant planet formation. 

\software{astropy \citep{astropy:2013, astropy:2018},
          scipy \citep{2020SciPy-NMeth},
          numpy \citep{numpy1, numpy2},
          pandas \citep{mckinney-proc-scipy-2010, reback2020pandas},
          \orbitcodename \citep{TimOrbitFitTemporary},
          \htofcodename \citep{htof_zenodo, MirekHTOFtemporary},
          corner \citep{corner},
          REBOUND \citep{rebound_2012_main},
          Jupyter (\url{https://jupyter.org/}).
          }

\section*{Acknowledgments}

This work made use of observations obtained at the W. M. Keck Observatory. The authors wish to recognize and acknowledge the very significant cultural role of the summit of Maunakea, upon which the W. M. Keck Observatory stands. 
It is also a pleasure to thank to Randy Campbell, Percy Gomez, Julie Renaud-Kim, and Tony Ridenour for Keck observing support.

The W. M. Keck Observatory is operated as a scientific partnership among the California Institute of Technology, the University of California and the National Aeronautics and Space Administration. The Observatory was made possible by the generous financial support of the W. M. Keck Foundation. 

G.~M. Brandt is supported by the National Science Foundation (NSF) Graduate Research Fellowship under grant no. 1650114.

B.P.B. acknowledges support from the National Science Foundation grant AST-1909209 and NASA Exoplanet Research Program grant 20-XRP20$\_$2-0119.

We thank Aaron Dotter, Lars Bildsten, and Emily Rickman for helpful discussions. We thank Andrew Howard for sharing their radial velocity data set with us.

The predicted positions of the companions presented in this work, and many others, can be found at \url{whereistheplanet.com}. We thank Jason Wang, Matas Kulikauskas, and Sarah Blunt for building and hosting this wonderful web service, and for incorporating our results.

We wish to thank the teams behind the calibrated HIRES \citep{TalOrHIRES} and HARPS \citep{TrifonovHarps} catalogs. These catalogs are an immense boon to the astronomical community.

This work has made use of data from the European Space Agency (ESA) mission {\it Gaia} (\url{https://www.cosmos.esa.int/Gaia}), processed by the {\it Gaia} Data Processing and Analysis Consortium (DPAC, \url{https://www.cosmos.esa.int/web/Gaia/dpac/consortium}). Funding for the DPAC has been provided by national institutions, in particular the institutions participating in the {\it Gaia} Multilateral Agreement.

This work made use of observations collected by the European Organisation for Astronomical Research in the Southern Hemisphere. This work also made use of observations made at Observatoire de Haute Provence (CNRS), France.

This work made use of the \orbitcodename code. We used version \orbitcodeversion.

This work made use of the \htofcodename code. We used version \htofversion \citep{htof_zenodo}.

\bibliographystyle{aasjournal}
\bibliography{refs.bib}

\newpage
\clearpage

\appendix
\section{Posteriors and Priors of the orbital fits.}

\newpage

\begin{deluxetable*}{ccccccc}
\tablewidth{0pt}
    \tablecaption{Posteriors of the Gl229 system.\label{tab:posteriors_Gl229}}
    \tablehead{
    \colhead{Parameter} & \colhead{Prior Distribution} & \multicolumn{2}{c}{{Posterior $\pm$1$\sigma$}}}
    \startdata
    Stellar mass & Uniform & \multicolumn{2}{c}{$0.579 \pm 0.007 \, \Msun$} \\
    Parallax ($\varpi$) & $173.574 \pm 0.017 \, \rm{mas}$ (\gaia eDR3) & \multicolumn{2}{c}{$173.574 \pm 0.013 \, \rm{mas}$} \\
    Barycenter Proper Motions\tablenotemark{a}  & Uniform & \multicolumn{2}{c}{ $\mu_{\alpha}=$$-145.46 \pm 0.13 \, \rm{mas/yr}$ \; \& \; $\mu_{\delta}=$$-705.83 \pm 0.15 \, \rm{mas/yr}$} \\
    HIRES RV Zero Point & Uniform & \multicolumn{2}{c}{$-7 \pmoffs{34}{36} \,\rm{m/s}$} \\
    HARPS RV Zero Point & Uniform & \multicolumn{2}{c}{$-10 \pmoffs{34}{36} \,\rm{m/s}$} \\
    Howard~et.~al.~2021 RV Zero Point & Uniform & \multicolumn{2}{c}{$0.59 \pmoffs{0.59}{0.15} \,\rm{m/s}$} \\
    RV jitter & Log-flat over [0,300 m/s] &  \multicolumn{2}{c}{$0.828 \pm 0.045 \, \rm{m/s}$} \\
    \hline
    Parameter & Prior Distribution & \multicolumn{2}{c}{Posterior $\pm$1$\sigma$} \\
    \hline
    $\sqrt{e} \sin \omega$ & Uniform over [-1, 1] & \multicolumn{2}{c}{$-0.14 \pmoffs{0.82}{0.21} \,$} \\
    $\sqrt{e} \cos \omega$ & Uniform over [-1, 1] & \multicolumn{2}{c}{$0.85 \pmoffs{0.05}{1.4} \,$} \\
    Semi-major axis ($a$) & $1/a$ (log-flat) & \multicolumn{2}{c}{$33.3 \pmoffs{0.4}{0.3} \,\rm{A.U.}$} \\
    Inclination ($i$) & $\sin i$ (geometric) & \multicolumn{2}{c}{$7.7 \pmoffs{7.6}{4.4} \,\rm{degrees}$} \\
    PA of ascending node & Uniform & \multicolumn{2}{c}{$-29 \pmoffs{13}{140} \,\rm{degrees}$} \\
    Mean Longitude at $t_{\rm ref}$ $(\lambda_{\rm ref})$ & Uniform & \multicolumn{2}{c}{$-57 \pmoffs{140}{13} \,\rm{degrees}$} \\
    BD Mass $(M)$ & $1/M$ (log-flat) & \multicolumn{2}{c}{$71.4 \pm 0.6 \, \Mjup$} \\
    \hline
    Eccentricity ($e$) & (derived quantity) & \multicolumn{2}{c}{$0.851 \pmoffs{0.002}{0.008} \,$} \\
    Argument of Periastron $(\omega)$ & (derived quantity) & \multicolumn{2}{c}{$-9 \pmoffs{140}{13} \,\rm{degrees}$} \\
    Periastron Time $(T_0)$ & (derived quantity) & \multicolumn{2}{c}{$2466912 \pmoffs{97}{63} \,\rm{days}$} \\
    Period & (derived quantity) & \multicolumn{2}{c}{$86909 \pmoffs{1900}{1700} \,\rm{days}$} \\
     &  & \multicolumn{2}{c}{$237.9 \pmoffs{5.1}{4.6} \,\rm{years}$} \\
    \hline
    \orbitcodename Reference Epoch $(t_{\rm ref})$ & 2455197.50 BJD & \nodata & \nodata  
    \enddata
        \tablecomments{Orbital elements all refer to orbit of the companion about the barycenter. The orbital parameters for the primary about each companion are identical except $\omega_{A} = \omega + \pi$. We use $\pm \sigma$ to denote the $1\sigma$ Gaussian error about the median when the posteriors are approximately symmetric. Otherwise, we denote the value by median$\pmoffs{u}{l}$ where $u$ and $l$ denote the  68.3\% confidence interval about the median. The reference epoch $t_{\rm ref}$ is not a fitted parameter and has no significance within the fit itself, it is the epoch at which the Mean Longitude $(\lambda_{\rm ref})$ is evaluated.}
        \tablenotetext{a}{$\mu_{\alpha}$ and $\mu_{\delta}$ refer to the proper motions in right-ascension and declination, respectively.}
\end{deluxetable*}

\begin{deluxetable*}{ccccccc}
\tablewidth{0pt}
    \tablecaption{Posteriors of the Gl758 system.\label{tab:posteriors_Gl758}}
    \tablehead{
    \colhead{Parameter} & \colhead{Prior Distribution} & \multicolumn{2}{c}{{Posterior $\pm$1$\sigma$}}}
    \startdata
    Stellar mass & Uniform & \multicolumn{2}{c}{$1.05 \pmoffs{0.25}{0.23} \,\Msun$} \\
    Parallax ($\varpi$) & $64.07 \pm 0.015 \, \rm{mas}$ (\gaia eDR3) & \multicolumn{2}{c}{$64.0703 \pm 0.00253 \, \rm{mas}$} \\
    Barycenter Proper Motions\tablenotemark{a}  & Uniform & \multicolumn{2}{c}{ $\mu_{\alpha}=$$81.05 \pmoffs{0.19}{0.27} \,\rm{mas/yr}$ \; \& \; $\mu_{\delta}=$$162.15 \pmoffs{0.51}{0.36} \,\rm{mas/yr}$} \\
    APF RV Zero Point & Uniform & \multicolumn{2}{c}{$110 \pmoffs{20}{25} \,\rm{m/s}$} \\
    Tull/McD RV Zero Point & Uniform & \multicolumn{2}{c}{$90 \pmoffs{20}{25} \,\rm{m/s}$} \\
    HIRES RV Zero Point & Uniform & \multicolumn{2}{c}{$0.9 \pmoffs{1.1}{0.58} \,\rm{m/s}$} \\
    RV jitter & Log-flat over [0,300 m/s] &  \multicolumn{2}{c}{$0.852 \pm 0.038 \, \rm{m/s}$} \\
    \hline
    Parameter & Prior Distribution & \multicolumn{2}{c}{Posterior $\pm$1$\sigma$} \\
    \hline
    $\sqrt{e} \sin \omega$ & Uniform over [-1, 1] & \multicolumn{2}{c}{$0.17 \pm 0.26 \, $} \\
    $\sqrt{e} \cos \omega$ & Uniform over [-1, 1] & \multicolumn{2}{c}{$-0.37 \pmoffs{0.31}{0.18} \,$} \\
    Semi-major axis ($a$) & $1/a$ (log-flat) & \multicolumn{2}{c}{$29.7 \pmoffs{5.3}{4.2} \,\rm{A.U.}$} \\
    Inclination ($i$) & $\sin i$ (geometric) & \multicolumn{2}{c}{$51.6 \pmoffs{4.4}{5.4} \,\rm{degrees}$} \\
    PA of ascending node & Uniform & \multicolumn{2}{c}{$180.6 \pmoffs{2.8}{3.9} \,\rm{degrees}$} \\
    Mean Longitude at $t_{\rm ref}$ $(\lambda_{\rm ref})$ & Uniform & \multicolumn{2}{c}{$48 \pmoffs{14}{11} \,\rm{degrees}$} \\
    BD Mass $(M)$ & $1/M$ (log-flat) & \multicolumn{2}{c}{$38.04 \pm 0.74 \, \Mjup$} \\
    \hline
    Eccentricity ($e$) & (derived quantity) & \multicolumn{2}{c}{$0.24 \pm 0.11 \, $} \\
    Argument of Periastron $(\omega)$ & (derived quantity) & \multicolumn{2}{c}{$155 \pmoffs{35}{56} \,\rm{degrees}$} \\
    Periastron Time $(T_0)$ & (derived quantity) & \multicolumn{2}{c}{$2470102 \pmoffs{2400}{4000} \,\rm{days}$} \\
    Period & (derived quantity) & \multicolumn{2}{c}{$56270 \pmoffs{23000}{14000} \,\rm{days}$} \\
     &  & \multicolumn{2}{c}{$154 \pmoffs{63}{39} \,\rm{years}$} \\
    \hline
    \orbitcodename Reference Epoch $(t_{\rm ref})$ & 2455197.50 BJD & \nodata & \nodata  
    \enddata
\end{deluxetable*}

\begin{deluxetable*}{ccccccc}
\tablewidth{0pt}
    \tablecaption{Posteriors of the HD13724 system.\label{tab:posteriors_HD13724}}
    \tablehead{
    \colhead{Parameter} & \colhead{Prior Distribution} & \multicolumn{2}{c}{{Posterior $\pm$1$\sigma$}}}
    \startdata
    Stellar mass & Uniform & \multicolumn{2}{c}{$0.95 \pmoffs{0.08}{0.07} \,\Msun$} \\
    Parallax ($\varpi$) & $23.016 \pm 0.018 \, \rm{mas}$ (\gaia eDR3) & \multicolumn{2}{c}{$23.0159 \pm 0.00214 \, \rm{mas}$} \\
    Barycenter Proper Motions\tablenotemark{a}  & Uniform & \multicolumn{2}{c}{ $\mu_{\alpha}=$$-31.474 \pm 0.019 \, \rm{mas/yr}$ \; \& \; $\mu_{\delta}=$$-67.765 \pm 0.016 \, \rm{mas/yr}$} \\
    COR07 RV Zero Point & Uniform & \multicolumn{2}{c}{$-20650.6 \pm 4.6 \, \rm{m/s}$} \\
    COR98 RV Zero Point & Uniform & \multicolumn{2}{c}{$-20605.9 \pmoffs{9.4}{10} \,\rm{m/s}$} \\
    COR14 RV Zero Point & Uniform & \multicolumn{2}{c}{$-20603 \pm 11 \, \rm{m/s}$} \\
    HARPS RV Zero Point & Uniform & \multicolumn{2}{c}{$0.8 \pmoffs{1.4}{0.63} \,\rm{m/s}$} \\
    RV jitter & Log-flat over [0,300 m/s] &  \multicolumn{2}{c}{$1.888 \pm 0.057 \, \rm{m/s}$} \\
    \hline
    Parameter & Prior Distribution & \multicolumn{2}{c}{Posterior $\pm$1$\sigma$} \\
    \hline
    $\sqrt{e} \sin \omega$ & Uniform over [-1, 1] & \multicolumn{2}{c}{$-0.029 \pm 0.021 \, $} \\
    $\sqrt{e} \cos \omega$ & Uniform over [-1, 1] & \multicolumn{2}{c}{$-0.578 \pm 0.022 \, $} \\
    Semi-major axis ($a$) & $1/a$ (log-flat) & \multicolumn{2}{c}{$12.4 \pmoffs{0.6}{0.5} \,\rm{A.U.}$} \\
    Inclination ($i$) & $\sin i$ (geometric) & \multicolumn{2}{c}{$45.1 \pmoffs{2}{1.8} \,\rm{degrees}$} \\
    PA of ascending node & Uniform & \multicolumn{2}{c}{$4.3 \pm 2 \, \rm{degrees}$} \\
    Mean Longitude at $t_{\rm ref}$ $(\lambda_{\rm ref})$ & Uniform & \multicolumn{2}{c}{$161.1 \pm 2 \, \rm{degrees}$} \\
    BD Mass $(M)$ & $1/M$ (log-flat) & \multicolumn{2}{c}{$36.2 \pmoffs{1.6}{1.5} \,\Mjup$} \\
    \hline
    Eccentricity ($e$) & (derived quantity) & \multicolumn{2}{c}{$0.34 \pm 0.03 \, $} \\
    Argument of Periastron $(\omega)$ & (derived quantity) & \multicolumn{2}{c}{$-177.1 \pm 2.1 \, \rm{degrees}$} \\
    Periastron Time $(T_0)$ & (derived quantity) & \multicolumn{2}{c}{$2456166 \pm 53 \, \rm{days}$} \\
    Period & (derived quantity) & \multicolumn{2}{c}{$16027 \pmoffs{830}{710} \,\rm{days}$} \\
     &  & \multicolumn{2}{c}{$43.9 \pmoffs{2.3}{1.9} \,\rm{years}$} \\
    \hline
    \orbitcodename Reference Epoch $(t_{\rm ref})$ & 2455197.50 BJD & \nodata & \nodata  
    \enddata
\end{deluxetable*}

\begin{deluxetable*}{ccccccc}
\tablewidth{0pt}
    \tablecaption{Posteriors of the HD19467 system.\label{tab:posteriors_HD19467}}
    \tablehead{
    \colhead{Parameter} & \colhead{Prior Distribution} & \multicolumn{2}{c}{{Posterior $\pm$1$\sigma$}}}
    \startdata
    Stellar mass & $0.953 \pm 0.022 \, \Msun$ & \multicolumn{2}{c}{$0.953 \pm 0.022 \, \Msun$} \\
    Parallax ($\varpi$) & $31.219 \pm 0.024 \, \rm{mas}$ (\gaia eDR3) & \multicolumn{2}{c}{$31.219 \pm 0.013 \, \rm{mas}$} \\
    Barycenter Proper Motions\tablenotemark{a}  & Uniform & \multicolumn{2}{c}{ $\mu_{\alpha}=$$-7.915 \pmoffs{0.087}{0.08} \,\rm{mas/yr}$ \; \& \; $\mu_{\delta}=$$-261.335 \pmoffs{0.095}{0.11} \,\rm{mas/yr}$} \\
    HIRES RV Zero Point & Uniform & \multicolumn{2}{c}{$-110 \pmoffs{170}{88} \,\rm{m/s}$} \\
    HARPS RV Zero Point & Uniform & \multicolumn{2}{c}{$2.9 \pmoffs{1.3}{0.37} \,\rm{m/s}$} \\
    RV jitter & Log-flat over [0,300 m/s] &  \multicolumn{2}{c}{$0.92 \pm 0.068 \, \rm{m/s}$} \\
    \hline
    Parameter & Prior Distribution & \multicolumn{2}{c}{Posterior $\pm$1$\sigma$} \\
    \hline
    $\sqrt{e} \sin \omega$ & Uniform over [-1, 1] & \multicolumn{2}{c}{$-0.67 \pmoffs{0.11}{0.09} \,$} \\
    $\sqrt{e} \cos \omega$ & Uniform over [-1, 1] & \multicolumn{2}{c}{$0.0 \pmoffs{0.3}{0.4} \,$} \\
    Semi-major axis ($a$) & $1/a$ (log-flat) & \multicolumn{2}{c}{$47 \pmoffs{18}{8.1} \,\rm{A.U.}$} \\
    Inclination ($i$) & $\sin i$ (geometric) & \multicolumn{2}{c}{$133 \pmoffs{14}{9.6} \,\rm{degrees}$} \\
    PA of ascending node & Uniform & \multicolumn{2}{c}{$31 \pmoffs{16}{98} \,\rm{degrees}$} \\
    Mean Longitude at $t_{\rm ref}$ $(\lambda_{\rm ref})$ & Uniform & \multicolumn{2}{c}{$186 \pmoffs{20}{120} \,\rm{degrees}$} \\
    BD Mass $(M)$ & $1/M$ (log-flat) & \multicolumn{2}{c}{$65.4 \pmoffs{5.9}{4.6} \,\Mjup$} \\
    \hline
    Eccentricity ($e$) & (derived quantity) & \multicolumn{2}{c}{$0.54 \pm 0.11 \, $} \\
    Argument of Periastron $(\omega)$ & (derived quantity) & \multicolumn{2}{c}{$-90 \pmoffs{24}{33} \,\rm{degrees}$} \\
    Periastron Time $(T_0)$ & (derived quantity) & \multicolumn{2}{c}{$2485464 \pmoffs{13000}{3100} \,\rm{days}$} \\
    Period & (derived quantity) & \multicolumn{2}{c}{$115283 \pmoffs{73000}{29000} \,\rm{days}$} \\
     &  & \multicolumn{2}{c}{$316 \pmoffs{200}{78} \,\rm{years}$} \\
    \hline
    \orbitcodename Reference Epoch $(t_{\rm ref})$ & 2455197.50 BJD & \nodata & \nodata  
    \enddata
\end{deluxetable*}

\begin{deluxetable*}{ccccccc}
\tablewidth{0pt}
    \tablecaption{Posteriors of the HD33632 system.\label{tab:posteriors_HD33632}}
    \tablehead{
    \colhead{Parameter} & \colhead{Prior Distribution} & \multicolumn{2}{c}{{Posterior $\pm$1$\sigma$}}}
    \startdata
    Stellar mass & $1.1 \pm 0.1 \, \Msun$ & \multicolumn{2}{c}{$1.086 \pm 0.092 \, \Msun$} \\
    Parallax ($\varpi$) & $37.895 \pm 0.026 \, \rm{mas}$ (\gaia eDR3) & \multicolumn{2}{c}{$37.8952 \pm 0.00547 \, \rm{mas}$} \\
    Barycenter Proper Motions\tablenotemark{a}  & Uniform & \multicolumn{2}{c}{ $\mu_{\alpha}=$$-144.935 \pmoffs{0.074}{0.067} \,\rm{mas/yr}$ \; \& \; $\mu_{\delta}=$$-134.99 \pmoffs{0.26}{0.24} \,\rm{mas/yr}$} \\
    Lick RV Zero Point & Uniform & \multicolumn{2}{c}{$1 \pmoffs{1.6}{0.74} \,\rm{m/s}$} \\
    RV jitter & Log-flat over [0,300 m/s] &  \multicolumn{2}{c}{$2.52 \pmoffs{0.19}{0.17} \,\rm{m/s}$} \\
    \hline
    Parameter & Prior Distribution & \multicolumn{2}{c}{Posterior $\pm$1$\sigma$} \\
    \hline
    $\sqrt{e} \sin \omega$ & Uniform over [-1, 1] & \multicolumn{2}{c}{$0.01 \pmoffs{0.2}{0.21} \,$} \\
    $\sqrt{e} \cos \omega$ & Uniform over [-1, 1] & \multicolumn{2}{c}{$0.1 \pm 0.37 \, $} \\
    Semi-major axis ($a$) & $1/a$ (log-flat) & \multicolumn{2}{c}{$23.6 \pmoffs{3.2}{4.5} \,\rm{A.U.}$} \\
    Inclination ($i$) & $\sin i$ (geometric) & \multicolumn{2}{c}{$45.2 \pmoffs{4.7}{11} \,\rm{degrees}$} \\
    PA of ascending node & Uniform & \multicolumn{2}{c}{$39.3 \pmoffs{5.7}{6.5} \,\rm{degrees}$} \\
    Mean Longitude at $t_{\rm ref}$ $(\lambda_{\rm ref})$ & Uniform & \multicolumn{2}{c}{$-158 \pmoffs{14}{9.5} \,\rm{degrees}$} \\
    BD Mass $(M)$ & $1/M$ (log-flat) & \multicolumn{2}{c}{$50 \pmoffs{5.6}{5} \,\Mjup$} \\
    \hline
    Eccentricity ($e$) & (derived quantity) & \multicolumn{2}{c}{$0.12 \pmoffs{0.18}{0.09} \,$} \\
    Argument of Periastron $(\omega)$ & (derived quantity) & \multicolumn{2}{c}{$-0 \pmoffs{86}{140} \,\rm{degrees}$} \\
    Periastron Time $(T_0)$ & (derived quantity) & \multicolumn{2}{c}{$2468815 \pmoffs{18000}{5800} \,\rm{days}$} \\
    Period & (derived quantity) & \multicolumn{2}{c}{$39178 \pmoffs{7900}{10000} \,\rm{days}$} \\
     &  & \multicolumn{2}{c}{$107 \pmoffs{21}{28} \,\rm{years}$} \\
    \hline
    \orbitcodename Reference Epoch $(t_{\rm ref})$ & 2455197.50 BJD & \nodata & \nodata  
    \enddata
\end{deluxetable*}

\begin{deluxetable*}{ccccccc}
\tablewidth{0pt}
    \tablecaption{Posteriors of the HD 33632 system from a joint fit including the stellar companion HD 33632 B.\label{tab:posteriors_HD33632_3body}}
    \tablehead{
    \colhead{Parameter} & \multicolumn{2}{c}{{Prior Distributions}} & \multicolumn{2}{c}{{Posteriors $\pm$1$\sigma$}}}
    \startdata
    Stellar mass & \multicolumn{2}{c}{$1.1 \pm 0.1 \, \Msun$} & \multicolumn{2}{c}{$1.084 \pm 0.085 \, \Msun$} \\
    Parallax ($\varpi$) & \multicolumn{2}{c}{$37.895 \pm 0.026 \, \rm{mas}$ (\gaia eDR3)} & \multicolumn{2}{c}{$37.896 \pm 0.026 \, \rm{mas}$} \\
    Barycenter Proper Motions\tablenotemark{a}  & \multicolumn{2}{c}{Uniform} & \multicolumn{2}{c}{ $\mu_{\alpha}=$$-144.88 \pmoffs{0.063}{0.059} \,\rm{mas/yr}$ \; \& \; $\mu_{\delta}=$$-135.71 \pmoffs{0.21}{0.2} \,\rm{mas/yr}$} \\
    Lick RV Zero Point & \multicolumn{2}{c}{Uniform} & \multicolumn{2}{c}{$3 \pmoffs{2.9}{1.7} \,\rm{m/s}$} \\
    RV jitter & \multicolumn{2}{c}{Log-flat over [0,300 m/s]} &  \multicolumn{2}{c}{$2.5 \pmoffs{0.2}{0.18} \,\rm{m/s}$} \\
    \hline \\
    \enskip
    Parameter & \multicolumn{2}{c}{Prior Distributions} & \shortstack{Posterior $\pm$1$\sigma$ \\ on BD companion} &  \shortstack{Posterior $\pm$1$\sigma$ \\ on stellar companion} \\
    \hline
    $\sqrt{e} \sin \omega$ & \multicolumn{2}{c}{Uniform over [-1, 1]} & $0 \pmoffs{0.18}{0.2} \,$ & $-0.32 \pmoffs{0.46}{0.22} \,$ \\
    $\sqrt{e} \cos \omega$ & \multicolumn{2}{c}{Uniform over [-1, 1]} & $0.18 \pmoffs{0.31}{0.36} \,$ & $-0.61 \pmoffs{0.4}{0.22} \,$ \\
    Semi-major axis ($a$) & \multicolumn{2}{c}{$1/a$ (log-flat)} & $23.3 \pmoffs{3}{3.9} \,\rm{A.U.}$ & $832 \pmoffs{250}{220} \,\rm{A.U.}$ \\
    Inclination ($i$) & \multicolumn{2}{c}{$\sin i$ (geometric)} & $44.8 \pmoffs{4.8}{10} \,\rm{degrees}$ & $74.6 \pmoffs{4.1}{11} \,\rm{degrees}$ \\
    PA of ascending node & \multicolumn{2}{c}{Uniform} & $39.4 \pm 5.3 \, \rm{degrees}$ &  $8.6 \pmoffs{7.4}{11} \,\rm{degrees}$ \\
    Mean Longitude at $t_{\rm ref}$ $(\lambda_{\rm ref})$ & \multicolumn{2}{c}{Uniform} & $-158 \pmoffs{10}{8.8} \,\rm{degrees}$ & $60 \pmoffs{32}{24} \,\rm{degrees}$ \\
    Mass $(M)$ & $1/M$ (log-flat) & $0.22 \pm 0.03 \, \Msun$ & $49.8 \pmoffs{5.5}{4.7} \,\Mjup$ & $0.215 \pm 0.029 \, \Msun$ \\
    \hline
    Eccentricity ($e$) & \multicolumn{2}{c}{(derived quantity)} & $0.12 \pmoffs{0.17}{0.09} \,$ &  $0.56 \pmoffs{0.27}{0.35} \,$ \\
    Argument of Periastron $(\omega)$ & \multicolumn{2}{c}{(derived quantity)} & $-1 \pmoffs{63}{120} \,\rm{degrees}$ &  $-1 \pmoffs{63}{120} \,\rm{degrees}$ \\
    Periastron Time $(T_0)$ & \multicolumn{2}{c}{(derived quantity)} & $2469028 \pmoffs{15000}{4900} \,\rm{days}$ &  $6934301 \pmoffs{4100000}{2800000} \,\rm{days}$ \\
    Period & \multicolumn{2}{c}{(derived quantity)} & $38522 \pmoffs{7300}{9000} \,\rm{days}$ &  $7715574 \pmoffs{3800000}{2900000} \,\rm{days}$ \\
     &  &  &  $105 \pmoffs{20}{25} \,\rm{years}$ &  $21124 \pmoffs{10000}{7900} \,\rm{years}$ \\
    \hline
    \orbitcodename Reference Epoch $(t_{\rm ref})$ & \multicolumn{2}{c}{2455197.50 BJD} & \nodata & \nodata  
    \enddata
\end{deluxetable*}

\begin{deluxetable*}{ccccccc}
\tablewidth{0pt}
    \tablecaption{Posteriors of the HD72946 system.\label{tab:posteriors_HD72946}}
    \tablehead{
    \colhead{Parameter} & \colhead{Prior Distribution} & \multicolumn{2}{c}{{Posterior $\pm$1$\sigma$}}}
    \startdata
    Stellar mass & $0.986 \pm 0.027 \, \Msun$ & \multicolumn{2}{c}{$0.987 \pm 0.026 \, \Msun$} \\
    Parallax ($\varpi$) & $38.981 \pm 0.041 \, \rm{mas}$ (\gaia eDR3) & \multicolumn{2}{c}{$38.9803 \pm 0.00707 \, \rm{mas}$} \\
    Barycenter Proper Motions\tablenotemark{a}  & Uniform & \multicolumn{2}{c}{ $\mu_{\alpha}=$$-133.965 \pm 0.091 \, \rm{mas/yr}$ \; \& \; $\mu_{\delta}=$$-136.888 \pm 0.041 \, \rm{mas/yr}$} \\
    SOPHIE RV Zero Point & Uniform & \multicolumn{2}{c}{$-29524 \pmoffs{12}{11} \,\rm{m/s}$} \\
    ELODIE RV Zero Point & Uniform & \multicolumn{2}{c}{$0.32 \pmoffs{0.11}{0.082} \,\rm{m/s}$} \\
    RV jitter & Log-flat over [0,300 m/s] &  \multicolumn{2}{c}{$2.64 \pm 0.10 \, \rm{m/s}$} \\
    \hline
    Parameter & Prior Distribution & \multicolumn{2}{c}{Posterior $\pm$1$\sigma$} \\
    \hline
    $\sqrt{e} \sin \omega$ & Uniform over [-1, 1] & \multicolumn{2}{c}{$0.6616 \pm 0.0080 \, $} \\
    $\sqrt{e} \cos \omega$ & Uniform over [-1, 1] & \multicolumn{2}{c}{$0.226 \pm 0.014 \, $} \\
    Semi-major axis ($a$) & $1/a$ (log-flat) & \multicolumn{2}{c}{$6.445 \pm 0.056 \, \rm{A.U.}$} \\
    Inclination ($i$) & $\sin i$ (geometric) & \multicolumn{2}{c}{$59.5 \pmoffs{1.2}{1.1} \,\rm{degrees}$} \\
    PA of ascending node & Uniform & \multicolumn{2}{c}{$167.9 \pm 2.6 \, \rm{degrees}$} \\
    Mean Longitude at $t_{\rm ref}$ $(\lambda_{\rm ref})$ & Uniform & \multicolumn{2}{c}{$23.93 \pm 0.66 \, \rm{degrees}$} \\
    BD Mass $(M)$ & $1/M$ (log-flat) & \multicolumn{2}{c}{$72.5 \pm 1.3 \, \Mjup$} \\
    \hline
    Eccentricity ($e$) & (derived quantity) & \multicolumn{2}{c}{$0.4889 \pm 0.0074 \, $} \\
    Argument of Periastron $(\omega)$ & (derived quantity) & \multicolumn{2}{c}{$71.1 \pm 1.2 \, \rm{degrees}$} \\
    Periastron Time $(T_0)$ & (derived quantity) & \multicolumn{2}{c}{$2455960 \pm 8 \, \rm{days}$} \\
    Period & (derived quantity) & \multicolumn{2}{c}{$5813 \pm 38 \, \rm{days}$} \\
     &  & \multicolumn{2}{c}{$15.92 \pm 0.1 \, \rm{years}$} \\
    \hline
    \orbitcodename Reference Epoch $(t_{\rm ref})$ & 2455197.50 BJD & \nodata & \nodata  
    \enddata
\end{deluxetable*}

\begin{deluxetable*}{ccccccc}
\tablewidth{0pt}
    \tablecaption{Posteriors of the HD 72946 system from a joint fit including the stellar companion HD 72945.\label{tab:posteriors_HD72945_3body}}
    \tablehead{
    \colhead{Parameter} & \multicolumn{2}{c}{{Prior Distributions}} & \multicolumn{2}{c}{{Posteriors $\pm$1$\sigma$}}}
    \startdata
    Stellar mass & \multicolumn{2}{c}{$0.986 \pm 0.027 \, \Msun$} & \multicolumn{2}{c}{$0.997 \pm 0.026 \, \Msun$} \\
    Parallax ($\varpi$) & \multicolumn{2}{c}{$38.981 \pm 0.041 \, \rm{mas}$ (\gaia eDR3)} & \multicolumn{2}{c}{$38.982 \pm 0.041 \, \rm{mas}$} \\
    Barycenter Proper Motions\tablenotemark{a}  & \multicolumn{2}{c}{Uniform} & \multicolumn{2}{c}{ $\mu_{\alpha}=$$-132.12 \pmoffs{0.13}{0.12} \,\rm{mas/yr}$ \; \& \; $\mu_{\delta}=$$-135.03 \pm 0.11 \, \rm{mas/yr}$} \\
    SOPHIE RV Zero Point & \multicolumn{2}{c}{Uniform} & \multicolumn{2}{c}{$-30347 \pmoffs{1800}{380} \,\rm{m/s}$} \\
    ELODIE RV Zero Point & \multicolumn{2}{c}{Uniform} & \multicolumn{2}{c}{$1.6 \pmoffs{2.3}{1.1} \,\rm{m/s}$} \\
    RV jitter & \multicolumn{2}{c}{Log-flat over [0,300 m/s]} &  \multicolumn{2}{c}{$2.65 \pmoffs{0.11}{0.097} \,\rm{m/s}$} \\
    \hline \\
    \enskip
    Parameter & \multicolumn{2}{c}{Prior Distributions} & \shortstack{Posterior $\pm$1$\sigma$ \\ on BD companion} &  \shortstack{Posterior $\pm$1$\sigma$ \\ on stellar companion} \\
    \hline
    $\sqrt{e} \sin \omega$ & \multicolumn{2}{c}{Uniform over [-1, 1]} & $0.651 \pm 0.00803 \, $ & $0.25 \pmoffs{0.21}{0.32} \,$ \\
    $\sqrt{e} \cos \omega$ & \multicolumn{2}{c}{Uniform over [-1, 1]} & $0.251 \pmoffs{0.012}{0.013} \,$ & $0.22 \pmoffs{0.5}{0.91} \,$ \\
    Semi-major axis ($a$) & \multicolumn{2}{c}{$1/a$ (log-flat)} & $6.514 \pm 0.056 \, \rm{A.U.}$ & $200 \pmoffs{52}{41} \,\rm{A.U.}$ \\
    Inclination ($i$) & \multicolumn{2}{c}{$\sin i$ (geometric)} & $62.8 \pm 1.4 \, \rm{degrees}$ & $95.5 \pmoffs{2.7}{1.2} \,\rm{degrees}$ \\
    PA of ascending node & \multicolumn{2}{c}{Uniform} & $174.2 \pmoffs{2.7}{2.9} \,\rm{degrees}$ &  $27 \pmoffs{180}{1.5} \,\rm{degrees}$ \\
    Mean Longitude at $t_{\rm ref}$ $(\lambda_{\rm ref})$ & \multicolumn{2}{c}{Uniform} & $23.01 \pmoffs{0.66}{0.61} \,\rm{degrees}$ & $190 \pmoffs{14}{160} \,\rm{degrees}$ \\
    Mass $(M)$ & $1/M$ (log-flat) & $1.1 \pm 0.1 \, \Msun$ & $72.1 \pmoffs{1.4}{1.3} \,\Mjup$ & $1.089 \pm 0.094 \, \Msun$ \\
    \hline
    Eccentricity ($e$) & \multicolumn{2}{c}{(derived quantity)} & $0.4871 \pmoffs{0.007}{0.00786} \,$ &  $0.44 \pmoffs{0.31}{0.26} \,$ \\
    Argument of Periastron $(\omega)$ & \multicolumn{2}{c}{(derived quantity)} & $68.9 \pm 1.1 \, \rm{degrees}$ &  $68.9 \pm 1.1 \, \rm{degrees}$ \\
    Periastron Time $(T_0)$ & \multicolumn{2}{c}{(derived quantity)} & $2455947.1 \pm 8.3 \, \rm{days}$ &  $2835898 \pmoffs{280000}{320000} \,\rm{days}$ \\
    Period & \multicolumn{2}{c}{(derived quantity)} & $5880 \pm 43 \, \rm{days}$ &  $715396 \pmoffs{300000}{210000} \,\rm{days}$ \\
     &  &  &  $16.1 \pm 0.12 \, \rm{years}$ &  $1959 \pmoffs{820}{570} \,\rm{years}$ \\
    \hline
    \orbitcodename Reference Epoch $(t_{\rm ref})$ & \multicolumn{2}{c}{2455197.50 BJD} & \nodata & \nodata  
    \enddata
\end{deluxetable*}

\end{document}